\documentclass[opre,nonblindrev]{informs4rad}

\newif\ifblind
\blindfalse  

\OneAndAHalfSpacedXI

\newif\iffigdraft

\figdraftfalse

\usepackage{xcolor}
\usepackage{mathtools}
\usepackage{amsmath}
\usepackage{subcaption}
\usepackage{tikz}
\usepackage{booktabs}
\usepackage{tabularx}
\usetikzlibrary{arrows.meta,shapes.misc,positioning,decorations.pathmorphing}
\usetikzlibrary{calc}

\definecolor{myred}{RGB}{200,0,0}
\definecolor{myblue}{RGB}{0,120,215}
\definecolor{mypurple}{RGB}{140,60,160}
\definecolor{mygreen}{RGB}{0,110,80}
\definecolor{darkblue}{HTML}{1A3A6E}
\definecolor{skyblue}{HTML}{2F80ED}
\definecolor{softblue}{HTML}{D9ECFF}
\definecolor{midblue}{HTML}{C6D4E8}
\definecolor{deepgreen}{HTML}{1B6B45}
\definecolor{okgreen}{HTML}{169B62}
\definecolor{softgreen}{HTML}{DFF5E8}
\definecolor{warmorange}{HTML}{D77836}
\definecolor{softorange}{HTML}{FBE7CC}
\definecolor{punchred}{HTML}{B5311E}
\definecolor{softred}{HTML}{FBE2DA}
\definecolor{maroon}{HTML}{800000}
\definecolor{muted}{HTML}{6B7280}
\definecolor{softgray}{HTML}{F2F4F7}
\definecolor{ink}{HTML}{1F2937}
\definecolor{parchment}{HTML}{EFE7D2}
\definecolor{antiquegold}{HTML}{8C6F3F}

\usepackage{mathtools, bm, enumitem}
\usepackage{xcolor}

\renewcommand{\qed}{$\hfill\square$}

\newenvironment{proofof}[1]{%
  \Trivlist
  \item[\hskip\labelsep {\it #1.}]\ignorespaces
}{\hfill \qed
\endTrivlist
\addvspace{0pt}
}

\TheoremsNumberedThrough 
\EquationsNumberedThrough

\usepackage{xfrac}
\usepackage[normalem]{ulem}
\usepackage{wasysym}
\usepackage{wrapfig}
\usepackage[english]{babel}
\usepackage[skip=2pt]{caption}
\usepackage{thm-restate}
\usepackage[ruled,vlined,linesnumbered]{algorithm2e}
\SetNoFillComment

\usepackage{scalefnt}

\usepackage[breakable,skins]{tcolorbox}
\newtcolorbox{myboxx}[3][]
{
  colframe = white,
  colback  = #2!10,
  left=0pt,
  right=0pt,
  top=0pt,
  bottom=0pt,
  enlarge left by=0mm,
  boxsep=10pt,
  arc=0pt,outer arc=0pt,
  #1,
}

\colorlet{shadecolor}{gray!50}

\usepackage{csquotes}

\makeatletter

\renewenvironment*{displayquote}
  {\begingroup\setlength{\leftmargini}{0cm}\csq@getcargs{\csq@bdquote{}{}}}
  {\csq@edquote\endgroup}
\makeatother

\usepackage[hypertexnames=false]{hyperref}
\hypersetup{
	colorlinks = true,
	linkcolor=royalazure,
	urlcolor=royalazure,
	citecolor=royalazure,
	linkbordercolor = {white}
}

\usepackage{mathtools}

\definecolor{cornellred}{rgb}{0.7, 0.11, 0.11}
\definecolor{maroon}{rgb}{0.52, 0, 0}
\definecolor{dgreen}{rgb}{0.0, 0.5, 0.0}
\definecolor{ballblue}{rgb}{0.13, 0.67, 0.8}
\definecolor{royalblue(web)}{rgb}{0.25, 0.41, 0.88}
\definecolor{bleudefrance}{rgb}{0.19, 0.55, 0.91}
\definecolor{royalazure}{rgb}{0.0, 0.22, 0.66}

\usepackage{cleveref}

\usepackage{enumitem}
\usepackage{multirow}

\usepackage{pgf,tikz,pgfplots}
\pgfplotsset{compat=1.15}

\usetikzlibrary{arrows,calc}
\usepackage{pgfplots}
\usetikzlibrary{patterns}
\usepgfplotslibrary{fillbetween}
\usetikzlibrary{intersections}

\usetikzlibrary{arrows, decorations.markings}

\tikzstyle{vecArrow} = [thick, decoration={markings,mark=at position
	1 with {\arrow[semithick]{open triangle 60}}},
	double distance=1.4pt, shorten >= 5.5pt,
	preaction = {decorate},
postaction = {draw,line width=1.4pt, white,shorten >= 4.5pt}]
\tikzstyle{innerWhite} = [semithick, white,line width=1.4pt, shorten >= 4.5pt]

	\usepackage{accents}

	\usepackage[authoryear,round]{natbib}

\usepackage[showdeletions]{color-edits}

\addauthor{MA}{brown}
\addauthor{RN}{maroon}

\newcommand*{\rom}[1]{\expandafter\romannumeral #1}
\newcommand{\Rom}[1]{\uppercase\expandafter{\romannumeral #1\relax}}

	\newcommand*{\R}{\mathbb{R}}

\newcommand{\prob}[2][]{\text{\bf Pr}\ifthenelse{\not\equal{}{#1}}{_{#1}}{}\!\left[{\def\givenn{\middle|}#2}\right]}
\newcommand{\expect}[2][]{\text{\bf E}\ifthenelse{\not\equal{}{#1}}{_{#1}}{}\!\left[{\def\givenn{\middle|}#2}\right]}

	\providecommand{\given}{}
	\DeclarePairedDelimiterX{\set}[1]\{\}{\renewcommand\given{\nonscript\:\delimsize\vert\nonscript\:\mathopen{}}#1}
	\let\Pr\relax
	\DeclarePairedDelimiterXPP{\Pr}[1]{\mathbb{P}}[]{}{\renewcommand\given{\nonscript\:\delimsize\vert\nonscript\:\mathopen{}}#1}
	\DeclarePairedDelimiterXPP{\Ex}[1]{\mathbb{E}}[]{}{\renewcommand\given{\nonscript\:\delimsize\vert\nonscript\:\mathopen{}}#1}

\newcolumntype{P}[1]{>{\centering\arraybackslash}c{#1}}

\usepackage{cleveref}

\usepackage{csquotes}
\makeatletter
\renewenvironment*{displayquote}
  {\begingroup\setlength{\leftmargini}{0.1cm}\csq@getcargs{\csq@bdquote{}{}}}
  {\csq@edquote\endgroup}
\makeatother

\newcommand{\revcolor}[1]{{\color{black}#1}}

\newcommand{\xhdr}[1]{\smallskip \noindent{\bf #1}}

\newcommand{\reward}{r}

\newcommand{\arrivalRate}{\lambda}

\newcommand{\PP}{\mathbb{P}}
\newcommand{\EE}{\mathbb{E}}
\newcommand{\Ind}[1]{\mathbf{1}\{#1\}}
\renewcommand{\given}{\;\vert\;}
\newcommand{\defeq}{\mathrel{\mathop:}=}

\newcommand{\ground}{E}                 %
\newcommand{\Fset}{\mathcal{F}}         %
\newcommand{\Ppoly}{\mathcal{P}}        %
\newcommand{\simplex}[1]{\Delta(#1)}    %

\newcommand{\xvec}{\bm{x}}              %
\newcommand{\pvec}{\bm{p}}              %
\newcommand{\wvec}{\bm{w}}              %
\newcommand{\thetavec}{\bm{\theta}}     %
\newcommand{\alphaStar}{\alpha^{*}} %

\newcommand{\Aset}{A}                    %
\newcommand{\Sset}{S}                    %
\newcommand{\dist}{\mu}                 %

\newcommand{\Entropy}{\mathsf{H}}        %
\newcommand{\Zpart}{\mathsf{Z}}          %
\newcommand{\ip}[2]{\langle #1,#2\rangle}

\newcommand{\Add}{\mathsf{Add}}          %
\newcommand{\Occ}{\mathsf{Matched}}          %

\newcommand{\eps}{\varepsilon}

\newcommand{\Sm}[2]{#1_{-#2}}            %

\providecommand{\conv}{\operatorname{conv}} %
\providecommand{\KL}{\mathrm{KL}}           %

\providecommand{\cM}{\mathcal{M}}           %
\providecommand{\Iset}{\mathcal{I}}         %
\providecommand{\Bset}{\mathcal{B}}         %
\providecommand{\Bpoly}{\mathcal{P}_{\Bset}}%

\providecommand{\RankOracle}[1]{\textrm{rank}_\cM(#1)}

\providecommand{\qvec}{\bm{q}}              %
\providecommand{\yvec}{\bm{y}}              %

\newcommand{\Match}{\mathcal{M}}

\providecommand{\Bpoly}{\mathcal{P}_{\Bset}}
\providecommand{\KL}{\mathrm{KL}}

\hypersetup{linktoc=all,bookmarksnumbered=true}
\providecommand{\rev}[1]{\textcolor{black}{#1}}

\MANUSCRIPTNO{}

\begin{document}

\RUNAUTHOR{Aminian, Niazadeh,  Nuti}

\RUNTITLE{Stationary Online Contention Resolution Schemes}

\TITLE{%
  \resizebox{\textwidth}{!}{Stationary Online Contention Resolution Schemes:}\\
  \resizebox{\textwidth}{!}{Theory and Applications to Bayesian Online Resource Allocation}%
}

\ARTICLEAUTHORS{%
  \AUTHOR{Mohammad Reza Aminian\qquad Rad Niazadeh\qquad Pranav Nuti}
  \AFF{The University of Chicago, Booth School of Business, Chicago, IL}
    
}

\renewcommand{\ABSfont}{%
  \EGT
  \fontsize{10}{14}\selectfont
  \setlength{\parskip}{4pt}%
}
\newpage
\ABSTRACT{

In many modern applications involving online platforms and digital marketplaces, the data-driven allocation of supply to demand involves new difficulties such as reusable inventory and complex combinatorial feasibility constraints governing simultaneous allocations. Such settings can be significantly harder than classical resource allocation problems because the state space is large—the platform potentially has to keep track of when each allocated product will return to inventory. Viewing these settings as instances of a Bayesian online allocation problem raises a natural question: might there be a universal technical framework for designing algorithmic solutions across all of these settings? Online contention resolution schemes (OCRSs) aim to provide such solutions by converting ex-ante relaxations into approximately optimal feasible online policies. Despite their importance---and relative universality---existing constructions of (near-)optimal OCRSs are often technically challenging. Many constructions rely on indirect reductions to prophet inequalities and LP duality, resulting in algorithms that are ad hoc and difficult to interpret or implement. Moreover, it is unclear whether such vanilla OCRSs can accommodate challenges such as resource reusability.

In this paper, we introduce \emph{stationary online contention resolution schemes (S-OCRSs)}, a permutation-invariant class of OCRSs in which the distribution of the selected feasible set is independent of the arrival order. We establish a distributional characterization of S-OCRSs and provide a universal online implementation, thereby reducing the design of online policies to the construction of suitable distributions over feasible sets. This reduction applies to arbitrary, potentially complex feasibility environments. We further show that S-OCRSs can be used to solve reusable resource allocation problems, and develop a general `maximum-entropy-based' approach for constructing and analyzing S-OCRSs across a variety of settings. Together, these results provide a universal and principled technical framework for designing simple and potentially improved OCRSs, yielding approximately optimal online allocation algorithms for a wide range of modern Bayesian online allocation problems.

We demonstrate the power of our new framework across several canonical feasibility environments. For bipartite matchings, we obtain an improved $(3-\sqrt{5})/2$-selectable OCRS, attaining the benchmark conjectured to be optimal for OCRSs. For knapsack constraints, we obtain a $1/4$-selectable S-OCRS. We also obtain a $1-\sqrt{2/(\pi k)}+O(1/k)$-selectable S-OCRS for $k$-uniform matroids; a simple, explicit $1/2$-selectable S-OCRS for weakly Rayleigh matroids, including all $\mathbb{C}$-representable matroids such as graphic and laminar matroids; a $1-\sqrt{2\log L/k}-O(1/\sqrt{k})$-selectable S-OCRS for rank-$L$ hypergraph matchings with capacity $k$ (also known as the network revenue management setting); and a $\Theta(1/\sqrt{N})$-selectable S-OCRS for general downward-closed constraints on sets of size $N$. These latter guarantees match the best-known bounds for OCRSs via simple and principled algorithms, in contrast to existing algorithms that are either ad-hoc, complex, or non-explicit. All of the aforementioned S-OCRS guarantees for different feasibility environments yield approximation factors for the Bayesian online allocation problems of reusable products in the same feasibility environments. Our results extend from the resource allocation setting to assortment optimization setting and our policies can be implemented using prices.

}

\maketitle

\clearpage
\begingroup
\SingleSpacedXI
\fontsize{11}{15}\selectfont
\setlength{\parskip}{0pt}
\hypersetup{linkcolor=black}
\newcommand{\TOCsectionnumberwidth}{1.8em}
\newcommand{\TOCsubsectionindent}{1.8em}
\newcommand{\TOCsubsectionnumberwidth}{2.6em}
\makeatletter
\let\ref\@refstar
\renewcommand{\@pnumwidth}{3em}
\renewcommand{\@tocrmarg}{4em}
\renewcommand{\@dotsep}{6.5}
\renewcommand*{\l@section}[2]{%
  \addpenalty{\@secpenalty}%
  \addvspace{0.7em}%
  \@dottedtocline{1}{0em}{\TOCsectionnumberwidth}%
    {{\bfseries\hypersetup{linkcolor=darkblue}#1}}%
    {{\hypersetup{linkcolor=darkblue}#2}}%
  \nobreak
}
\renewcommand*{\l@subsection}[2]{%
  \@dottedtocline{2}{\TOCsubsectionindent}{\TOCsubsectionnumberwidth}%
    {#1}{{\hypersetup{linkcolor=darkblue}#2}}%
}
\renewcommand{\tableofcontents}{%
  \section*{Table of \contentsname}%
  \noindent{\sffamily\small\color{muted}Main paper}\par
  \vspace{0.3em}%
  \@starttoc{toc}%
}
\makeatother
\newcommand{\TOCcompanionbreak}{%
  \clearpage
  \renewcommand{\TOCsectionnumberwidth}{3.5em}%
  \renewcommand{\TOCsubsectionindent}{1.5em}%
  \renewcommand{\TOCsubsectionnumberwidth}{4.5em}%
  \pdfbookmark[2]{Electronic companion contents}{contents-ec}%
  \section*{Table of Contents (continued)}%
  \noindent{\sffamily\small\color{muted}Electronic companion}\par
  \vspace{0.3em}%
}
\pdfbookmark[1]{Contents}{contents}
\tableofcontents
\clearpage
\endgroup

\setlength{\parskip}{4pt plus 1pt minus 1pt}
\raggedbottom

\section{Introduction}
\label{sec:intro}
\providecommand{\rev}[1]{\textcolor{black}{#1}}
In many modern applications involving online platforms and digital marketplaces, the data-driven, real-time allocation of supply to demand gives rise to  challenges that require novel technical treatments. These challenges often stem from constraints on feasible allocations---typically imposed by marketplace rules or by compatibility restrictions between the demand and supply sides of the platform---or from specific properties of the resources being allocated. One class of examples comes from \emph{combinatorial feasibility environments}, including matching, knapsack, network revenue management, and complex matroid constraints. Such environments arise in applications such as two-sided matching platforms, crowdsourcing marketplaces, and revenue management over complex networks. Another class of examples involves environments with \rev{\emph{reusable} (or \emph{recurring}) resources}, where each resource unit is occupied for some duration before becoming available again \rev{to serve future requests}. Models with reusable resources arise prominently in online revenue management and assortment optimization~\citep{rusmevichientong2023revenue,rusmevichientong2020dynamic,feng2024near,goyal2025asymptotically,feng2025robustness,ekbatani2025online}, and are motivated by applications including cloud and edge computing platforms that allocate compute capacity or GPU memory to jobs, digital rental marketplaces, volunteer matching platforms, and, more generally, service systems in which capacity cycles between busy and idle states. 

Viewing these settings as instances of a Bayesian online resource allocation problem raises a natural question: might there be a \emph{universal} technical framework for designing algorithmic solutions across all these settings? A recurring algorithmic theme in operations research and related work on Bayesian online decision making---particularly Bayesian online resource allocation---is to approach problems such as those described above through the following (potentially universal) design recipe. First, one computes an \emph{ex-ante} feasible plan that satisfies the relevant feasibility constraints---arising from the combinatorial structure of the allocations, the nature of the resources, or both---only in expectation. One then converts this plan into an online policy that maintains \emph{ex-post} feasibility in every realization while incurring only a small loss, relative to the ex-ante plan, in the expected reward of the allocation. This ex-ante-to-ex-post conversion is central to problems in dynamic resource allocation, revenue management, and mechanism design, where decisions must be made sequentially subject to hard feasibility constraints.\footnote{It also underlies several well-studied problem classes at the intersection of operations research, economics, and computer science, including prophet inequalities~\citep{krengel1978semiamarts,samuel1984comparison, kleinberg2012matroid, dutting2020prophet, ezra2022prophet} (see \cite{lucier2017economic} for a comprehensive survey), Bayesian online resource allocation~\citep{alaei2012online,alaei2014bayesian,feng2022near,anari2019nearly}, sequential posted-price mechanisms~\citep{chawla2010multi,correa2017posted,yan2011mechanism, pollner2022improved}, and, more broadly, dynamic resource-constrained reward collection models~\citep{balseiro2024survey}.}

Online contention resolution schemes (OCRSs)---studied by \citet{alaei2014bayesian} for the special case of $k$-uniform matroids and later introduced for general feasibility environments by \citet{feldman2016online,feldman2021online}---provide a principled framework for carrying out this conversion. Fix a downward-closed feasibility family $\Fset\subseteq 2^{\ground}$ over a ground set $\ground$, and let $
\Ppoly\defeq \conv\{\mathbf{1}_S\in\Fset\}$
denote its associated polytope. Given a fractional point $\xvec\in\Ppoly$, each element $e\in\ground$ is independently ``active'' with probability $\PP[e\text{ active}]=x_e$, and the activations are revealed online in an arrival order that may be adversarial. An OCRS is an online policy that irrevocably selects a feasible set $\Sset\in\Fset$ of active elements while seeking to guarantee sufficient \emph{selectability} for every element. Specifically, the scheme is $\alpha$-selectable if, for every arrival order and every element $e$,
\[
\PP[e\in\Sset \mid e\text{ is active}]\ \ge\ \alpha,
\quad\text{or, equivalently,}\quad
\PP[e\in\Sset]\ \ge\ \alpha x_e.
\]
An OCRS can therefore be viewed as a black-box online rounding primitive that preserves each marginal probability up to a common factor. This is precisely the type of guarantee needed to convert an ex-ante solution---for example, an LP-based allocation rule for a resource allocation problem that is feasible only in expectation---into an implementable online allocation.\footnote{This perspective also explains the close connection between OCRSs and prophet-inequality-style guarantees obtained through ex-ante relaxations~\citep{lee2018optimal}.} Owing to their versatility, OCRSs have served as a modular tool in Bayesian online resource allocation and \rev{mechanism design} since their introduction.

Designing OCRSs, however, is often technically delicate, and two major issues arise when one attempts to use them as the algorithmic engine of the universal framework envisioned above:
\begin{enumerate}[leftmargin=2em, label=(\Roman*)]
    \item The first concerns the design process itself. A fundamental obstruction is the intrinsic asymmetry of online information: elements arriving early must compete against an unknown future, whereas elements arriving late face a system that has already committed some of its capacity. Consequently, many OCRS designs and analyses depend essentially on the arrival order, and the resulting algorithms can be \rev{complex, ad hoc,} and specific to the underlying feasibility environment\rev{---rather than following a universal design template}. Moreover, for certain feasibility environments, such as matroids, the best-known guarantees are obtained only indirectly through computationally intensive implicit algorithms derived via reductions to prophet inequalities with respect to ex-ante relaxations~\citep{lee2018optimal}.\footnote{These reductions and implicit constructions rely on duality and minimax arguments, together with algorithms for solving the resulting minimax problems. They typically produce a randomized online policy by applying first-order optimization procedures, such as multiplicative-weights updates, over the \emph{space of online policies}; see \cite{lee2018optimal} for details.} Although such reductions are powerful, they can obscure the eventual rounding rule and make it difficult to characterize the distribution over feasible outcomes that the algorithm implicitly implements. From an operations research perspective, this lack of transparency can limit interpretability when the OCRS is used as a building block within a larger online control policy.
    \item The second issue concerns the reach of the OCRS abstraction itself. Standard OCRSs are inherently \emph{one-shot}: each element arrives exactly once, in some possibly adversarial order, and, once selected, occupies its share of the resources until the end of the horizon. In the reusable environments highlighted at the outset, by contrast, allocated capacity is released over time as new requests continue to arrive. Thus, there is no single arrival order to speak of: the pattern of contention among requests keeps changing as resource units are occupied and released. It is therefore unclear whether vanilla OCRSs, whose selectability guarantees are stated with respect to a single arrival of each element, can accommodate resource reusability.
\end{enumerate}

\subsection{Our conceptual contribution: ``Stationary OCRS''}
This paper explores a novel design principle for OCRSs by introducing and studying a symmetry requirement that directly addresses the ``first vs.\ last'' asymmetry \rev{(and, as we will see, the reusability challenge)}: \emph{permutation invariance} of the output distribution, that is, the distribution of the selected set should be independent of the online arrival order of the elements. Formally, for every fixed $\xvec\in\Ppoly$, we require the existence of a single distribution $\dist^{(\xvec)}$ over feasible sets $S\in\Fset$ such that, under every arrival order, the selected set $\Sset$ is distributed according to $\dist^{(\xvec)}$. We call an OCRS satisfying this symmetry requirement a \emph{stationary OCRS (S-OCRS)}. Stationarity can be interpreted as requiring the scheme to ``treat every element as if it were the last''.

A natural question is why stationary OCRSs merit study. Beyond their conceptual appeal, stationarity directly addresses the two issues discussed above. First, we posit that even though imposing symmetry through stationarity restricts the class of admissible policies, it introduces additional structure that facilitates the design of simple---and, in some cases, improved or (near-)optimal---OCRS algorithms that arise from a universal design recipe. Second, it turns out that stationarity is an extremely useful property to address the reusable-resource settings introduced at the outset. Recent work uses OCRSs as admission-control subroutines to convert solutions to ex-ante LP benchmarks into ex-post feasible and approximately optimal online policies in these settings~\citep{feng2022near,feng2024near,baek2022bifurcating}. In such models, as noted earlier, a single resource may repeatedly encounter dynamically reordered contention processes, making permutation-invariant admission-control primitives a natural choice; see \Cref{fig:chain-ocrs} for an illustration of this contention pattern.\footnote{Besides the main reasons we study S-OCRS, we anticipate our framework has further applications. For instance, another consideration that recurs in the theory and applications of Bayesian online resource allocation is fairness and robustness when agents can influence \emph{timing}. Order-dependent rules can create incentives for groups of friendly agents to manipulate their arrival times, because their joint distribution of allocations is a function of the arrival order. In contrast, a stationary OCRS removes any advantage that is purely due to the arrival times. This is in spirit related to time-strategy-proofness or order incentive compatibility notions studied, e.g., in the context of secretary problems and online auctions~\citep{buchbinder2009secretary,hajiaghayi2004adaptive}.}

Our perspective gives rise to a set of fundamental technical questions, which we address in this paper:

\smallskip
\begin{displayquote}\itshape

~~~~(i)~Can we design stationary OCRSs with constant-factor selectability for natural feasibility environments, such as matching, matroid, and knapsack constraints? In important special cases, does imposing stationarity lead to simpler or even improved OCRSs?

\smallskip
(ii) Is there a universal recipe for designing stationary OCRSs across a broad class of feasibility environments? Can the resulting schemes be made explicit---that is, can they yield concrete, implementable online algorithms rather than merely indirect existence proofs?

\smallskip
(iii) Can stationary OCRSs help address resource reusability? Specifically, is there a universal recipe for designing policies for the Bayesian online allocation of reusable products (under stationary request arrivals) in any feasibility environment (e.g., $k$-uniform matroids, matchings, or general matroids)---for instance, by reducing the problem to the design of stationary OCRSs in the same feasibility environment?
\end{displayquote}

\smallskip

\subsection{Our main technical contributions}
We introduce a new conceptual and technical framework for designing and analyzing S-OCRSs and OCRSs. Our results provide affirmative answers to all the questions posed above, with a particular emphasis on explicit distributions and explicit (near-)optimal algorithms\rev{---culminating in a formal reduction from Bayesian online allocation with reusable products to the design of S-OCRSs}.

\xhdr{Distributional LP characterization \& meta-algorithm.}
We begin by giving a precise characterization of stationary OCRSs via a linear feasibility problem over distributions on feasible sets, which we call the \emph{stationary OCRS LP}. In addition to the marginal (selectability) constraints—requiring $\PP_{\Sset\sim\dist^{(\xvec)}}[e\in\Sset]\ge \alpha x_e$ for all $e$—this LP imposes a strengthened \emph{stationary implementability} condition. Unlike standard OCRSs, where implementability is defined by conditioning on an arrival prefix, stationarity requires conditioning on the selection outcomes of \emph{all other} elements. This provides a mathematical formalization of the principle of ``treating every element as if it might be the last.'' We show that the stationary OCRS LP is exact: a distribution is feasible in this LP if and only if it can be implemented by a stationary OCRS.

A key algorithmic consequence of our LP characterization is a universal implementation procedure. Given any feasible witness distribution $\dist^{(\xvec)}$ for the stationary OCRS LP, we provide a simple \emph{simulate-then-replace} meta-algorithm (sharing similarities with a Gibbs sampling update or single-site Glauber dynamics, see \cite{levin2017markov} for instance) that implements $\dist^{(\xvec)}$ for \emph{every} arrival order. 

The algorithm maintains a random set $\widehat{S}$ (initialized as a draw from $\dist^{(\xvec)}$ at time $0$) whose distribution is preserved throughout the execution. $\widehat{S}$ consists of a “real” part containing the elements selected so far and an “imaginary” part representing a random subset of future elements. When an element arrives, its membership in $\widehat{S}$ is resampled using the appropriate conditional probability under $\dist^{(\xvec)}$ and in consistency with its activation. 

As a result, once a distribution $\dist^{(\xvec)}$ is specified, the online policy is essentially fixed, and the design of a stationary OCRS reduces to finding a ``good'' feasible solution to the stationary OCRS LP. Accordingly, most of the paper focuses on constructing explicit witness distributions with constant selectability factors.

\xhdr{Maximum-entropy template \& the addability analysis.}
Our main distributional construction is based on a maximum-entropy principle. Fixing target marginals $\pvec=\alpha\xvec$, where $\alpha\in(0,1)$ is the selectability factor of interest, we consider the maximum-entropy distribution $\dist^*$ over feasible sets with these marginals. For downward-closed systems, this choice leads to a crucial structural simplification: the resulting distribution over feasible sets is a \emph{Gibbs distribution} (defined later; cf.~\cite{singh2014entropy, gharan2011randomized}), and has a product form with weights. In particular, the conditional inclusion probability of an element depends only on its weight and on whether the element is \emph{addable} to the current set. As a result, feasibility of $\dist^*$ for the stationary OCRS LP with parameter $\alpha$ reduces to verifying a single, intuitive condition—namely, lower bounding the probability that a random feasible set drawn from $\dist^*$ allows for the addition of any given element with probability $\alpha$. This  reduction converts the analysis of a stationary OCRS into a static probabilistic statement, and it is the main technical engine behind our results for matchings and uniform matroids.

\xhdr{Comparison with greedy OCRSs.}
The greedy OCRSs of \citet{feldman2016online,feldman2021online} are also analyzed using a related notion of addability. They fix a random downward-closed subfamily of feasible sets, start from the empty set, and accept an active element whenever it can be added to the set selected so far. Our approach uses the same feasibility test, but changes the state on which it is applied---the simulate-then-replace policy starts from a simulated imaginary state $\widehat{S}$ drawn from a target distribution rather than the empty set. Furthermore, unlike that paper, we allow the probability with which we discard feasible elements to depend on our past decisions. For Gibbs distributions, our algorithm can be viewed as a greedy policy (accepting elements with a fixed probability when they can be added to $\widehat{S}_{-e}$) initialized at a good imaginary state.\footnote{We will always assume that the order in which we observe elements is not influenced by this initially sampled imaginary state.} Our results will explain why greedy behavior performs significantly better when started from the correct stationary state.

\xhdr{Stationary OCRSs for matchings \& uniform matroids.}
We instantiate the maximum-entropy framework in several canonical feasibility environments and obtain simple, explicit stationary OCRSs. For (non-bipartite) matchings, we obtain a clean $\tfrac{1}{3}$-selectable S-OCRS (and extend it to rank-$L$ hypergraph matchings to obtain a $\tfrac{1}{L+1}$-selectable S-OCRS, matching the usual benchmark for this environment\rev{;} see \cite{ma2026online, correa2022optimal}). For bipartite matchings, we obtain an explicit  selectability constant 
\[
\alpha^* = \frac{3-\sqrt{5}}{2},
\]
and we show that this constant is optimal within the stationary class. This result provides an improved OCRS for this environment, and therefore settles an open problem in \cite{ma2024onlineM}. The factor is also optimal for a natural class of OCRSs (satisfying a concentration property), and is conjectured to be optimal in general. This result also provides the best known competitive ratio for the bipartite matching prophet inequality problem\rev{,} improving on \cite{gravin2023prophet,ezra2022prophet, macrury2025random}. Our proof highlights the advantage of the distributional viewpoint over previous approaches. In our analysis, the addability of an edge is the same as both endpoints being unmatched, and the result follows from a classical correlation inequality for Gibbs measures over matchings.

For the $k$-uniform matroid (selecting at most $k$ elements from $E$), also known as the \emph{magician’s problem}~\citep{alaei2014bayesian}, we also provide an explicit Gibbs distribution achieving the optimal stationary selectability. The optimal constant has a simple description using Poisson CDFs:
\[
\alpha_k
\ \defeq\
\PP[Q<k \given Q\le k]=\frac{\PP[Q<k]}{\PP[Q\leq k]}\approx 1-\sqrt{\frac{2}{\pi k}}+\mathcal{O}\left(\frac{1}{k}\right), \qquad \text{where } Q\sim \mathrm{Poisson}(k).
\]
The associated simulate-then-replace algorithm fixes the greedy algorithm (with independent rejection w.p. $\revcolor{\mathcal{O}(\sqrt{\log k/k})}$) by initializing $\widehat{S}$ at a Gibbs distribution. This greedy algorithm is known to achieve a sub-optimal selectability of $1-\mathcal{O}(\sqrt{\log k/k})$.

\xhdr{S-OCRSs for hypergraph matchings with capacity (network revenue management).} We further develop the previous insight by designing an S-OCRS distribution for rank-$L$ hypergraph matchings in which vertices have capacity $k$. Rather than using the solution of a maximum-entropy convex program, we use an explicit Gibbs distribution where every hyperedge is selected independently with probability $\gamma x_e$ for \rev{$\gamma= 1-\mathcal{O}(\sqrt{\log L / k})$}, after which we condition on obtaining a feasible set. This leads to a greedy-with-random-discarding algorithm started from a stationary distribution that is a (near-optimal) \rev{$1-\sqrt{2\log L / k} -\mathcal{O}(1/\sqrt{k})$}-selectable OCRS.\footnote{It is well known that the factor of $1-\mathcal{O}\left(\sqrt{\frac{\log L}{k}}\right)$ cannot be beaten, even for offline contention resolution schemes on uniform multi-hypergraphs.} The dependence on $L$ and $k$ here matches large-capacity results in several related online resource allocation models. See for instance the work of \citet{gupta2016experts, kesselheim2018primal}\rev{,} who obtain an analogous result for random-order online packing LPs, and \citet{ghuge2025single} for single sample prophet inequalities. We also provide a different OCRS for this setting with a slightly improved selectability, which uses our earlier $k$-uniform S-OCRS as a black-box and has a provable polynomial-time implementation.

\xhdr{Further extensions beyond the maximum-entropy approach.} Finally, we delineate the limits of the maximum-entropy approach and explore how the approach can be modified to yield S-OCRSs in other feasibility environments. While these approaches are different from our standard maximum-entropy approach, they rely on similar ideas or are slightly modified versions of this approach.

\underline{Weakly Rayleigh matroids: relative entropy maximization \& thinning.} For general matroids, maximum-entropy distributions over independent sets can concentrate on configurations with poor addability, and therefore fail to yield strong S-OCRSs. To overcome this barrier, we develop a modified template for a broad class of matroids with negative dependence structure, which we call \emph{weakly Rayleigh matroids}. Our construction replaces entropy maximization with relative entropy maximization with respect to a Rayleigh base measure, followed by a simple thinning step. This yields an explicit $\tfrac{1}{2}$-selectable S-OCRS, matching the implicit OCRS in \cite{lee2018optimal}. Note that $\tfrac{1}{2}$ is the best factor possible for this class.

\underline{Knapsack constraints: Heavy \& light items.} For knapsack constraints, \rev{as with matroids}, there is a barrier to applying the maximum entropy approach directly (also faced by~\citet{feldman2016online,feldman2021online} in designing deterministic greedy OCRSs). However, there is still a $1/4$-selectable S-OCRS using a modified maximum entropy approach. We obtain this guarantee by using the technique of splitting the items into light and heavy types \citep{feldman2016online,feldman2021online, dutting2020prophet}, and then using a mixture of two different maximum entropy S-OCRSs. We note that our OCRS for the knapsack constraint is not optimal, since ~\citet{jiang2025tight} give an OCRS with selectability $1/(3+e^{-2})$. This is not just a limitation of our analysis, since in their tight instance, we can show that no S-OCRS can obtain a selectability larger than $3/10 < 1/(3+e^{-2})$.

\underline{General downward-closed feasibility environments.} For general downward closed feasibility environments, we show that there is a  $\Theta(1/\sqrt{|E|})$-selectable S-OCRS, and no (even ordinary) OCRS can do better. Our results in this general setting are thus sharp. To the best of our knowledge, these are the first results for OCRSs in general downward closed constraints, though the impossibility result is closely related to the \citet{kleinberg2012matroid} proof of impossibility of obtaining a competitive ratio better than $O\left(\frac{1}{\sqrt q}\right)$ for prophet inequalities for the intersection of $q$ matroids.\footnote{Note that while logarithmic competitive ratios are known for prophet inequalities in this general setting, OCRSs are dual to \textit{ex-ante} prophet inequalities rather than the usual prophet inequalities, which explains why our results do not contradict the existing literature.}

\xhdr{Extension to $\boldsymbol{\xvec\in b\Ppoly}$.} Our results extend to the case where $\xvec$ belongs to $b\Ppoly$, a version of $\Ppoly$ scaled by a parameter $b\geq 0$ (we must still assume that $x_e \in [0, 1]$). For rank-$L$ hypergraph matchings, we obtain a $\tfrac{1}{1+bL}$-selectable S-OCRS. For bipartite matchings, we obtain a $\tfrac{2b+1 -\sqrt{4b+1}}{2b^2}$-selectable S-OCRS. For $k$-uniform matroids, we obtain \rev{an} $\alpha_k(b)$-selectable S-OCRS, where \[
\alpha_k(b)
\ \defeq\
\PP[Q<k \given Q\le k], \qquad \text{and } Q\sim \mathrm{Poisson}(bk).
\]
For weakly Rayleigh matroids, we obtain a $\frac{1}{1+b}$-selectable S-OCRS. Finally, for knapsack constraints, we obtain a $\frac{1}{2+2b}$-selectable S-OCRS. 

\xhdr{Algorithmic tractability\rev{.}}
Across all environments, we complement the distributional results with a systematic discussion of polynomial-time implementation. This emphasis on implementability is particularly important for applications in \rev{operations research}. In particular, in \Cref{appendix-polytime}, we discuss how to sample (exactly or approximately) from these distributions in polynomial time and how to compute the required conditional expectations, enabling a computationally efficient implementation of our algorithms in several feasibility environments. Our S-OCRSs are fully explicit and implementable for matchings, $k$-uniform matroids, knapsack constraints, $\mathbb{C}$-representable matroids, and, more generally, whenever we can efficiently find a maximum-entropy distribution or KL-projected distribution. In addition, we also provide an OCRS for the network revenue management setting that runs in polynomial time. 

\xhdr{\rev{Reducing Bayesian online allocation of reusable products to S-OCRS design.}} Finally, we return to our second motivation for studying S-OCRSs and show that they help handle reusable resources. We consider a Bayesian online allocation model with reusable products over an arbitrary feasibility environment (and with stationary request arrivals). We let $\ground$ be a ground set of all \textit{products}, and $\mathcal{N}$ a family of multisets of products that can be feasibly allocated at any particular time (so multiple units of the same product can also be simultaneously allocated; $\Fset \subseteq \mathcal{N}$). Requests \rev{of different types} $i\in\mathcal I$ arrive according to an independent homogeneous Poisson process with constant rate $\arrivalRate_i$.  When a type-$i$ request arrives, the algorithm may reject it or allocate one product $e\in\ground$.  Such an allocation earns reward $\reward_{ie}\ge 0$ and occupies product $e$ for an exponentially distributed duration with mean $\ell_{ie}>0$, after which the product becomes available again.\footnote{The exponential duration distribution assumption is not necessary, and is only made for expositional clarity. Our guarantees extend to arbitrary duration distributions with finite means by applying Theorem 2 in \citet{zachary2007note}, using the detailed-balance equations established in \cref{apx-recurring-implies-admission}.}

For this general model, we establish a formal reduction from the Bayesian online allocation of reusable products in any feasibility environment to the design of S-OCRSs in the same environment: any $\alpha$-selectable S-OCRS yields an online policy that obtains an $\alpha$-fraction of the long-run average reward of the optimal omniscient policy (which knows all future arrivals). The reduction proceeds in two steps.

First, starting from an ex-ante fluid LP relaxation of the reusable product model, we use a propose-then-admit argument to reduce the construction of good policies to the construction of admission-control primitives. Second, we show that an S-OCRS for a particular feasibility environment, through its witness distribution as a feasible solution to the S-OCRS LP, can be used to design admission control primitives.  Combined with the S-OCRSs in this paper, this recipe yields approximately optimal online allocation algorithms for reusable products across all of our feasibility environments---thereby answering question (iii) in the affirmative. 

We note that the result also extends from the allocation setting to the assortment optimization setting where each arriving request is shown a set of products, and the request chooses which product to accept based on a choice model that is type dependent and satisfies weak substitutability. This extension follows the ideas of \citet{feng2024near}, especially on sub-assortment sampling. We also note that our policies can be implemented using prices in a straight-forward way.

In a companion paper \ifblind\citep{anonymous2026optimal}\else\citep{aminian2026optimal}\fi
, we focus on the special feasibility environment of $k$-uniform matroids, and allow for non-homogeneous Poisson process arrivals of requests, together with a finite horizon. By leveraging the ideas related to stationarity developed in the current paper, we solve a long-standing open problem in revenue management with reusable resources. More precisely, it turns out that in the vanishing regime for $k$-uniform matroids, where each of the $x_e$ is tiny, there is an alternative description of the solution to the maximum-entropy convex program. This description leads to a simpler optimal S-OCRS construction with selectability factor $\alpha_k$ that can be applied to non-homogeneous Poisson arrivals. This yields an $\alpha_k$-selectable policy for Bayesian online allocation with reusable resources under heterogeneous arrivals, improving upon the previously best known bound of $1-\Theta(\sqrt{\log k/k})$ in \cite{feng2022near} and matching the best explicit bound known for the non-reusable setting~\cite{wang2018online}.

\Cref{tab:summary-results} summarizes our selectability guarantees for various feasibility environments, together with the principal contribution of each result. Our work is also related to several lines of research in operations research and computer science, which we briefly summarize in \Cref{apx:further-related}.
\begin{table}[htb]
\caption{Summary of our explicit S-OCRS guarantees and their main contributions.}
\label{tab:summary-results}
\centering
\small
\renewcommand{\arraystretch}{1.5}
\setlength{\tabcolsep}{4pt}

\begin{tabularx}{\textwidth}{@{}
    >{\raggedright\arraybackslash}p{0.22\textwidth}
    >{\centering\arraybackslash}p{0.19\textwidth}
    >{\raggedright\arraybackslash}X
    >{\centering\arraybackslash}p{0.075\textwidth}
@{}}
\toprule
\textbf{Feasibility environment}
&
\textbf{Selectability}
&
\textbf{Main contribution}
&
\textbf{Section}
\\
\midrule

Bipartite matchings
&
$\frac{3-\sqrt{5}}{2}\approx 0.382$
&
Improved OCRS; optimal  S-OCRS.
&
\labelcref{sec:matching-uniform}
\\

Rank-$L$ hypergraph matchings
&
$\frac{1}{L+1}$
&
Stationary scheme matching basic OCRS benchmark.
&
\labelcref{sec:matching-uniform}
\\

$k$-uniform matroids
&
$1-\sqrt{\frac{2}{\pi k}}+ \mathcal{O}\left(\frac{1}{k}\right).$
&
Optimal S-OCRS; matches best explicit OCRS upper bound asymptotically.
&
\labelcref{sec:matching-uniform}
\\

Rank-$L$ hypergraphs with capacity $k$ (NRM)
&
$1-\mathcal{O}\left(\sqrt{\frac{\log L}{k}}\right)$
&
Optimal even among offline CRSs for $k \gg \log L$.
&
\labelcref{sec:NRM}
\\

Weakly Rayleigh matroids
&
$\frac{1}{2}$
&
Optimal even among OCRSs.
&
\labelcref{sec:rayleigh-matroids}
\\

Knapsack constraints
&
$\frac{1}{4}$
&
Stationary scheme matching basic OCRS benchmark.
&
\labelcref{sec:knapsack-constraints}
\\

General downward-closed
&
$\Theta\left(1/\sqrt{|E|}\right)$
&
First OCRS guarantee; optimal among OCRSs.
&
\labelcref{sec:downward-closed-constraints}
\\

\midrule
\multicolumn{4}{@{}p{\textwidth}@{}}{%
\emph{Consequence for reusable resources.}
For every environment above, an $\alpha$-selectable S-OCRS yields an
$\alpha$-approximation for the Bayesian online allocation of reusable
products against the long-run omniscient benchmark
(\Cref{sec:reusable-product-application}).}
\\
\bottomrule
\end{tabularx}
\end{table}

\section{Preliminaries and Notation}
\label{sec:prelim}
We begin by formalizing the general setting studied in this paper and the notion of an online contention resolution scheme (OCRS), introduced by \cite{feldman2016online,feldman2021online}. We also introduce the notation and provide some details for the specific feasibility environments we focus on.

\xhdr{General setting.}
Throughout this paper, we work with a finite ground set of elements $\ground$ and a downward-closed family of feasible sets
$\Fset\subseteq 2^{\ground}$, i.e., for every $S\in\Fset$ and every $T\subseteq S$ we have $T\in\Fset$. We refer to $(\ground,\Fset)$ as a \emph{feasibility environment}. Let
\[
\Ppoly \defeq \conv\{\mathbf{1}_S : S\in\Fset\} \subseteq [0,1]^{\ground}
\]be the corresponding feasibility polytope, where $\mathbf{1}_S\in\{0,1\}^{\ground}$ is the incidence vector of $S$. For a (deterministic or random) set $S\subseteq\ground$ and an element $e\in\ground$, we write
$\Sm{S}{e}$ for the set obtained by removing $e$ if $e$ is present in $S$, i.e., $\Sm{S}{e}=S\setminus\{e\}$. Note that if $S\in\Fset$ then $\Sm{S}{e}\in\Fset$ as well. 

Given a feasibility environment $(\ground,\Fset)$, we fix a point $\xvec\in\Ppoly$ and interpret $x_e$ as the marginal \emph{activation} probability of element $e$. The \emph{active set} $\Aset\subseteq\ground$ is then a random set generated by including each element $e\in\ground$ independently with probability $x_e$. We will always assume $x_e > 0$ for all $e \in \ground$ (in all our arguments, if $x_e$ were ever equal to 0, we could always replace $\ground$ with $\ground \setminus \{e\}$).
We then consider an online setting in which elements arrive sequentially and are revealed to an online algorithm that makes irrevocable decisions. An \emph{arrival order} is a permutation $\pi \in \mathrm{Sym}(\ground)$, that is, a bijection $\pi:{1,\ldots,|\ground|}\to\ground$. With this setup\footnote{In our positive results, we will allow for an `online' (or `adaptive') adversary who can change the order in which future elements are revealed based on the algorithm's decisions in the past, but does not have access to the activations of unrevealed elements, or the random bits the algorithm uses.}, we recall the following standard definitions (cf.~\cite{feldman2016online,feldman2021online}).
\begin{definition}[OCRS]
\label{def:ocrs}
An online contention resolution scheme for $(\ground,\Fset)$ is an online algorithm that, for every $\xvec\in\Ppoly$ (known by the online algorithm upfront) and every permutation $\pi$, processes elements sequentially in the order $\pi(1),\pi(2),\dots,\pi(|\ground|)$. When an element $e$ is revealed, the algorithm learns whether $e\in\Aset$ and must irrevocably decide whether to accept or reject $e$. In the end, the algorithm must output a random set $\Sset$ satisfying $\Sset\subseteq \Aset$ and $\Sset\in\Fset$.
\end{definition}

\begin{definition}[$\alpha$-selectability]
\label{def:selectability}
An OCRS is \emph{$\alpha$-selectable} (also called \emph{$\alpha$-balanced}) if for every $\xvec\in\Ppoly$, for every arrival order $\pi$ and every $e\in\ground$,
\[
\PP[e\in \Sset] \ge \alpha x_e.
\]
Equivalently, since $e\in\Sset$ implies $e\in\Aset$, we can also write this as
$\PP[e\in \Sset \given e\in\Aset]\ge \alpha$.\footnote{Unlike in \citep{feldman2016online, feldman2021online}, we consider the \textit{selectability} of schemes that are not necessarily greedy, and our schemes are not tailored to an almighty adversary.}
\end{definition}

Central to our technical framework in this paper, we also rely on a distributional view of OCRSs. Fixing $\xvec$ and an arrival order $\pi$, any OCRS induces a distribution over feasible sets; we denote by $\dist^\pi\in\simplex{\Fset}$ the distribution of the final selected set $\Sset$. In general, $\dist^{\pi}$ may depend on~$\pi$.

\xhdr{Feasibility environments studied in our paper.}
We will instantiate $(\ground,\Fset)$ with the following concrete feasibility environments (and use the corresponding notation throughout the paper):
\begin{enumerate}[leftmargin=2em, label=(\roman*)]
    \item \textbf{Rank-$\boldsymbol{L}$ hypergraph matchings (\Cref{sec:matching}, \Cref{sec:NRM},  \Cref{apx:hypergraph}, and \Cref{apx:constant-scaling-k-uniform}):} $\ground$ is the edge set of a hypergraph $H=(V,\ground)$ of rank at most $L$ (each $e\in\ground$ satisfies $|e|\le L$), and $S\subseteq\ground$ is feasible iff it is a matching, i.e., its hyperedges are pairwise vertex-disjoint, and hence 
    $$\Fset  \defeq \Bigl\{S\subseteq \ground: \forall v\in V, \bigl|\{e\in S: v\in e\}\bigr|\le 1\Bigr\}.$$
    We allow parallel edges. \revcolor{For a vertex $v$, let $\delta(v)\defeq\{e\in\ground:v\in e\}$ denote its set of incident edges.} Note that for every $\xvec\in \Ppoly$, we must necessarily have:
\begin{equation}
\label{eq:hg-vertex-packing}
\xvec\in [0,1]^E\text{ and }\sum_{e\in \delta(v)} x_e \le 1 \qquad \forall v\in V.
\end{equation}
The special case $L=2$ corresponds to (generally non-bipartite) graph matchings. We will also be interested in the case where each vertex has a capacity $k$. In this case, 
   $$\Fset  \defeq \Bigl\{S\subseteq \ground: \forall v\in V, \bigl|\{e\in S: v\in e\}\bigr|\le k\Bigr\},$$ and for every $\xvec\in \Ppoly$, we have
\begin{equation}
\label{eq:k-capacity-hg-vertex-packing}
\xvec\in [0,1]^E\text{ and }\sum_{e\in \delta(v)} x_e \le k \qquad \forall v\in V.
\end{equation}
This model captures \emph{network revenue management (NRM)} with products that consume up to $L$ resources and each resource has $k$ available units: vertices represent resources, hyperedges represent products (bundles of resources), and a feasible allocation corresponds to choosing  hyperedges that are collectively incident to each vertex at most $k$ times; see, e.g., \cite{ma2026online}.

    \item \textbf{Bipartite matchings (Section~\ref{sec:bipartite}):} $\ground$ is the edge set of a bipartite graph $G=(U\cup V,\ground)$, and $S\subseteq\ground$ is feasible iff it is a (not-necessarily-perfect) matching. We allow parallel edges.  Every $\xvec\in\Ppoly$ is a \emph{fractional matching}, that is, it belongs to the following integral (and totally unimodular) polytope:
\begin{equation}
\label{eq:fractional-matching-constraints}
\xvec\in [0,1]^E:~~~\sum_{e\in \delta(z)} x_e  \le 1
\qquad\forall z\in U\cup V.
\end{equation}
    \item \textbf{$\boldsymbol{k}$-selection or $\boldsymbol{k}$-uniform matroid  (Section~\ref{sec:uniform}):} $\ground$ is an arbitrary ground set and $S\subseteq\ground$ is feasible iff $|S|\le k$. Therefore $\Fset \defeq \bigl\{S\subseteq \ground : |S|\le k\bigr\}$ and $\xvec\in\Ppoly$ is characterized by a simple \emph{capacity constraint}:
    \begin{equation}
    \xvec\in [0,1]^E:~~~\sum_{e\in\ground} x_e \le k.
    \end{equation}
    \item \textbf{(Weakly Rayleigh) matroids (Section~\ref{sec:rayleigh-matroids}):} A \emph{matroid} is a pair $\cM=(\ground,\Iset)$ where $\Iset\subseteq 2^{\ground}$ is a family of \emph{independent sets} satisfying:
    (i) $\emptyset\in\Iset$;
    (ii) if $A\in\Iset$ and $B\subseteq A$ then $B\in\Iset$;
    (iii) if $A,B\in\Iset$ and $|A|<|B|$, then there exists $e\in B\setminus A$ such that $A\cup\{e\}\in\Iset$.
    A set $S$ is feasible iff it is independent, i.e., $\Fset=\Iset$. A matroid can equivalently be described using its set of \emph{bases} $\Bset$, consisting of all maximal independent sets in $\Iset$. A matroid has \emph{rank} $r$ iff all bases have size $r$. Given a matroid $\cM$, the \emph{rank function} (also known as the \emph{rank oracle}) of the matroid is defined as $\RankOracle{S}:=\max_{I\in\Iset: I\subseteq S}\lvert I \rvert$. Given the rank function, the feasibility polytope $\Ppoly=\conv\{\mathbf{1}_S:S\in\Fset\}$, called the \emph{independence polytope}, is given by:
\begin{equation}
\label{eq:matroid-ind-poly}
\Ppoly  =  \Bigl\{\xvec\in [0,1]^E: \sum_{e\in T}x_e \le \RankOracle{T},~~~~\forall T\subseteq \ground\Bigr\}.
\end{equation}
It will also be convenient to define the \emph{base polytope} of a rank-$r$ matroid:
\begin{equation}
\label{eq:base-poly}
\Bpoly
 \defeq\
\Bigl\{\qvec\in [0,1]^E: \sum_{e\in T}q_e \le \RankOracle{T}  \forall T\subseteq \ground,  \sum_{e\in\ground}q_e=r\Bigr\}.
\end{equation}
It is a standard result that there is a polynomial-time equivalence between the rank and separation oracles for the independence (or base) polytope; see, e.g., \cite{Oxley2011MatroidTheory,schrijver2003combinatorial}. We further focus on the subclass of \emph{weakly Rayleigh} matroids that admit distributions over bases with a certain type of negative dependence (see Definitions~\ref{def:rayleigh-measure} and~\ref{def:weakly-rayleigh}). Many commonly encountered classes of matroids are weakly Rayleigh, including linear matroids over $\mathbb{C}$ (hence, graphic, laminar, and regular matroids).\footnote{Weakly Rayleigh matroids are a common weakening of Weak Half-Plane Property (WHPP) matroids and Rayleigh matroids. See \cite{choe2004homogeneous,choe2006rayleigh} for more details on the differences between these matroid classes.}

\item \textbf{Knapsack constraints (Section~\ref{sec:knapsack-constraints}):} $\ground$ is a set of items with sizes $a_e\in(0,1]$, and $S\subseteq\ground$ is feasible iff $\sum_{e\in S}a_e\le1$. For every $\xvec \in \Ppoly$, we necessarily have that:
    \begin{equation}
    \label{eq:knapsack-relaxation}
    \xvec\in[0,1]^E\text{ and }\sum_{e\in\ground}a_ex_e\le1.
    \end{equation}

\end{enumerate}

We note that our positive results will also work when we only assume that $\xvec$ belongs to a standard relaxation of the feasibility polytope $\mathcal{P}$, rather than the actual convex hull of feasibility sets. In particular, for hypergraph matchings, we only need to assume:
\[
   \xvec \in \widetilde \Ppoly =\left\{\xvec\in[0,1]^E:\sum_{e\in \delta(v)} x_e \le k \quad \forall v\in V\right\}.
\]
and for knapsack constraints, we only need to assume:
\[
   \xvec \in \widetilde \Ppoly =\left\{\xvec\in[0,1]^E:\sum_{e\in E}a_ex_e\le1\right\}.
\]

\section{Stationary OCRS: Definitions \& LP Formulation}
\label{sec:socrs-basics-LP}
This section formalizes the main mathematical object studied in this paper: the \emph{Stationary Online Contention Resolution Scheme (S-OCRS)}. We start by defining S-OCRS and providing intuition for this notion in \Cref{sec:socrs-def}. We then characterize the problem of designing S-OCRS in \Cref{sec:socrs-LP} via an LP over the space of distributions, which we call \emph{the stationary OCRS LP}. This reduction is constructive and yields a meta S-OCRS algorithm that can implement any feasible solution of this LP---and hence plays the role of a template for all S-OCRS algorithms in the rest of the paper.

\subsection{Definition and motivation}
\label{sec:socrs-def}
One motivation to study stationarity (to be formalized soon) comes from a variant of the standard OCRS problem that naturally appears in Bayesian online allocation models with \emph{reusable resources}~\citep{feng2024near,feng2022near,rusmevichientong2020dynamic}. In such models, one often reduces the design of an online policy to an OCRS-style \emph{admission} subroutine, abstracting away the arrival process and rewards from the admission decisions. In that reduction, the input to the admission subroutine is stochastic requests generated in an online fashion by sampling the solution to an ex-ante offline fluid linear program for the original problem. The generated requests can be feasibly fulfilled in expectation. The goal of the subroutine is then to maintain the feasibility of the set of admitted requests, while ensuring that each request is selected with a large enough probability, à la OCRS selectability in Definition~\ref{def:selectability}.

In reusable settings, an allocated agent (or request) uses our resources for a duration of time and releases the resource afterwards.  A stylized abstraction of the reduced model that arises is the following recurring variant of the OCRS problem, which is a generalization of the standard OCRS problem when the requests (or elements) come back repeatedly. 

\smallskip
\begin{definition}[The recurring OCRS problem]
\label{def:recurring-ocrs}
Fix a downward-closed feasibility environment $(\ground,\Fset)$ with corresponding feasibility polytope $\Ppoly=\conv(\Fset)$ and a point $\xvec\in\Ppoly$ (known by the algorithm upfront).  An adversary fixes an infinite sequence $e_1,e_2,\ldots$ with $e_t\in\ground$ that is unknown to the algorithm. At time $t$, the algorithm observes $e_t$, and a fresh independent coin determines whether $e_t$ is \emph{active}, with probability $x_{e_t}$. An $\alpha$-selectable recurring OCRS is an online algorithm that starts with $S_0=\emptyset$ and, after each observation, maintains a random set $S_t$ with the following properties:
\begin{enumerate}[leftmargin=2em,label=(\roman*)]
    \item If  $e_t$ is inactive, then $S_t=S_{t-1}\setminus\{e_t\}$; if $e_t$ is active, then $S_t\in\{S_{t-1}\setminus\{e_t\},S_{t-1}\cup\{e_t\}\}$.
    \item $S_t\in\Fset$ at all times $t$.
    \item $\PP[e_t\in S_t]\ge \alpha x_{e_t}$, where the probability is over the activity coins and the algorithm's randomness.
\end{enumerate}
As in standard OCRSs, the goal is to maximize the selectability factor $\alpha$.
\end{definition}

\begin{figure}[htb] 
\centering 
\vspace{-2mm}
\begin{tikzpicture}[x=0.8cm,y=1cm,line cap=round,line join=round, scale=0.95]
\newcommand{\timearc}[4]{%
  \draw[#4] (#1) .. controls ($(#1)!0.25!(#2)+(0,#3)$) and ($(#1)!0.75!(#2)+(0,#3)$) .. (#2);
}

\foreach \i in {0,...,15} {
  \coordinate (p\i) at (\i,0);
}

\draw[line width=0.9pt] (-0.45,0) -- (15.45,0);
\foreach \i in {0,...,15} {
  \draw[line width=0.9pt] (p\i) -- ++(0,-0.28);
}

\coordinate (ex1) at (16.0,1.35);
  \coordinate (ex2) at (16.0,1);
\coordinate (ex3) at (16.0,0.75);
\node[]    at (ex1) {};
\node[]    at (ex2) {};
\node[]    at (ex3) {};

\draw[line width=0.6pt] (9.5,-0.8) -- (9.5,3.4);
\node[below=9pt] at (8.8,0) {\footnotesize time $t$};

\timearc{p3}{p5}{1.05}{mygreen, line width=1.1pt}
\timearc{p5}{p11}{3.10}{mygreen, line width=1.1pt}
\timearc{p11}{ex1}{0.5}{mygreen, line width=1.1pt}

\timearc{p1}{p4}{1.25}{myblue, line width=1.1pt}
\timearc{p4}{p7}{1.35}{myblue, line width=1.1pt}
\timearc{p7}{p13}{2.15}{myblue, line width=1.1pt}
\timearc{p13}{ex2}{0.35}{myblue, line width=1.1pt}

\timearc{p2}{p10}{2.55}{mypurple, line width=1.1pt}
\timearc{p10}{p12}{1.05}{mypurple, line width=1.1pt}
\timearc{p12}{p14}{0.95}{mypurple, line width=1.1pt}
\timearc{p14}{ex3}{0.2}{mypurple, line width=1.1pt}

\timearc{p0}{p6}{2.25}{myred, line width=1.1pt}
\timearc{p6}{p8}{1.10}{myred, line width=1.1pt}
\timearc{p8}{p9}{0.65}{myred, line width=1.1pt}
\timearc{p9}{p15}{2.25}{myred, line width=1.1pt}

\node[anchor=west] at (9.75,2.9) {\footnotesize $\displaystyle
\xvec \in \Ppoly~~\& ~~ \text{at each time } t,~~S_t\in \mathcal{F}$};

\foreach \col [count=\i from 0] in
  {myred,myblue,mypurple,mygreen,myblue,mygreen,
   myred,myblue,myred,myred,mypurple,mygreen,
   mypurple,myblue,mypurple,myred}
{
  \filldraw[draw=black, line width=0.35pt, fill=\col] (p\i) circle (3pt);
}
\end{tikzpicture} 
\vspace{-4pt}
\caption{A recurring OCRS instance. Each color corresponds to one element $e\in\ground$. Consecutive arcs depict successive ``epochs" of the same element; at the beginning of each epoch, $e$ is independently active with probability $x_e$. If an active epoch is accepted, $e$ remains selected until the end of that arc.} 
\label{fig:chain-ocrs} 
\vspace{-3mm}
\end{figure}

Informally, we may consider, for each element $e\in\ground$, \textit{epochs} which are the intervals between its two consecutive appearances in the sequence. A \emph{renewal event} occurs when an epoch begins. At each renewal of element $e$, a fresh independent coin determines whether $e$ is \emph{active} in the next epoch. A recurring OCRS must decide at every renewal whether to select an active element $e$. If the algorithm selects $e$ at the beginning of an active epoch, then $e$ remains selected for the entire epoch and becomes unselected when it ends. The algorithm must maintain feasibility at all times, and the goal is to be \emph{$\alpha$-selectable} at \emph{every} renewal of \emph{every} element.

Unlike a standard OCRS instance---where each element appears once and the adversary chooses a single arrival permutation---in the recurring OCRS problem the algorithm repeatedly faces a new contention pattern as requests depart and re-arrive. Since epoch lengths can be arbitrary, the stream of renewal events can interleave in a highly irregular way. From the viewpoint of the a policy, this means that the relative order in which elements are encountered keeps reshuffling over time (Figure~\ref{fig:chain-ocrs}), and still the algorithm must remain feasible and maintain  $\alpha$-selectability at each renewal. 

The above feature suggests that, in reusable settings, the exact permutation in which renewals are processed should be treated as incidental rather than as a meaningful part of the model. This motivates focusing on OCRSs whose output distribution is invariant to the arrival permutation: for each fixed $\xvec\in\Ppoly$, all arrival orders should induce the same distribution over the selected feasible set. Formally, we have the following definition.

\begin{definition}[Stationary OCRS (S-OCRS)]
\label{def:stationary-ocrs}
An OCRS is \emph{stationary} if, for every $\xvec\in\Ppoly$, the distribution of the output set is independent of the arrival order. Formally, for every permutations $\pi,\pi'$ of $\ground$:
\[
\dist^{(\xvec,\pi)}=\dist^{(\xvec,\pi')}\equiv \dist^{(\xvec)},~
\]
where $\dist^{(\xvec,\pi)}\in\simplex{\Fset}$ (resp. $\dist^{(\xvec,\pi')}\in\simplex{\Fset}$) is the distribution of the output set of the OCRS algorithm for $\xvec\in\Ppoly$ under arrival order $\pi$ (resp. $\pi'$). We write $\dist^{(\xvec)}\in\simplex{\Fset}$ for the common distribution of $\Sset$ (and drop $\xvec$ when it is clear from the context). 
\end{definition}

The following proposition formalizes the conceptual reason for why stationarity is the ``right'' abstraction for the recurring OCRS problem:

\begin{proposition}[S-OCRS $\Rightarrow$ Recurring OCRS]
\label{prop:socrs-to-chain}
Fix $\xvec\in\Ppoly$.
If there exists an $\alpha$-selectable stationary OCRS for $(\ground,\Fset)$ at $\xvec$ with common output distribution $\dist\in\simplex{\Fset}$, then there exists an online policy for the recurring OCRS problem with activation probabilities $\xvec$ that is $\alpha$-selectable.
\end{proposition}
We postpone the proof to \Cref{apx:socrs-recurring-equivalence}, since it depends on concepts developed in the next section, where we identify an alternative characterization of S-OCRS via a certain linear program. This characterization is accompanied by a meta algorithm, which can be used to prove the reduction from recurring OCRS to S-OCRS.

\subsection{Characterization via LP, and a ``Simulate-then-Replace'' meta algorithm}
\label{sec:socrs-LP}
We start by introducing the following LP, where the decision variables are distributions over $\Fset$.
\begin{definition}[The stationary OCRS LP]
\label{def:socrs-LP}
Fix $\xvec\in\Ppoly$. In the stationary OCRS LP, we seek the largest $\alpha\in[0,1]$ such that there exists a distribution
$\dist\in\simplex{\Fset}$ satisfying
\begin{align}
\tag{\textsc{Selectability}}
\PP_{\Sset\sim\dist}[e\in\Sset] &\ge \alpha\,x_e
&&\forall e\in\ground,
\label{eq:soCRS-selectability}\\
\tag{\textsc{Stationary-Implementability}}
\PP_{\Sset\sim\dist}\big[e\in\Sset \given \Sm{\Sset}{e}=T\big] &\le x_e
&&\forall e\in\ground,\ \forall T\in\Fset\ \text{with }e\notin T.
\label{eq:soCRS-stationary}
\end{align}
\end{definition}
A few comments are in order. Constraint~\eqref{eq:soCRS-selectability} is exactly $\alpha$-selectability in the distributional view.
Constraint~\eqref{eq:soCRS-stationary} is a \emph{stationary implementability} condition: even after conditioning on the
selection status of \emph{all} other elements, the conditional probability of selecting $e$ cannot exceed its activation
probability~$x_e$. Note that when we condition on an event of the type $S_{-e} = T$, we always implicitly assume $\PP_{\Sset\sim\dist}\big[\Sm{\Sset}{e}=T\big] > 0$.  We now have the following proposition, demonstrating an equivalence between stationary OCRS algorithms and feasible solutions to the LP:
\begin{proposition}[S-OCRS $\Longleftrightarrow$ stationary OCRS LP]
\label{prop:stationary-equivalence}
Fix $\xvec\in\Ppoly,~\alpha\in[0,1]$.
There exists an $\alpha$-selectable stationary OCRS with activation probabilities $\xvec$ if and only if the stationary OCRS LP (Definition~\ref{def:socrs-LP}) is feasible.
\end{proposition}
The full proof of the proposition is straightforward, and we postpone to \Cref{apx:socrs-LP-equivalence}. Instead, it is convenient for our purposes to go over an alternative constructive proof of \emph{(LP $\Rightarrow$ S-OCRS)} using a randomized policy we call \emph{``simulate-then-replace.''} Assume we are given $\dist\in\simplex{\Fset}$ that satisfies~\eqref{eq:soCRS-stationary}.
The simulate-then-replace policy---formalized in \Cref{alg:sample-then-replace}---begins by simulating a feasible \emph{simulated set} $\widehat{\Sset}\sim\dist$ of all elements. Then, once each element arrives, the policy
\emph{replaces} the simulated membership of that element with an activation-feasible decision (i.e., accepting the active element $e$ with a target conditional probability $q_e/x_e=
\PP_{\Sset\sim\dist}[e\in\Sset \given \Sm{\Sset}{e}=T]/x_e$), updating $\Sset_e$, while keeping the remaining allocation $T=\Sm{\widehat{\Sset}}{e}$ intact. See \Cref{fig:STR-evolution} for an illustration of how set $\widehat{S}$ evolves as we run \Cref{alg:sample-then-replace}.\footnote{Our policy shares similarities with single-site Glauber dynamics (see for instance Chapter 3 in \cite{levin2017markov}, or \cite{Vigoda:PhD}), but we start at (and therefore stay at) the stationary distribution, the order in which we update sites is arbitrary, and updates are `gated' by the activation.}

To see why this algorithm is an $\alpha$-selectable S-OCRS, note that due to Constraint \ref{eq:soCRS-stationary}, we have  $q_e\leq x_e$, which is exactly what makes the acceptance probability $q_e/x_e$ well-defined. Moreover, if $T\cup\{e\}\notin \Fset$ then $q_e=0$; therefore, the algorithm  maintains the feasibility of the selected set of elements throughout (as $T$ is a superset of actually selected elements). Furthermore, if we consider the arrival time of element $e$ and condition on $T=\Sm{\widehat{\Sset}}{e}$, the updated membership of $e$ is $\mathrm{Bernoulli}(q_e)$, and the rest of the simulated set remains equal to~$T$. Therefore, the one-step update leaves the joint distribution of $\widehat{\Sset}$ equal to~$\dist$.
Since this holds for every processed element, after the full arrival sequence we still have that $\widehat{\Sset}$ has the distribution $\dist$, and hence $\Sset$ has the distribution $\dist$. Finally, because of Constraint~\ref{eq:soCRS-selectability}, we have $\PP_{\Sset\sim\dist}[e\in\Sset] \ge \alpha x_e$, which is exactly the $\alpha$-selectability.

\begin{algorithm}[htb]
\caption{\textsc{Simulate-then-Replace} policy for distribution $\dist$}
\label{alg:sample-then-replace}
\KwIn{Activation vector $\xvec\in\Ppoly$ and target distribution $\dist\in\simplex{\Fset}$ satisfying \eqref{eq:soCRS-stationary}}
\smallskip

Sample a feasible simulated set  $\widehat{\Sset}\sim\dist$\tcp*[r]{\textcolor{blue}{\footnotesize{initialize simulated state with target distribution}}}

\ForEach{$e\in\ground$ revealed online (in the given order)}{
  $T\gets \Sm{\widehat{\Sset}}{e}$\tcp*[r]{\textcolor{blue}{\footnotesize{forget the membership of $e$ in the simulated set}}}
  \If{$e$ is active}{
     $q_e\gets \PP_{\Sset\sim\dist}\!\big[e\in\Sset \mid \Sm{\Sset}{e}=T\big]$\tcp*[r]{\textcolor{blue}{\footnotesize{calculate target conditional probability under $\dist$}}}
    $T\gets T\cup\{e\}$ with probability $q_e/x_e$\tcp*[r]{\textcolor{blue}{\footnotesize{randomly replace the membership of $e$ if active}}}
    Accept element $e$ if $e\in T$}
   \Else{Reject element $e$ \tcp*[r]{\textcolor{blue}{\footnotesize{reject element $e$ if inactive}}}}
   $\widehat{\Sset}\gets T$
 }

\Return{$\Sset\gets \widehat{\Sset}$}\tcp*[r]{\textcolor{blue}{\footnotesize final set has distribution $\dist$}}
\end{algorithm}

\begin{figure}[h]
    \centering
\begin{tikzpicture}[scale=0.9]
  \node at (-0.55,0.55) {\Large $\widehat{S}\;=\;\Big($};
  \foreach \i/\v in {1/1,2/0,3/1,4/0,5/1}{%
    \node[draw,fill=okgreen!20,minimum width=0.7cm,minimum height=0.7cm] at (\i*0.8,0.55) {\v};
  }
  \foreach \i/\v in {6/0,7/1,8/0,9/1}{%
    \node[draw,fill=warmorange!25,minimum width=0.7cm,minimum height=0.7cm] at (\i*0.8,0.55) {\v};
  }
  \node at (8.0,0.55) {\Large $\Big)$};
  \node[anchor=west] at (8.4,0.55) {\large $\sim\;\mu$};
  \draw[decorate,decoration={brace,amplitude=5pt,mirror},okgreen,thick] (0.5,0.05) -- (4.3,0.05);
  \node[align=center,text width=3.6cm,font=\small,okgreen] at (2.4,-0.65) {actual (already decided)\\ \footnotesize{coordinates $1,\dots,i$}};
  \draw[decorate,decoration={brace,amplitude=5pt,mirror},warmorange,thick] (4.5,0.05) -- (7.5,0.05);
  \node[align=center,text width=3.6cm,font=\small,warmorange] at (6.0,-0.65) {simulated\\ \footnotesize{coordinates $i+1,\dots,n$}};
\end{tikzpicture}
  \caption{Evolution of set $\widehat{S}$ throughout the execution of \Cref{alg:sample-then-replace}.}
    \label{fig:STR-evolution}
    \vspace{-2mm}
\end{figure}

We end this section by noting that although the reduction established in Proposition~\ref{prop:stationary-equivalence} conceptually simplifies the design of S-OCRS algorithms and reduces the design problem to solving the stationary OCRS LP, it is highly unclear how to solve this (exponential-size) LP or even find an appropriate feasible solution for this program. We address this point in the next section.

\section{S-OCRS via the Maximum-Entropy Principle}
\label{sec:socrs-template}
In order to find good feasible solutions for the stationary OCRS LP, in this section we introduce a general framework that utilizes the maximum-entropy method~\citep{gharan2011randomized, singh2014entropy,straszak2019maximum}. \revcolor{The framework is motivated by our desire to make the probabilities $\PP_{\Sset\sim\dist}\!\big[e\in\Sset \mid \Sm{\Sset}{e}=T\big]$ used in \cref{alg:sample-then-replace} as simple as possible, while simultaneously ensuring that $\PP_{\Sset\sim\dist}[e\in\Sset] \ge \alpha x_e$.} For the design and analysis of our S-OCRS algorithms, we follow the following template with four steps:

\smallskip
\begin{enumerate}[label=(\roman*), align=left, labelwidth=1em,  
  labelsep=1em,         %
  leftmargin=!,         %
  itemsep=0.3em]
    \item  Pick a target selectability parameter $\alpha\in (0,1)$ and define scaled marginals $\pvec\defeq \alpha\xvec$,
    \item Consider the \emph{maximum-entropy distribution} $\dist^*$ over feasible sets $\Fset$ with marginals $\pvec$, which is the solution of a particular concave program (formally defined later),
    \item Prove that under $\dist^*$, the stationary implementability constraint is always satisfied,
    \item Run the simulate-then-replace policy (\Cref{alg:sample-then-replace}) with $\dist^*$ as input.
\end{enumerate}
\smallskip

We formalize our template in \Cref{sec:maxent}\revcolor{, calculating the values of $\PP_{\Sset\sim\dist}\!\big[e\in\Sset \mid \Sm{\Sset}{e}=T\big]$}. We then show in \Cref{sec:addability} that step~(iii) is equivalent to proving a certain combinatorial statement related to adding an element to a given set. In the following section (\Cref{sec:matching-uniform}), we will show how to apply this method to three feasibility environments: matchings in (hyper-)graphs, bipartite matchings, and $k$-uniform matroids. 

\subsection{Maximum-entropy with prescribed marginals}
\label{sec:maxent}
Fix any target marginal vector $\pvec$ that is in the relative interior of $\Ppoly$, which is the case in our applications after scaling by
$\alpha<1$. Consider the entropy maximization problem:
\begin{align}
\label{eq:maxent-primal-template}
\tag{\textsc{Max-Entropy}}
\max_{\dist\in\simplex{\Fset}}\quad & \Entropy(\dist)\ \defeq\ -\sum_{S\in\Fset}\dist(S)\log \dist(S)\\
\text{s.t.}\quad & \PP_{\Sset\sim\dist}[e\in\Sset] \ =\ p_e &&\forall e\in\ground,\\
& \revcolor{\sum_{S\in\Fset}\dist(S)=1.}
\end{align}

Note that this optimization problem is indeed a concave optimization over the space $\simplex{\Fset}$. As Slater's condition holds (cf. \cite{boyd2004convex}), the optimizer is unique. Moreover, it is well known in combinatorics and statistical physics that the solution to this problem has a product form, which is often referred to as a \emph{Gibbs distribution}~\citep{jaynes1957information,georgii2011gibbs}.

\begin{definition}[Gibbs distribution on a feasibility set]
\label{def:gibbs}
Fix a feasibility environment $(\ground, \Fset)$ and a vector of weights
$\wvec=\{w_e\}_{e\in\ground}$ with $w_e>0$ for all $e$.
The \emph{Gibbs distribution} on $\Fset$ with parameters $\wvec$
is the probability distribution $\dist_{\wvec}\in\simplex{\Fset}$ defined by
\begin{equation}
\label{eq:gibbs-def}
\dist_{\wvec}(S)
\ \defeq\
\frac{1}{\Zpart(\wvec)}\Big(\prod_{e\in S} w_e\Big)\,\Ind{S\in\Fset},
\qquad
\Zpart(\wvec)\ \defeq\ \sum_{T\in\Fset}\prod_{e\in T} w_e.
\end{equation}
\end{definition}

We now have the following standard lemma, for which we include a proof for completeness.

\begin{lemma}[\cite{vishnoi2021algorithms}, for instance]
\label{lem:gibbs-addability-template}
Let $\pvec$ be in the relative interior of $\Ppoly$ and let $\dist^{*}$ be an optimal solution of~\eqref{eq:maxent-primal-template}.
Then there exists a parameter vector $\thetavec\in\mathbb{R}^{\ground}$ such that $\dist^{*}$ is a Gibbs distribution on $\Fset$ with parameters $w_e=\exp(\theta_e)$. Moreover, for each $e\in\ground$, define
\begin{equation}
\label{eq:rho-def}
\rho_e\ \defeq\ \frac{w_e}{1+w_e}\ \in(0,1).
\end{equation}
Then for every feasible $T\in\Fset$ with $e\notin T$,
\begin{equation}
\label{eq:stationary-conditional-template}
\displaystyle
\PP_{\Sset\sim\dist^{*}}\big[e\in\Sset \given \Sm{\Sset}{e}=T\big]
\ =\ \frac{\dist^{*}(T\cup\{e\})}{\dist^{*}(T)+\dist^{*}(T\cup\{e\})}=
\begin{cases}
\rho_e, & \text{if}~~T\cup\{e\}\in\Fset,\\
0, & \text{otherwise.}
\end{cases}
\end{equation}
\end{lemma}

\begin{proofof}{Proof}
Introduce a Lagrange multiplier $\gamma\in\mathbb{R}$ for the normalization constraint and Lagrange multipliers
$\{\theta_e\}_{e\in\ground}$ for the marginal constraints.
The Lagrangian of~\eqref{eq:maxent-primal-template} is
\[
\mathcal{L}(\dist,\gamma,\thetavec)
=
-\sum_{S\in\Fset}\dist(S)\log\dist(S)
+\gamma\Big(\sum_{S\in\Fset}\dist(S)-1\Big)
+\sum_{e\in\ground}\theta_e\Big(\sum_{S\ni e}\dist(S)-p_e\Big).
\]
Since $\pvec$ lies in the relative interior of $\Ppoly$, Slater’s condition holds~\citep{boyd2004convex}. Therefore  the optimum is unique, attained in the interior, and satisfies the KKT conditions (strong duality).
In particular, letting $(\thetavec,\gamma)$ be the optimal duals of the concave program, for each $S\in\Fset$, the first-order condition gives
\[
0=\frac{\partial \mathcal{L}}{\partial \dist(S)}
=-(\log\dist(S)+1)+\gamma+\sum_{e\in S}\theta_e~\Longrightarrow \dist^{*}(S)=\exp(\gamma-1)\exp\!\Big(\sum_{e\in S}\theta_e\Big)~.
\]
Normalizing over $S\in\Fset$ yields the Gibbs form~ in \eqref{eq:gibbs-def}, with $w_e\defeq \exp(\theta_e)>0$.

Fix any $e\in\ground$ and condition on $\Sm{\Sset}{e}=T$.
As $\Fset$ is downward-closed, the only feasible sets compatible with this conditioning are $T$ and (if feasible) $T\cup\{e\}$.
Using the multiplicative form $\dist^{*}(S)\propto\prod_{f\in S}w_f$, we have
\[
\frac{\dist^{*}(T\cup\{e\})}{\dist^{*}(T)}=w_e
\qquad\text{whenever }T\cup\{e\}\in\Fset.
\]
Therefore the conditional probability $\PP_{\Sset\sim\dist^{*}}\big[e\in\Sset \given \Sm{\Sset}{e}=T\big]$ is
$\dist^{*}(T\cup\{e\})/(\dist^{*}(T)+\dist^{*}(T\cup\{e\}))=w_e/(1+w_e)=\rho_e$ when $T\cup\{e\}\in\Fset$,
and $0$ otherwise, proving~\eqref{eq:stationary-conditional-template}.
\end{proofof}

\xhdr{Using the maximum-entropy solution.} 
Consider setting the target marginals to $\pvec=\alpha\xvec$ in the concave program~\eqref{eq:maxent-primal-template}, where $\alpha\in(0,1)$ is a target selectability parameter. Given the Gibbs distribution $\dist^*$ on $\Fset$ with parameters $\wvec=(w_e)_{e\in\ground}$---as the solution to the concave program---Lemma~\ref{lem:gibbs-addability-template} shows that the simulate-then-replace meta-algorithm for $\dist^*$ takes a particularly simple form. Once an initial sample $\widehat{\Sset}\sim\dist^*$ is drawn, each arriving element $e$, if active, can be processed by (i) testing whether $T\cup\{e\}\in\Fset$ and (ii), if $T\cup\{e\}$ is feasible, sampling a Bernoulli random variable with parameter $\rho_e/x_e$ to decide whether to accept or reject $e$. See Algorithm~\ref{alg:str-gibbs} for a detailed description.\footnote{Once again, we note the similarity to Glauber dynamics~\citep{Vigoda:PhD, levin2017markov}} 

By the guarantee of the simulate-then-replace meta-algorithm (Proposition~\ref{prop:stationary-equivalence}), if $\dist^*$ is feasible for the stationary OCRS LP, then Algorithm~\ref{alg:str-gibbs} implements $\dist^*$ for every arrival order, and therefore yields an $\alpha$-selectable S-OCRS. Checking constraint~\ref{eq:soCRS-stationary} for a general distribution $\dist$ is not always easy, but Lemma~\ref{lem:gibbs-addability-template} is precisely what makes this step tractable for the special case of Gibbs distribution: For a Gibbs witness $\dist^*$, the  conditional probability
$\PP_{\Sset\sim\dist^*}\big[e\in\Sset \given \Sm{\Sset}{e}=T\big]$ depends only on whether $e$ is \emph{addable} to $T$, and in that case equals a constant $\rho_e$ that does not depend on the particular choice of $T$. We formalize this statement in the next subsection.

\medskip
\begin{algorithm}[H]
\caption{\textsc{Simulate-then-Replace} for a Gibbs distribution $\dist^*$}
\label{alg:str-gibbs}
\DontPrintSemicolon
\KwIn{Activation vector $\xvec\in\Ppoly$ and Gibbs distribution $\dist^*\in\simplex{\Fset}$ with parameters $\wvec\in \mathbb{R}_{>0}^\ground$ }
Sample a feasible simulated set $\widehat{\Sset}\sim \dist^*$

\ForEach{$e\in\ground$ revealed online (in the given order)}{
  $T\gets \Sm{\widehat{\Sset}}{e}$\tcp*[r]{\textcolor{blue}{\footnotesize{remove $e$ from the simulated set}}}
  \If{$e$ is active and $T\cup\{e\}\in\Fset$}{
    Accept element $e$ with probability $\frac{\rho_e}{x_e}:=\frac{w_e}{x_e(1+w_e)}$, and set $T\gets T \cup \{e\}$\tcp*[r]{\textcolor{blue}{\footnotesize{randomly accept $e$ if active and addable}}}}
   \Else 
   {Reject element $e$ \tcp*[r]{\textcolor{blue}{\footnotesize{reject element $e$ if inactive or adding it creates infeasibility}}}
  }
$\widehat{\Sset}\gets T$
}
\end{algorithm}

\subsection{Addability as the sufficient condition for stationarity}
\label{sec:addability}

\revcolor{For a (possibly random)  feasible set $T\in\Fset$ and an element $e\in\ground$, define the \emph{addability event}}
\begin{equation}
\label{eq:addability-event}
\revcolor{\Add_T(e)\ \defeq\ \{T\cup\{e\}\in\Fset\}.}
\end{equation}
\revcolor{When $T\sim\dist^*$ is drawn from the distribution under discussion, we simply drop the subscript and write $\Add(e)$. Note that since $\Sm{\Sset}{e}$ simply removes $e$ if it is present, the event $\{\Sm{\Sset}{e}\cup\{e\}\in\Fset\}$ is the same as $\{\Sset\cup\{e\}\in\Fset\}$.}
Under $\dist^{*}$, combining Lemma~\ref{lem:gibbs-addability-template} with the law of iterated expectations yields the factorization
\begin{equation}
\label{eq:marginal-factorization-template}
p_e
=\PP[e\in\Sset]
=\EE\!\left[\PP_{\Sset\sim\dist^{*}}\big[e\in\Sset \given \Sm{\Sset}{e}\big]\right]
=\PP[\Add(e)]\cdot \rho_e.
\end{equation}
Thus, once we set marginal probabilities $p_e=\alpha x_e$, \eqref{eq:soCRS-stationary}---which is equivalent to the constraint $\rho_e\leq x_e$---reduces to proving the following lower bound on addability:
\begin{equation}
\tag{\textsc{Addability}}
\label{eq:addability-lower-template}
\PP_{\Sset\sim\dist^{*}}[\Add(e)]\ \ge\ \alpha
\qquad\forall e\in\ground.
\end{equation}
Therefore, if~\eqref{eq:addability-lower-template} holds, $\dist^{*}$ provides a feasible solution to the stationary OCRS LP satisfying
Constraints~\eqref{eq:soCRS-selectability}--\eqref{eq:soCRS-stationary} with selectability $\alpha$.

\xhdr{Analysis via addability.}
Putting everything together, we have reduced the construction of a stationary OCRS with selectability $\alpha$ to establishing the addability lower bound~\eqref{eq:addability-lower-template} for the maximum-entropy distribution with marginals $\alpha\xvec$. Once established for an environment $(\ground,\Fset)$, \Cref{alg:str-gibbs} would be a $\alpha$-selectable S-OCRS for that environment.

\section{Maximum-Entropy for Matchings \& Uniform Matroids}
\label{sec:matching-uniform}
We now investigate each of the three promised feasibility environments. We begin with a warm-up in \Cref{sec:matching} for non-bipartite matchings (easily extendable to hypergraph matchings). We then turn to bipartite matchings in \Cref{sec:bipartite}, where we prove an optimal S-OCRS for this environment. Finally, in \Cref{sec:uniform}, we study $k$-uniform matroids and obtain an optimal S-OCRS for this setting.

The underlying approach is common across all three environments and follows the four-step maximum-entropy template introduced in \Cref{sec:socrs-template}. In particular, all algorithms are instantiations of Algorithm~\ref{alg:str-gibbs} for the corresponding Gibbs distribution. The only place where problem-specific structure is used is in establishing the addability lower bound in each case.

Finally, we note that a polynomial-time implementation of these algorithms requires two ingredients: (i) computing the Gibbs parameters $\wvec$ by solving~\eqref{eq:maxent-primal-template} (or its dual), and (ii) sampling (exactly or approximately) an initial set $\widehat{S}\sim\dist^*$ in polynomial time. We show that both steps admit polynomial-time routines for (bipartite and general) matchings and uniform matroid feasibility environments studied in the current section, and we defer the corresponding results and technical details to \Cref{apx:time-complexity}. 

\vspace{-2mm}
\subsection{Warm-up: matchings in non-bipartite  graphs}
\label{sec:matching}
Recall from \Cref{sec:prelim} the (non-bipartite) matching environment defined on a graph $G=(V,\ground)$. Our first application of the maximum-entropy template gives a simple S-OCRS for this setting. Our result matches the most basic analysis of the algorithm of \cite{ezra2022prophet}.

\begin{theorem}[S-OCRS for matchings]
\label{thm:graph-socrs}
For the matching environment $(\ground,\Fset)$ and every $\xvec\in\Ppoly$, there exists a Gibbs distribution $\dist^*$ such that Algorithm~\ref{alg:str-gibbs}, when instantiated with $\dist^*$, is a $\tfrac{1}{3}$-selectable S-OCRS.
\end{theorem}

\begin{proofof}{Proof}
Fix any $\xvec\in\Ppoly$ and set $\alpha\defeq \tfrac{1}{3}$ and $\pvec\defeq \alpha\,\xvec$.
Let $\dist^{*}$ be the maximum-entropy distribution---that is, solution to \eqref{eq:maxent-primal-template}---with marginals $\pvec$. As we discussed in \Cref{sec:addability}, we only need to verify the  condition in \eqref{eq:addability-lower-template}.
Fix an edge $e=\{u,v\}\in\ground$. The addability event
\[
\Add(e)\ =\ \{\Sm{\Sset}{e}\cup\{e\}\in\Fset\}
\]
is exactly the event that neither endpoint of $e$ is matched by $\Sm{\Sset}{e}$. Now, for each vertex $v\in V$, we define the following ``matched'' event:
\[
\Occ(v)\ \defeq\ \Bigl\{\exists f\in \Sset\ \text{with}\ v\in f\Bigr\}.
\]
Since $\Sset$ is always a matching, for any fixed $v$ the events $\{f\in\Sset\}$ over distinct $f\in\delta(v)$ are disjoint.
Therefore,

\begin{equation}
\label{eq:graph-occ-bound}
\PP_{\Sset\sim\dist^{*}}\!\big[\Occ(v)\big]
\ =\ \sum_{f\in\delta(v)}\PP_{\Sset\sim\dist^{*}}[f\in\Sset]
\ =\ \sum_{f\in\delta(v)} p_f.
\end{equation}
Using $p_f=\alpha x_f$ and~\eqref{eq:hg-vertex-packing}, we get
\begin{equation}
\label{eq:graph-occ-alpha}
\PP_{\Sset\sim\dist^{*}}\!\big[\Occ(v)\big]
\ =\ \alpha\sum_{f\in\delta(v)} x_f
\ \le\ \alpha.
\end{equation}
Finally, $\Add(e)$ fails only if $\Occ(u)$ or $\Occ(v)$ occurs, so by a union bound and~\eqref{eq:graph-occ-alpha},
\[
\PP_{\Sset\sim\dist^{*}}[\Add(e)]
\ \ge\ 1-\PP[\Occ(u)]-\PP[\Occ(v)]
\ \ge\ 1-2\alpha
\ =\ \alpha.
\]
This is exactly~\eqref{eq:addability-lower-template}. The conclusion follows.
\end{proofof}

We should highlight that we use the same style of the above simple analysis to extend our results to other feasibility environments in the following subsection. Also, an almost identical argument to the proof of \Cref{thm:graph-socrs} shows the existence of $\frac{1}{L+1}$-selectable S-OCRS for rank-$L$ hypergraph matching. We postpone the details to \Cref{apx:hypergraph}. This bound matches the common benchmark in this environment (and the known upper-bound on the best possible factor achievable by slightly restricted OCRS algorithms, c.f.  \cite{ma2026online}).

\vspace{-2mm}
\subsection{Bipartite matchings}
\label{sec:bipartite}
Recall from \Cref{sec:prelim} the bipartite matching environment defined on a graph $G=(U\cup V,\ground)$. Our next application of the maximum-entropy template yields an optimal S-OCRS for this setting.
\begin{theorem}[Optimal S-OCRS for bipartite matchings]
\label{thm:bipartite-tight}
For the bipartite matching environment $(\ground,\Fset)$ and every fractional matching $\xvec\in\Ppoly$, there exists a Gibbs distribution $\dist^*$ such that Algorithm~\ref{alg:str-gibbs}, when instantiated with $\dist^*$, is an $\alphaStar$-selectable S-OCRS with $\alphaStar=\frac{3-\sqrt{5}}{2}\approx 0.382$. Moreover, $\alphaStar$ is the best possible selectability achievable by any S-OCRS algorithm.
\end{theorem}
Before proceeding to the proof, a few remarks are in order. As we show in \Cref{apx:time-complexity}, our algorithm runs in polynomial time. Consequently, it resolves an open problem  in \cite{ma2024onlineM} by providing an improved OCRS for bipartite matchings in the non-vanishing regime, and also results in a prophet inequality with the same competitive ratio. Prior to our work, \cite{gravin2023prophet} obtained a $1/3$-prophet inequality via a simple pricing-based policy. This factor was subsequently improved by \cite{ezra2022prophet} to $0.337$, and then by  \cite{macrury2025random} to $0.349$. Our work improves all of these results.

Not only is $\alpha^*=(3-\sqrt{5})/2$ the best bound achievable by any S-OCRS algorithm, but it is also the best bound attainable by a broad class of OCRS policies satisfying the so-called \emph{concentration property}, and is conjectured to be optimal for general matching OCRSs~\citep{ma2024onlineM}.

\smallskip
\xhdr{Positive result.} The proof of the positive result follows the same addability argument as in \Cref{sec:matching}, but it relies on the following well-known \emph{positive correlation} lemma for Gibbs measures over matchings in bipartite graphs (see Theorem 6.3 in \cite{heilmann1972theory} or Lemma 22.15 in \cite{molloy2002hard}), which has been extensively studied in combinatorics and statistical physics, sometimes under the name ``monomer-dimer systems.''
To state this lemma, for $z\in U\cup V$, define the following event (similar to \Cref{sec:matching})
\vspace{-1mm}
\[
\Occ(z)\ \defeq\ \Bigl\{\exists f\in \Sset\ \text{with}\ z\in f\Bigr\}.
\]

\begin{lemma}[Gibbs positive correlation for bipartite matchings]
\label{correlation_lemma}
When $S\subseteq E$ is sampled from a Gibbs distribution on matchings of a bipartite graph $G=(U\cup V,\ground)$, the events $\Occ(u)$ and $\Occ(v)$ (equivalently, their complements) for $u\in U$ and $v\in V$ are positively correlated.
\end{lemma}

We first prove Theorem~\ref{thm:bipartite-tight} assuming Lemma~\ref{correlation_lemma}, and then provide a combinatorial proof of the lemma for completeness in \Cref{apx:monomer-dimer}.

\begin{proofof}{Proof of Theorem~\ref{thm:bipartite-tight}}
Fix any $\xvec\in\Ppoly$ and set $\alpha\defeq \alphaStar$ and $\pvec\defeq \alpha\,\xvec$.
Let $\dist^{*}$ be the maximum-entropy distribution---that is, the solution to~\eqref{eq:maxent-primal-template}---with marginals $\pvec$.
Based on the discussion in \Cref{sec:addability}, we only need to verify the condition in~\eqref{eq:addability-lower-template}.
Fix an edge $e=\{u,v\}\in\ground$ with $u\in U$ and $v\in V$.
The addability event~\eqref{eq:addability-event} is
\[
\Add(e)\ =\ \bigl\{\Sm{\Sset}{e}\ \text{contains no edge incident to $u$ or $v$}\bigr\}.
\]
As before, for any $z\in U\cup V$ the events $\{f\in\Sset\}$ over distinct $f\in\delta(z)$ are disjoint. Therefore, we have that
\[
\PP_{\Sset\sim\dist^{*}}\!\big[\Occ(z)\big]
\ =\ \sum_{f\in\delta(z)}\PP_{\Sset\sim\dist^{*}}[f\in\Sset]
\ =\ \sum_{f\in\delta(z)} p_f.
\]
Using $p_f=\alpha x_f$ and feasibility in expectation in \eqref{eq:fractional-matching-constraints}, we obtain
\begin{equation}
\label{eq:bip-occ-alpha}
\PP_{\Sset\sim\dist^{*}}\!\big[\Occ(z)\big]
\ =\ \alpha\sum_{f\in\delta(z)} x_f
\ \le\ \alpha.
\end{equation}
If $u$ and $v$ are simultaneously unmatched, then $\Add(e)$ occurs. Hence
\begin{align*}
\PP_{\Sset\sim\dist^{*}}[\Add(e)]
&\ge \PP_{\Sset\sim\dist^{*}}\big[\overline{\Occ(u)}\cap\overline{\Occ(v)}\big] \\
&\ge \PP\big[\overline{\Occ(u)}\big]\,\PP\big[\overline{\Occ(v)}\big] \qquad \text{(by Lemma~\ref{correlation_lemma})}\\
&= (1-\PP[\Occ(u)])(1-\PP[\Occ(v)]) \\
&\ge (1-\alpha)^2
\ =\ \alpha. \qquad \text{(by the choice of $\alpha=\alphaStar=\frac{3-\sqrt{5}}{2}$)}
\end{align*}
This is exactly~\eqref{eq:addability-lower-template} and finishes the proof of $\alphaStar$-selectability.\footnote{We \revcolor{believe that our proof should extend} to the setting of non-bipartite matchings in the vanishing regime studied in \cite{ma2024onlineM} due to correlation decay arguments from the literature implying that Lemma \ref{correlation_lemma} holds approximately.}
\end{proofof}

\xhdr{Negative results.} To show that $\alphaStar=\frac{3-\sqrt{5}}{2}$ is the best possible selectability constant, we prove the following lemma---making essential use of our LP characterization in Proposition~\ref{prop:stationary-equivalence}---by constructing a counterexample for the stationary OCRS LP. We defer the proof to \Cref{apx:bipartite-impossiblity}.

\begin{lemma}[Bipartite matching upper-bound]
\label{lemma:bipartite-impossibility}
For every $\eps\in(0,1/2)$ there exists a bipartite graph $G$ and a fractional matching $\xvec\in\Ppoly$ such that any distribution $\dist\in\simplex{\Fset}$ satisfying the stationary constraints~\eqref{eq:soCRS-selectability}--\eqref{eq:soCRS-stationary} has selectability at most $\alphaStar+O(\eps)$, where $\alphaStar=\frac{3-\sqrt{5}}{2}$. In particular, letting $\eps\to 0$ implies that every S-OCRS has selectability at most $\alphaStar$.
\end{lemma}

We conclude this subsection by highlighting a limitation of the maximum-entropy approach. We have shown that this approach achieves selectability $\alphaStar=(3-\sqrt{5})/2$ for \emph{bipartite} matchings. A natural question is whether the same approach can yield an $\alphaStar$-selectable S-OCRS for general (non-bipartite) matchings.\footnote{\citet{ma2024onlineM} conjecture that in the general matching environment no OCRS can be better than $\alphaStar$-selectable, even in the vanishing regime with tree graphs, and that $\alphaStar$ should be achievable by some (ideally polynomial-time) OCRS algorithm.} Unfortunately, we show that the maximum-entropy approach \emph{does not} extend to general matchings to obtain $\alphaStar$ (recall that we also showed this approach yields a $\frac{1}{3}$-selectable S-OCRS in Theorem~\ref{thm:graph-socrs}). In particular, already on a four-vertex example (also studied in section 2.3 of \cite{macrury2025random}), the maximum-entropy distribution fails to provide selectability better than $(\sqrt{3}-1)/2<\alphaStar$. We conjecture that the maximum-entropy distribution always provides an S-OCRS with selectability $(\sqrt{3}-1)/2$, which would beat the best known OCRS for this environment; we leave this as an open question. 
(formalized impossibility result in \Cref{apx:K4-maxent-barrier}) Evidence suggests that, in fact, \textit{no S-OCRS} can obtain a selectability of $\alphaStar$ for general matchings.\footnote{Let $V=\{a,b,c,u,v,h\}$ and $G=K_6$. Define the fractional matching $x^\varepsilon$ by $x^\varepsilon_e= \frac13-\frac{2\varepsilon}{3}$ for $e\in\{ab,ac,bc\}$, $\frac16+\frac{\varepsilon}{6},$ for $e\in\{au,av,bu,bv,cu,cv\}$, 
$\frac12-\frac{3\varepsilon}{2},$ for $e\in\{hu,hv\},$ and 
$\varepsilon$ for $e\in\{uv,ah,bh,ch\}.$ Numerical evidence with the S-OCRS LP as
$\varepsilon\to 0$ suggests that we cannot obtain a
selectability of $(3-\sqrt5)/2$ on this instance.}

\subsection{The \texorpdfstring{$\mathbf{k}$}{k}-selection problem (\texorpdfstring{$\mathbf{k}$}{k}-uniform matroid)}
\label{sec:uniform}
Recall from \Cref{sec:prelim} the $k$-uniform matroid environment on a finite ground set $\ground$, where a set $S\subseteq \ground$ is feasible iff $|S|\le k$. Our final application of the maximum-entropy template in this section yields an optimal S-OCRS  with selectability factor $\alpha_k$ for this environment, where 
\begin{equation}
\label{eq:alpha-k-def}
\alpha_k
\ \defeq\
\PP[Q<k \given Q\le k]=\frac{\PP[Q<k]}{\PP[Q\leq k]}\approx 1-\sqrt{\frac{2}{\pi k}}+\mathcal{O}\left(\frac{1}{k}\right), \qquad \text{where } Q\sim \mathrm{Poisson}(k).
\end{equation}

\begin{theorem}[Optimal S-OCRS for the $\mathbf{k}$-uniform matroid]
\label{thm:uniform-matroid}
For the $k$-uniform matroid environment $(\ground,\Fset)$ and every $\xvec\in\Ppoly$, there exists a Gibbs distribution $\dist^*$ such that Algorithm~\ref{alg:str-gibbs}, when instantiated with $\dist^*$, is an $\alpha_k$-selectable S-OCRS. Moreover, $\alpha_k$ is the best possible selectability achievable by any S-OCRS.
\end{theorem}

Before turning to the proof, we briefly place Theorem~\ref{thm:uniform-matroid} in context. The $k$-uniform matroid (selecting at most $k$ active elements) is one of the most extensively studied OCRS environments in both computer science and operations research. As mentioned in the introduction, the seminal \emph{magician's algorithm} of \citet{alaei2014bayesian} is optimal for this setting; it is obtained by solving an appropriate dynamic program for the best selectability via LP duality. The analysis in \citet{alaei2012online} established the explicit guarantee $1-\sqrt{1/(k+3)}$, and \citet{jiang2025tightness} later made this analysis tight, proving optimality and characterizing the optimal selectability through a factor-revealing LP. The best explicit asymptotic bound currently available is due to \citet{wang2018online}, who analyze the DP-based optimal policy in the vanishing regime (infinitesimal activation probabilities) and obtain the explicit factor $1-\sqrt{2/(\pi k)}+\mathcal{O}(1/k)$; since the vanishing regime is worst-case for the standard OCRS problem, this bound also applies to general instances. Finally, \citet{dinev2024simple} propose a simpler adaptive, combinatorial scheme that achieves a near-optimal factor of $1-\mathcal{O}(\sqrt{1/k})$, but this algorithm is ad-hoc and somewhat difficult to analyze.

In contrast to these DP/LP-based approaches, our policy is simple and highly interpretable. Algorithm~\ref{alg:str-gibbs} is almost static: it only maintains a simulated set $\widehat{S}$ as an \emph{imaginary state} of the system, and makes local acceptance decisions so as to preserve the target distribution of $\widehat{S}$. For the uniform matroid, this can be viewed as a minor modification of the \emph{greedy} rule: we run greedy on the imaginary state $\widehat{S}$, but accept an active element $e$ only with a fixed probability $\rho_e/x_e$ (rather than with probability $1$). This yields the explicit selectability $\alpha_k$ for all $k$ (and, in particular, recovers the optimal factor $1/2$ when $k=1$). Moreover, as we show in our companion paper~\ifblind\citep{anonymous2026optimal}\else\citep{aminian2026optimal}\fi, the stationarity property of our S-OCRS approach can be leveraged to obtain the same factor $\alpha_k$ for the \emph{reusable} variant of the OCRS problem (and the reusable Bayesian online allocation problem), thereby improving upon the previously best known bound of $1-\mathcal{O}\left(\sqrt{\log(k)/k}\right)$ due to \citet{feng2022near}. In that companion paper, we also further simplify the algorithm and its analysis.

\smallskip
\xhdr{Positive result.} The proof of selectability in \Cref{thm:uniform-matroid} again follows the addability argument from \Cref{sec:matching}. The main additional ingredient is the following technical lemma (more general results are known in the literature, see \cite{marsiglietti2026concentration}, and the citations therein), comparing truncated Poisson and truncated Poisson--binomial distributions and establishing a consequence of a second-order stochastic dominance relationship between them.

\begin{lemma}[Poisson comparison]
\label{lem:poisson-comparison}
Let $\revcolor{W}=\sum_{e\in\ground} Y_e$ be a sum of independent Bernoulli random variables (with arbitrary success probabilities), and let $Q$ be a Poisson random variable (with arbitrary mean). If $~\EE[\revcolor{W} \given \revcolor{W}\le k] \le \EE[Q \given Q\le k]$, then
\begin{equation}
\label{eq:ratio-lowerbound}
\PP[\revcolor{W} < k \given \revcolor{W} \le k]\ \ge\ \PP[Q < k \given Q\le k].
\end{equation}
\end{lemma}

We first prove Theorem~\ref{thm:uniform-matroid} assuming Lemma~\ref{lem:poisson-comparison}, and then give a self-contained proof of the lemma in \Cref{apx:poisson-comparison} for the sake of completeness, using the \emph{ultra-log-concavity} of the Poisson--binomial distribution~\citep{pemantle2000towards}.

\begin{proofof}{Proof of Theorem~\ref{thm:uniform-matroid}}
Fix any $\xvec\in\Ppoly$ and set $\alpha\defeq \alpha_k$ and $\pvec\defeq \alpha\,\xvec$.
Once again, let $\dist^{*}$ be the maximum-entropy distribution---that is, the solution to~\eqref{eq:maxent-primal-template}---with marginals $\pvec$, and draw $\Sset\sim\dist^{*}$.
Based on the discussion in \Cref{sec:addability}, it suffices to verify the addability lower bound~\eqref{eq:addability-lower-template} for $\alpha=\alpha_k$. For the $k$-uniform matroid, the addability event~\eqref{eq:addability-event} is
\begin{equation}
\label{eq:addability-uniform}
\Add(e)\ =\ \bigl\{\ |\Sm{\Sset}{e}| < k\ \bigr\}.
\end{equation}
In particular, $\Add(e)$ holds whenever $|\Sset|<k$. Since $\dist^{*}$ is a Gibbs distribution, we may write $\dist^{*}(S)\propto \prod_{e\in S} w_e$ over all $S$ with $|S|\le k$, for some weights $w_e>0$. Define $\rho_e\defeq w_e/(1+w_e)$, and let $(Y_e)_{e\in\ground}$ be independent Bernoulli random variables with $\PP[Y_e=1]=\rho_e$. Then, for every $S\subseteq \ground$,
\[
\PP\big[\{e:Y_e=1\}=S\big]
=
\prod_{e\notin S} (1-\rho_e)\cdot \prod_{e\in S} \rho_e
=
\Big(\prod_{e\in\ground}(1-\rho_e)\Big)\cdot \prod_{e\in S} w_e,
\]
so conditioning on the event $\{\revcolor{W}\le k\}$, where $\revcolor{W}\defeq\sum_{e\in\ground} Y_e$, yields the Gibbs distribution with parameters $\wvec$ on feasible sets (sets of size at most $k$). Equivalently, $\Sset\sim\dist^{*}$ has the same distribution as $\{e:Y_e=1\}$ conditioned on $\{\revcolor{W}\le k\}$.

We now verify the mean condition in Lemma~\ref{lem:poisson-comparison}. On the one hand,
\[
\EE[\revcolor{W} \given \revcolor{W}\le k]
= \EE[|\Sset|]
= \sum_{f\in\ground}\PP[f\in\Sset]
= \sum_{f\in\ground} p_f
= \alpha_k\sum_{f\in\ground} x_f
\le k\alpha_k~,
\]
where the last inequality uses feasibility of $\xvec$.
On the other hand, for $Q\sim\mathrm{Poisson}(k)$\footnote{If $\xvec \in b\Ppoly$, we may let $Q\sim\mathrm{Poisson}(bk)$.}, \revcolor{we know that $t\,\PP[Q=t]=k\,\PP[Q=t-1]$ for $t\ge1$, so}
\[
\EE[Q\Ind{Q\le k}]
= \sum_{t=1}^k t\,\PP[Q=t]
= \sum_{t=1}^k k\,\PP[Q=t-1]
= k\,\PP[Q<k],
\]
and hence
\[
\EE[Q\given Q\le k]
= \frac{\EE[Q\Ind{Q\le k}]}{\PP[Q\le k]}
= k\,\PP[Q<k\given Q\le k]
= k\alpha_k.
\]
Therefore $\EE[\revcolor{W}\given \revcolor{W}\le k]\le \EE[Q\given Q\le k]$.
Applying Lemma~\ref{lem:poisson-comparison} gives
\[
\PP_{\Sset\sim\dist^{*}}\big[\,|\Sset|<k\,\big]
= \PP[\revcolor{W}<k\given \revcolor{W}\le k]
\ge \alpha_k.
\]
Since $\Add(e)$ holds whenever $|\Sset|<k$, the above inequality implies $\PP[\Add(e)]\ge \alpha_k$ for every $e$, which is exactly~\eqref{eq:addability-lower-template} for $\alpha=\alpha_k$---hence finishing the proof.
\end{proofof}

\xhdr{Negative results.} We next show that $\alpha_k$ is the best possible selectability constant in Theorem~\ref{thm:uniform-matroid}.
\begin{lemma}[Upper bound approaching $\alpha_k$]
\label{prop:uniform-tightness}
Fix $k\ge 1$. For every $n\ge k$, consider the symmetric instance $|\ground|=n$ and $x_e=q\defeq k/n$ for all $e\in\ground$.
Any distribution $\dist\in\simplex{\Fset}$ satisfying the stationary OCRS LP constraints~\eqref{eq:soCRS-selectability}--\eqref{eq:soCRS-stationary} has selectability at most
\begin{equation}
\label{eq:binomial-ratio}
\frac{\PP\big[\mathrm{Bin}(n-1,q) < k\big]}{\PP\big[\mathrm{Bin}(n,q)\le k\big]}.
\end{equation}
In particular, letting $n\to\infty$ implies that every S-OCRS has selectability at most $\alpha_k$.
\end{lemma}

We defer the proof of Lemma~\ref{prop:uniform-tightness} to \Cref{apx:uniform-tightness}.

\section{Beyond Maximum-Entropy: Network Revenue Management}
\label{sec:NRM}
In this section, we study a feasibility environment that shares similarities with both hypergraph matchings and $k$-uniform matroids, being a common generalization of both. As mentioned earlier, this class is commonly known as the \emph{network revenue management (NRM)} environment, which is mathematically equivalent to a capacitated version of the hypergraph matching environment. While, as in the previous section, we will use a Gibbs distribution to solve the S-OCRS LP, we will not solve a maximum-entropy convex program.

Recall from \Cref{sec:prelim} the NRM environment, i.e., the rank-$L$ hypergraph matching environment with vertices with capacity $k$, defined on a hypergraph $H = (V, E)$. In this case, we have: 
   $$\Fset  \defeq \Bigl\{S\subseteq \ground: \forall v\in V, \bigl|\{e\in S: v\in e\}\bigr|\le k\Bigr\},$$ and for every $\xvec\in \Ppoly$, we have
\[
\xvec\in [0,1]^E\text{ and }\sum_{e\in \delta(v)} x_e \le k \qquad \forall v\in V.\]

\smallskip
\xhdr{A simpler algorithm without maximum-entropy.} We show that for this feasibility environment, the Gibbs distribution with $\rho_e = \gamma x_e$ with $\gamma = 1-O(\sqrt{\log L / k})$, yields a selectability of $1-O(\sqrt{\log L / k})$. While this choice of $\rho_e$ does not arise as a solution to a max-entropy program\footnote{We do not claim that the max-entropy approach does not work in this feasibility environment; our choice of Gibbs distribution gives a simple analysis and therefore we only consider studying this choice.}, we can still run Algorithm~\ref{alg:str-gibbs}. The advantage of this choice is that we accept each active element with the same constant probability $\gamma$, and the resulting algorithm is an even simpler greedy-with-homogeneous-randomized-discarding algorithm (started from the stationary distribution). Furthermore, \eqref{eq:soCRS-stationary} is satisfied automatically. 

To formally state the result, fix $\varepsilon\in [0,1)$. Let
\[
Q\sim\operatorname{Poisson}((1-\varepsilon)k),\quad
\beta:=\PP[Q=k\given Q\le k],\quad
\gamma:=(1-\varepsilon)(1-\beta).
\]
Let $\dist$ be the Gibbs distribution on $\Fset$ with parameters $\rho_e=\gamma x_e$. 

\begin{theorem}[S-OCRS for rank-$L$ hypergraph matchings with capacity $k$]
\label{thm:constant-scaling-k-uniform}
For the rank-$L$ hypergraph matchings with capacity $k$ environment $(\ground,\Fset)$ and every $\xvec\in\Ppoly$, the Gibbs distribution $\dist$ with $\rho_e = \gamma x_e$ and $\gamma$ as defined above is such that Algorithm~\ref{alg:str-gibbs}, when instantiated with $\dist$, is a  $\gamma(1-L\beta)$-selectable S-OCRS. In particular, for $k > 2\log L$, taking $\varepsilon=\sqrt{2\log L/k}$ gives selectability
\[
1-\sqrt{\frac{2\log L}{k}} -O\left(\sqrt{\frac{1}{k}}\right).
\]
\end{theorem}

The full proof of Theorem~\ref{thm:constant-scaling-k-uniform} is deferred to~\Cref{apx:constant-scaling-k-uniform}. The main idea is that to establish $\PP_{\Sset\sim\dist}[e\in\Sset] \ge \gamma (1 - L\beta)x_e,$ it suffices to show that $
 \PP[\mathrm{Add}(e)] \ge 1-L\beta.$ We prove this by using \Cref{lem:poisson-comparison} to show that each individual vertex is at capacity with probability at most $\beta$, and then applying a union bound over the $L$ vertices that $e$ includes.

We note that there is also an ordinary OCRS with a slightly stronger selectability guarantee. First, independently thin each active edge with probability $1-\varepsilon$. Then, at every vertex $v$, run a separate virtual $k$-uniform-matroid S-OCRS on the thinned edges in $\delta(v)$, and accept an edge $e$ precisely when all vertices $v\in e$ accept it in their virtual schemes. Since $\sum_{e\in\delta(v)}(1-\varepsilon)x_e\le(1-\varepsilon)k$, the proof of Theorem~\ref{thm:uniform-matroid} shows that each virtual scheme is $(1-\beta)$-selectable, and hence rejects a thinned edge with probability at most $\beta$. By a union bound,
\[
 \PP[e\text{ is accepted}\given e\text{ is active}]
 \ge (1-\varepsilon)(1-|e|\beta)
 \ge (1-\varepsilon)(1-L\beta).
\]
We show in \Cref{apx:time-complexity} that the uniform-matroid S-OCRS is polynomial-time implementable, so this OCRS is polynomial-time implementable as well. Although every virtual scheme is stationary, their intersection need not be, so this construction is not an S-OCRS.

\section{Application to Bayesian Online Allocation of Reusable Products}
\label{sec:reusable-product-application}
We now demonstrate the application of S-OCRS to Bayesian online allocation problems with reusable products under homogeneous Poisson arrivals of types. This reduction, at a high level, follows the same propose-then-admit framework as our companion paper~\ifblind\cite{anonymous2026optimal}\else\cite{aminian2026optimal}\fi---inspired by earlier related work in the revenue management literature~\citep{feng2024near,feng2022near}. 

\subsection{A Bayesian reusable-product model}
\label{sec:roduct-model}
Fix a set of \textit{products} $\ground$, and $\mathcal{N} \subseteq \mathbb Z_{\geq 0}^{\ground}$ a family of multisets of products that can be feasibly allocated at any particular time. Note that we allow for multiple units of the same product to be simultaneously allocated, but we will always assume that at most $N$ units can be allocated, so $\mathcal{N} \subseteq [N]^{\ground}$. We will also assume that $\mathcal{N}$ is downward closed, i.e., if $n\in\mathcal{N}$ and $n'\le n$ coordinatewise, then $n'\in\mathcal{N}$. Let $\Ppoly$ be the convex hull of $\mathcal{N}$.

Let $\mathcal I$ be a finite set of request types.  Requests of type $i\in\mathcal I$ arrive according to an independent homogeneous Poisson process with constant rate $\arrivalRate_i > 0$.  When a type-$i$ request arrives, the algorithm may reject it or allocate one unit of a product $e\in\ground$.  Such an allocation earns reward $\reward_{ie}\ge 0$ and occupies the unit of $e$ for a duration that is independently exponentially distributed with mean $\ell_{ie}>0$, after which the unit becomes available again.  The rates, rewards, duration means, and feasibility environment are known to the algorithm.  The allocation decision is immediate and irrevocable.  If $N_e(t)$ is the number of units of product $e$ occupied at time $t$, the feasibility requirement is
\[
(N_e(t))_{e \in \ground}\in\mathcal{N}\quad\text{for every }t\ge 0.
\]
For an online policy $\pi$, let $N_{ie}^{\pi}(T)$ denote the number of type-$i$ requests allocated a unit of product $e$ during $[0,T]$.  The long-run average reward of the policy is:
\begin{equation}
\operatorname{R}(\pi)
 \defeq
\liminf_{T\to\infty}\frac{1}{T}
\EE\left[\sum_{i\in\mathcal I}\sum_{e\in\ground}\reward_{ie}N_{ie}^{\pi}(T)\right].
\label{eq:long-run-reward}
\end{equation}
We write $\operatorname{OPT}$ for the supremum of~\eqref{eq:long-run-reward} over all (online and offline) policies, including those policies that have knowledge of all future arrivals at the time of making an allocation decision.\footnote{We assume all policies are ignorant of the time at which an allocated unit will return to the system, until the unit actually returns.} Thinking of $\operatorname{OPT}$ as the benchmark, we would like to identify an online (and ideally polynomial-time implementable) policy $\pi$ for which $\operatorname{R}(\pi) \geq \alpha \operatorname{OPT}$ holds for all problem instances, for a constant $\alpha\in(0,1]$ as large as possible---referred to as the \emph{competitive ratio} of the online policy $\pi$. 

\subsection{The static fluid relaxation and proposal step}
\label{sec:fluid}
Consider the following static fluid relaxation LP for the problem:
\begin{align}
\operatorname{FLP} \defeq\
\max_{y_{ie}\ge 0}\quad
&\sum_{i\in\mathcal I}\sum_{e\in\ground}\reward_{ie}y_{ie}
\label{eq:fluid-objective}\\
\textnormal{s.t.}\quad
&\sum_{e\in\ground}y_{ie}\le \arrivalRate_i
&&\forall i\in\mathcal I,
\label{eq:fluid-arrivals}\\
&x_e=\sum_{i\in\mathcal I}\ell_{ie}y_{ie}
&&\forall e\in\ground,
\label{eq:fluid-load}\\
&\xvec\in\Ppoly.
\label{eq:fluid-feasibility}
\end{align}
Here, we interpret $y_{ie}$ as the long-run average rate at which type-$i$ requests are allocated a unit of product $e$ by a policy, and constraint \eqref{eq:fluid-load} gives the corresponding long-run average load on product $e$. The long-run reward then equals \eqref{eq:fluid-objective}. The constraint \eqref{eq:fluid-arrivals} is a consequence of the fact that request of type $i$ can only be allocated one unit each,  while \eqref{eq:fluid-feasibility} follows from averaging the fact that at each time $t$, the set of allocated goods must be feasible. We therefore have the following proposition:

\begin{proposition}
\label{prop:fluid-upper}
For the reusable-product allocation problem, $\operatorname{OPT}\le \operatorname{FLP}$. In particular, to identify a policy with $\operatorname{R}(\pi) \geq \alpha \operatorname{OPT}$, it suffices to identify a policy with $\operatorname{R}(\pi) \geq \alpha \operatorname{FLP}$. 

\end{proposition}

Note that $\operatorname{FLP}$ can be solved in polynomial time as long as there is a polynomial-time separation oracle for $\Ppoly$~\citep{gls88}. In some settings, we may choose to replace $\Ppoly$ with a relaxation $\widetilde\Ppoly$ to ensure $\operatorname{FLP}$ is solvable in polynomial time.

In order to use the static fluid relaxation LP to design an online policy, let $\{y_{ie}^*\}$ be an optimal solution to this fluid LP. Consider an online policy that upon the arrival of a type-$i$ request, \textit{proposes} each product $e$ (mutually exclusively) with probability $y_{ie}^*/\arrivalRate_i$, and makes no proposal with the remaining probability. Then, if the proposal is \textit{admitted} by the product $e$, the request is allocated a unit of product $e$; else the request is rejected with no product allocated.

The proposals for every pair $(i,e)$ form independent homogeneous Poisson processes of rate $y_{ie}^*$.  If every proposal were admitted, the reward of the policy would be $\operatorname{FLP}$, with each product receiving a load of $x_e$. Unfortunately, it is not  possible to admit every request since we must have $(N_e(t))_{e \in \ground} \in \mathcal{N}$. However, if we can show that every request in the long run can be admitted with probability $\alpha$, this provides a way to turn the ex-ante feasible proposals each product receives into an ex-post feasible allocation, at a cost of only a factor $\alpha$ for the rate which allocations are made. Thus, if we can find an admission rule that ensures that every proposal in the long run is accepted with probability $\alpha$, we obtain a policy such that $\operatorname{R}(\pi) \geq \alpha \operatorname{OPT}$.

\subsection{S-OCRSs are sufficient for admission control}
\label{sec:recurring-sufficient-application}

For the purpose of the analysis, it is more convenient to work in a setting where the rate of proposals for every pair $(i, e)$ is small; this is because to apply recurring OCRSs, we need consecutive requests for a product not to overlap. To achieve this goal, we can split each unit of a product into virtual clones, and decide when a proposal arrives which of the virtual clones it actually proposes to. This splitting also allows us to apply recurring OCRSs designed for feasibility families of \textit{sets} to the setting with the multiset $\mathcal{N}$.

Formally speaking, an \emph{$n$-split} of $(\ground,\mathcal{N})$ replaces every product $e$ by a finite nonempty set $C_e^{(n)}\defeq\{e^1,\ldots,e^n\}$ of virtual clones.  Writing $\ground_n=\bigsqcup_{e\in\ground}C_e^{(n)}$, the feasible family of sets associated with $E_n$ is:
\[\Fset_n
\defeq
\left\{S \subseteq\ground_n: (|S\cap C_e^{(n)}|)_e \in\mathcal{N}\right\},\]
and we let $\Ppoly_n\defeq\conv(\Fset_n).$ For a sufficiently large $n$ where each product is split into $n$ virtual clones, it turns out that an S-OCRS may be used as an approximate admission rule. We formalize this in the following lemma:

\begin{lemma}
\label{lem:recurring-implies-admission}
Suppose that for every $n$-split of  $(\ground,\mathcal{N})$, there is an $\alpha$-selectable S-OCRS. Then, for every $\varepsilon>0$, there is an admission rule which ensures that $\lim_{T\to\infty}\frac{1}{T}
\EE\left[N_{ie}^{\pi}(T)\right]\geq (\alpha-\varepsilon) y_{ie}$.
\end{lemma}

\revcolor{Although the construction can also be implemented using the recurring OCRS definition, the formal argument is simpler using S-OCRSs.}
\begin{proofof}{Proof} Let $\xvec \in \Ppoly$ be the vector of loads for the products in $E$ for which we desire an admission rule, so:
\[x_e=\sum_{i\in\mathcal I}\ell_{ie}y_{ie},\]
where $y_{ie}$ are rates of proposals. Note that $\|\xvec\|_\infty \leq N$. For $n \geq N \cdot \lceil \frac{1}{\varepsilon}\rceil$, consider the $n$-split $(\ground_n, \Fset_n)$. For $e\in\ground$, and $ j\in[n]$, let $ x_{e^j}^{(n)}
:= \frac{x_e}{n+x_e},$
and note that $\mathbf{x}^{(n)} \in \Ppoly_n$ (because $\frac{\mathbf{x}}{n}\in \Ppoly_n$, and $\frac{x_e}{n+x_e}\leq \frac{x_e}{n}$). From our assumption that there is an $\alpha$-selectable S-OCRS for $(\ground_n, \Fset_n)$ and $\mathbf{x}^{(n)}$, we know that there is a distribution $\mu_n$ such that  for every clone $e^j$ and every $T \in\Fset_n$ with $e^j \notin T$:
\[\frac{\mu_n (T \cup \{e^j\})}{\mu_n (T) +\mu_n (T \cup \{e^j\})} \leq  \frac{x_e}{n+x_e} \implies \mu_n (T \cup \{e^j\}) \leq \frac{x_e}{n} \mu_n (T).\]
Furthermore, note also that: 
\[\PP_{S\sim \mu_n}[e^j \in S] \geq \alpha \frac{x_e}{n + x_e} \geq (\alpha - \varepsilon)\frac{x_e}{n}.\]
We now define the admission rule we consider. For each arriving proposal for a pair $(i, e)$, assign it uniformly at random to one of the $n$ clones $e^1, e^2, \ldots, e^n$. For every time $t$, let ${S}(t) \in \Fset_n$ be the set of currently admitted proposals just before time $t$ with $S(0) = \emptyset$. Suppose that a type-$(i,e)$ proposal arriving at time $t$ is assigned to clone $e^j$. If $e_j \in S(t)$ or $S(t) \cup\{e^j\}\notin\Fset_n$, reject the proposal and let $f(S(t), e^j)  := 0$. Otherwise, admit it with probability
\[
f(S(t), e^j) \defeq \frac{n\mu_n(S(t)\cup\{e^j\})}{x_e\mu_n(S(t))} \leq 1,\]
and then remove $e^j$ from $S(t)$ after an exponentially distributed time with mean $\ell_{ie}$ passes. Note that since $S(t)$ is always in $\Fset_n$, the vector of the number of admitted proposals of each product $N_e(t) = |S(t)\cap C_e^{(n)}|$ is always in $\mathcal{N}$, so this is a valid admission rule. We claim that with our admission rule, $S(t)$ converges in distribution to $\mu_n$. We postpone the proof of this fact to \Cref{apx-recurring-implies-admission}.

Now fix a type $i$, a product $e$, and a clone $e^j$. The
type-$(i,e)$ proposals assigned to $e^j$ form a homogeneous Poisson process of rate $y_{ie}/n$. Let $N_{ie^j}^{\pi}(T)$ denote the number of type $(i,e)$ proposals assigned to
$e^j$ that are admitted during $[0,T]$. Then, by the Campbell-Mecke formula (see equation (17) in \citet{coeurjolly2017tutorial}):
\[ \EE\left[N_{ic}^{\pi}(T)\right]
=
\frac{y_{ie}}{n}
\int_0^T\EE\left[f(S(t), e^j)\right]dt.\]
Since $S(t)$ converges to $\mu_n$, we know that:
$\EE[f(S(t), e^j)] \to \sum_{T\in\Fset_n}\mu_n(T)f(T, e^j).$
Therefore,
\begin{align*}
\lim_{T\to\infty}
\frac{1}{T}\EE\left[N_{ie^j}^{\pi}(T)\right]
&=
\frac{y_{ie}}{n}
\sum_{T\in\Fset_n}\mu_n(T)f(T, e^j)
\\
&=
\frac{y_{ie}}{n}
\frac{n}{x_e}
\sum_{\substack{T\in\Fset_n\\e^j\notin T}}
\mu_n(T\cup\{e^j\})
\\
&=
\frac{y_{ie}}{x_e}
\PP_{S\sim\mu_n}[e^j\in S]
\\
&\geq
(\alpha-\varepsilon)\frac{y_{ie}}{n}.
\end{align*}
The first equality is sometimes referred to as PASTA, or Poisson arrivals see time averages; see \citet{wolff1982poisson}. Summing over all $n$ clones of product $e$ gives $\liminf_{T\to\infty}
\frac{1}{T}\EE\left[N_{ie}^{\pi}(T)\right]
\geq
(\alpha-\varepsilon)y_{ie},$ as desired.\end{proofof}

Note that we may remove the $\varepsilon$ loss if we wish by projecting the distribution $\mu_n$ onto a distribution on $\mathcal{N}$, and finding a subsequence of the projections that converges to a distribution $\mu$. For the concrete Gibbs distribution we work with, this limit can sometimes be found directly.

Furthermore, the assumption that the distribution of the duration for which a unit is occupied is exponential is not essential, and is made only for the sake of expositional ease in \cref{apx-recurring-implies-admission}. \Cref{lem:recurring-implies-admission} also holds for general duration distributions with means $\ell_{ie}$, since the detailed-balance equations in \cref{apx-recurring-implies-admission} satisfy \citet[Theorem~2]{zachary2007note}. 

Finally, we note that our policy can be implemented using \textit{posted pricing} in a straightforward way, as discussed in \ref{apx-posted-pricing}.

\subsection{Assortment optimization}
\label{sec:reusable-assortments}

We now extend our reusable-product allocation model to the assortment optimization setting. In this setting, each type (along with being associated with an arrival rate $\lambda_i$, rewards $r_{ie}$, and durations $\ell_{ie}$) is also associated with a \textit{choice model} $\varphi_{ie}$. When a request of type $i$ arrives, rather than being allocated a product, it is offered an assortment $A\in\mathcal A_i\subseteq
2^{\ground}$. The request then chooses a product $e \in A$ with probability $\varphi_{ie}(A)$, choosing no product with the remaining probability. The chosen product must then be
 allocated immediately. We assume that each $\mathcal A_i$ is nonempty and downward closed, and that the choice model is \emph{weakly substitutable}:
\[
\varphi_{ie}(B)\geq\varphi_{ie}(A)
\quad\text{whenever }B\subseteq A\in\mathcal A_i\text{ and }e\in B.
\]
Let $\operatorname{OPT}_{\mathrm A}$ be the supremum of
\eqref{eq:long-run-reward} over all assortment policies, including policies that know future arrivals but observe choices only after offering an assortment and returns only when they occur. Following \Cref{sec:fluid}, we may define a fluid relaxation for the problem:
\begin{align*}
\operatorname{FLP_A} \defeq\
\max_{\ z_{iA}\ge 0}\quad
&\sum_{i\in\mathcal I}\sum_{e\in\ground}\reward_{ie}y_{ie}\\
\textnormal{s.t.}\quad
&\sum_{A\in\mathcal A_i}z_{iA}\le \arrivalRate_i
&&\forall i\in\mathcal I,\\
&y_{ie}=\sum_{A\in\mathcal A_i}z_{iA}\varphi_{ie}(A)
&&\forall i\in\mathcal I,\ e\in\ground,\\
&x_e=\sum_{i\in\mathcal I}\ell_{ie}y_{ie}
&&\forall e\in\ground,\\
&\xvec\in\Ppoly.
\end{align*}
Here, $z_{iA}$ is the rate at which type-$i$ requests are offered
assortment $A$, $y_{ie}$ is the induced rate at which type $i$ requests are allocated product $e$, and $x_e$ is the corresponding average load on product $e$. The same time-averaging and rate-conservation argument as in
\cref{prop:fluid-upper} gives $\operatorname{OPT}_{\mathrm A}\leq\operatorname{FLP_A}$.

\begin{theorem}
\label{prop:reusable-assortments}
Suppose that for every $n$-split of $(\ground,\mathcal N)$, there is an
$\alpha$-selectable S-OCRS. Then, for every $\eps > 0$, there
is an assortment policy $\pi$ such that
\[
\operatorname{R}(\pi) \geq(\alpha-\eps)\operatorname{OPT}_{\mathrm A}.
\]
\end{theorem}

\begin{proofof}{Proof}
Let $(y^*,z^*,\xvec^*)$ be an optimal solution to the fluid LP. Use the same $n$-split and admission probabilities
$f(S,e^j)$ from the proof of \cref{lem:recurring-implies-admission} for proposal rates $y^*_{ie}$. Upon the arrival of a type-$i$ request, sample a set of products $A$ with probability
$z^*_{iA}/\arrivalRate_i$ (offering nothing with the remaining probability) and independently for each $e\in A$, select a uniformly random clone $e^{j_e}$. Let $S$ be the currently occupied set of clones, and let $B\defeq\{e\in A:f(S,e^{j_e})>0\}$. For every $e\in B$, weak
substitutability gives
\[
0\leq\varphi_{ie}(A)f(S,e^{j_e})\leq\varphi_{ie}(B).
\]
By the sub-assortment sampling result of
\citet[Proposition~3]{feng2024near}, conditional on $S$, $A$, and
the sampled clones, we can sample $\widetilde A\subseteq B$ so that each $e\in A$ is chosen by the arriving request with probability $\varphi_{ie}(A)f(S,e^{j_e})$. Offer $\widetilde A$ and allocate the sampled clone of the chosen
product, if any. With this policy, when the set of occupied clones is $S$, the rate of type-$i$
allocations to an unoccupied clone $e^j$ is therefore
\[
\sum_{A\in\mathcal A_i}z^*_{iA}\varphi_{ie}(A)
\frac{f(S,e^j)}{n}
=\frac{y^*_{ie}}{n}f(S,e^j).
\]
These rates match those in \Cref{lem:recurring-implies-admission} exactly, so the allocation processes in the assortment optimization setting has the same distribution as in \Cref{lem:recurring-implies-admission}, and the result then follows immediately.
\end{proofof}

To solve $\operatorname{FLP_A}$ in polynomial time, like in \Cref{sec:fluid}, we may first have to replace $\Ppoly$ with a relaxation $\widetilde\Ppoly$ which has polynomial-time separation oracle. We must also additionally assume that that for every type $i$ and nonnegative weights $w_e$, we have a polynomial-time oracle for the offline assortment optimization problem:
\[
\max_{A\in\mathcal A_i}\sum_{e\in\ground}w_e\varphi_{ie}(A).
\]
This is Assumption 3 in \citet{feng2024near}, and is necessary to obtain any non-trivial guarantee. Note that commonly used choice models like unconstrained multinomial logit (MNL)~\citep{talluri2004revenue}, cardinality-constrained MNL~\citep{rusmevichientong2010dynamic}, and unconstrained Markov-chain choice~\citep{blanchet2016markov} satisfy the assumption. The equivalence of optimization and separation~\citep{gls88}, applied to the convex hulls of the choice vectors and to $\widetilde\Ppoly$, then let us calculate an optimal solution to $\operatorname{FLP_A}$ in polynomial time, with polynomially many nonzero $z_{iA}$. See also the argument in \citet[Proposition~2]{feng2024near}.

\subsection{Applications to environments in this paper}
In case the constraints on the set of products $\ground$ that can be allocated are weakly Rayleigh matroids, bipartite matchings, or general graph matchings, we may take $\mathcal{N} \subseteq \{0, 1\}^{\ground}$ to be the family of feasible sets $\Fset$ in these environments, as defined in \cref{sec:prelim}. In these cases, an $n$-split of $(\ground, \mathcal{N})$ results in a feasibility environment $(\ground_n, \mathcal{F}_n)$ that is once again a weakly Rayleigh matroid, bipartite matching, or a matching constraint, respectively, so we obtain algorithms for the reusable product problem with competitive ratios $1/2, (3-\sqrt{5})/2,$ and $1/3$, respectively. As discussed in \revcolor{\Cref{app:polytime-weakly-rayleigh,app:polytime-maxent-template}}, the associated S-OCRSs can all be implemented in polynomial time. The associated polytopes all have polynomial time separation oracles, so the fluid relaxations can be solved in polynomial time as well.

Suppose that we now have resources in a set $V$, each of which have $k$ identical units, and requests are for products $e \in E$, each of which consume at most one unit of at most $L$ different resources. This is the classic NRM problem, and in this case:

\[\mathcal N
=
\left\{
m\in\mathbb Z_{\ge0}^{\ground}:
\sum_{e\ni v}m_e\le k
\quad\forall v
\right\}.\]
For the fluid LPs, we use the relaxation
\[
\widetilde\Ppoly=\left\{\xvec\in[0,k]^{\ground}:\sum_{e\ni v}x_e\le k\quad\forall v\in V\right\},
\]and they can then be solved in polynomial time. Every $n$-split on $(E, \mathcal{N})$ results in a feasibility environment that is a rank-$L$ hypergraph matching with vertices with capacity $k$, so we obtain an algorithm for the reusable product problem with competitive ratio $1 - \sqrt{2 \log L/ k}-\mathcal{O}(1/\sqrt{k})$. In case $L = 1$, we get a competitive ratio of $\alpha_k = \PP[Q<k\mid Q\le k]$, where $Q\sim\operatorname{Poisson}(k),$ and our algorithm runs in polynomial time (as discussed in \cref{apx:time-complexity}). In case $L > 1$, we may apply the trick used in \cref{sec:NRM} where we scale the rate of proposals down by $1 -\mathcal{O}(\sqrt{\log L/ k})$, and each individual resource uses its own admission control to obtain an alternative polynomial time algorithm.

Finally, suppose we have products $e \in E$ with  many units, each with size $a_e$, and at any point of time, we can allocate units with total size at most $1$. In this case,
\[
\mathcal N
=
\left\{
m\in\mathbb Z_{\ge0}^{\ground}:
\sum_{e\in\ground}a_em_e\le1
\right\}.
\]
For the fluid LPs, to enable polynomial time solvability we use the relaxation
\[
\widetilde\Ppoly=\left\{\xvec\in[0,N]^{\ground}:\sum_{e\in\ground}a_ex_e\le1\right\}.
\]
Any $n$-split gives an ordinary knapsack constraint, so our $1/4$-selectable knapsack S-OCRS yields a $\frac14$-competitive algorithm for the reusable product problem. Since our knapsack S-OCRS can be implemented in polynomial time (once again, see \revcolor{\cref{app:polytime-knapsack}}), as long as 
\[N
=
\max\left\{m_e:m\in\mathcal N, e\in\ground\right\}
=
\max_{e\in\ground}\left\lfloor\frac1{a_e}\right\rfloor
\]
is polynomially bounded, the admission rule can be implemented in polynomial time.

\section{Conclusion \& Future Directions}
\label{sec:conclusion}

Motivated by online Bayesian resource allocation problems, we introduced the stationary OCRS (S-OCRS) framework which gives us OCRSs that satisfy an arrival order invariance property. This structural restriction enabled the design of explicit OCRS policies across various feasibility environments, including matchings, hypergraph matchings, bipartite matchings, and $k$-uniform matroids in a unified manner using a maximum-entropy approach. Furthermore, for environments where maximum entropy is insufficient—specifically, knapsack constraints and weakly Rayleigh matroids—we developed alternative approaches that modify the maximum entropy approach. We studied the computational efficiency of all our policies and showed that for many common feasibility environments, they admit polynomial-time execution. 

A significant open problem is to determine whether it is possible to find a $1/2$-selectable S-OCRS for general matroids. We conjecture that the answer is yes. As discussed in the introduction, while the existence of a $1/2$-selectable OCRS is known via indirect duality arguments, it would be interesting to find a more explicit OCRS. The KL-divergence minimization based approach is not the only way to sample a distribution over bases of a matroid with the correct marginals; perhaps it is possible to use other methods and thin the sampled base in a correlated way to obtain a witness distribution for the stationary OCRS LP. 

Another open problem is to the optimal selectability of an S-OCRS for matchings, or at the very least, determine whether the maximum-entropy approach gives $(\sqrt{3}-1)/2$-selectable S-OCRS. Once again, we conjecture that the answer to this question is yes. Also open is the optimal selectability for an S-OCRS for knapsack constraints.

One problem arising from our work is to understand the gap between OCRS and S-OCRS better. We should clarify that although one may conjecture that S-OCRSs can obtain the same selectability as OCRSs, that would be false---for instance, we can show that the best S-OCRS for 2-uniform matroids has a selectability of $3/5$, while the best OCRS has a selectability $>0.61$, and the best S-OCRS for the knapsack relaxation has a selectability at most $3/10$, while the best OCRS has selectability $1/(3+ e^{-2})$ \citep{jiang2025tight}. Understanding this gap better, and possibly characterizing it in different feasibility environments is a compelling avenue for future research. 

Finally, OCRSs have seen significant applications in revenue management. In a companion paper~\ifblind\cite{anonymous2026optimal}\else\cite{aminian2026optimal}\fi, we develop an application of S-OCRSs to Bayesian online allocation of reusable resources, originally studied in \cite{rusmevichientong2020dynamic} and further developed in \cite{feng2022near}. It would be useful to find other applications of S-OCRSs beyond the reusable setting. We leave the development of connections between the concept of S-OCRS and applications in revenue management, as well as queueing theory, as interesting directions for future research. 


\bibliographystyle{plainnat}
\OneAndAHalfSpacedXI
{\footnotesize
\let\originalbibsection\bibsection
\renewcommand{\bibsection}{%
  \originalbibsection
  \addcontentsline{toc}{section}{References}%
}
\bibliography{refs}
}

\newpage
\ECSwitch
\ECDisclaimer
\renewcommand{\theHchapter}{A\arabic{chapter}}
\renewcommand{\theHsection}{A\arabic{section}}

\addtocontents{toc}{\protect\TOCcompanionbreak}

\section{Further related work}
\label{apx:further-related}
Our work is related to several lines of research in operations research and computer science, which we briefly summarize below.

\smallskip
\xhdr{Contention resolution schemes and order of arrivals.}
Contention resolution schemes (CRSs) were introduced as offline rounding primitives for the multilinear
relaxation in submodular maximization; see the framework of \citet{chekuri2014submodular}.
OCRSs extend this primitive to online settings with independently active elements. The notion of OCRS for general downward-closed systems was first introduced by \citet{feldman2016online,feldman2021online} and has since become a standard tool for converting ex-ante
solutions into online policies. Since their introduction, OCRSs have also been studied in alternative arrival models, e.g., random-order \citep{AdamczykWlodarczyk2018RandomOrderCRS} and free-order \citep{ma2024onlineM} variants. Our work is complementary to these models: we impose a stronger symmetry requirement---permutation invariance of the \emph{output distribution}---and study the resulting stationary OCRS policies, in contrast to the order-based standard OCRS policies.

\xhdr{CRSs for matchings, and applications.} The contention resolution scheme problem for the matching feasibility system has been extensively studied. In the offline setting, initial results include those of \citet{guruganesh2018understanding,BruggmannZenklusen2022BipartiteMonotoneCRS}, and the best known results may be found in \citet{nuti2023towards}.  The online setting (under edge and vertex arrivals) was first studied by \citet{ezra2022prophet}, and their results were subsequently improved by \citet{macrury2025random}. Random order arrival models have also been studied in \citet{pollner2022improved} and \citet{macrury2025random}. The first of these papers also discusses applications to pricing. In addition, vertex arrival models have also been explored: \citet{ezra2022prophet,macrury2024random,fu2021random}. A special regime of interest is when activation probabilities are tiny; this case has been studied by \citet{nuti2023towards, ma2024onlineM}. These results are connected to applications in revenue-management. Rank-$L$ hypergraph matchings provide a canonical abstraction for network revenue management,
and \citet{ma2026online} develop OCRS-type guarantees tailored to that model.

\smallskip
\xhdr{\rev{Bayesian online allocation of reusable resources.}}
\rev{Motivated by the applications highlighted at the outset (e.g., cloud computing, rental marketplaces, and service systems), a growing line of work studies Bayesian online allocation and assortment optimization with \emph{reusable} resources, allowing for possibly random usage durations. Earlier OR and CS literature on reusable capacity includes LP-based approaches for revenue management with reusable resources and reservations~\citep{levi2010provably,chen2017revenue,owen2018price}, online Bayesian bipartite matching with reusable resources~\citep{dickerson2018allocation}, and dynamic assortment optimization for rental products with random usage durations~\citep{rusmevichientong2020dynamic}; see also \citep{rusmevichientong2023revenue,ao2024twostage} for richer models with advance bookings and reservations, \citep{sumita2022online,shanks2023approximation,dong2021efficient} for stochastic task assignment and matching with reusable agents, and \citep{huang2024joint} for joint inventory and assortment planning. Related Bayesian/LP-guided policies appear for flexible products and network revenue management~\citep{gallego2016online,ma2020approximation,baek2022bifurcating,anari2019nearly}, for admission control in overloaded loss networks with logarithmic regret~\citep{xie2025dynamic}, and for variants with volunteer nudging or recovering rewards~\citep{manshadi2022online,sumida2025dynamic}. Closest to our work, \citet{feng2022near,feng2024near} study Bayesian online assortment/allocation of reusable resources under non-stationary arrivals and obtain a $1-O(\sqrt{\log k/k})$ guarantee in the large-inventory regime via an LP-guided random-greedy (admission-control) policy. Our reduction in \Cref{sec:reusable-product-application} complements this literature by showing that, for any downward-closed feasibility environment, designing an $\alpha$-selectable S-OCRS immediately yields an $\alpha$-approximate policy for the Bayesian online allocation of reusable products in the same environment but under stationary arrivals (i.e., homogeneous Poisson processes) for each type; our companion paper~\ifblind\citep{anonymous2026optimal}\else\citep{aminian2026optimal}\fi\ leverages this viewpoint to tackle the problem under non-stationary arrivals (i.e., non-homogeneous Poisson processes). The main result of this companion paper solves an open problem in the $k$-uniform case, obtaining an $\alpha_k$-selectable policy that removes the logarithmic loss in \citet{feng2022near} and matches the known lower-bound guarantee for the multi-unit, non-reusable counterpart~\citep{alaei2014bayesian,wang2018online,jiang2025tight}.}

\smallskip
\xhdr{\rev{Reusable resources under adversarial arrivals \& online learning.}}
\rev{A complementary line of work studies online allocation of reusable resources when the request sequence is adversarial, under both matching and assortment models, generalizing classical adversarial models such as single-leg revenue management~\citep{ball2009toward}. In a semi-stationary model---where rewards are time-invariant and usage durations are i.i.d.\ or fixed---the greedy policy is $1/2$-competitive even for the more general assortment problem~\citep{gong2022online}, its ratio provably improves with the degree of reusability~\citep{baek2023leveraging}, and LP/fluid-guided designs achieve asymptotically optimal competitive ratios as capacities grow~\citep{feng2025robustness,goyal2025asymptotically}; see also \citep{simchilevi2025greedy} for decaying rewards and \citep{delong2024online} for online bipartite matching with small capacities, motivated by ride-sharing applications. Beyond the semi-stationary model, recent work studies multi-class arrivals and customer-dependent usage durations~\citep{huo2023multiclass,huo2023assortment}, online job assignment and its preemptive variant~\citep{ekbatani2025online,aminian2026online}, and---closest in spirit to our framework---online rounding schemes for $k$-rental problems, which separate the ex-ante fractional plan from an online feasibility-preserving rounding step~\citep{nekouyan2025online}. Finally, when arrival or usage statistics are unknown, online learning and blind policies have been developed for nonstationary environments~\citep{zhang2025online}, assortment planning~\citep{wu2025blind}, and fair allocation of reusable resources~\citep{liu2025fair}.}

\smallskip
\xhdr{CRSs for multiple selection (or the uniform matroid).}
The $k$-uniform matroid (a.k.a. ``multiple selection'' or the ``magician'' setting) has a long history in
Bayesian online allocation and mechanism design.
The seminal OCRS constructions of \citet{alaei2014bayesian} characterize the
optimal policy via a dynamic-programming/LP-duality viewpoint.
Subsequent work sharpened the guarantees and refined the optimality story;
see \citet{wang2018online} for the analysis in the vanishing regime, and the tight analysis in \citet{jiang2025tightness,jiang2025tight} using factor-revealing and LP duality. More recently, \citet{dinev2024simple} propose a simpler adaptive and combinatorial algorithm with near-optimal asymptotic performance. Still, our algorithm for $k$-uniform matroid has (arguably) a simpler form and the analysis is quite short (with the right setup). Also,
our stationary viewpoint is particularly compatible with online allocation of reusable resources, which is a natural extension of $k$-uniform setting and is extensively studied in the revenue management literature under both Bayesian and adversarial arrivals \rev{(see the dedicated discussion of reusable resources above)}.

The CRS problem for uniform matroids has also been studied in other related settings. In the offline setting, \citet{yan2011mechanism} established the optimal correlation gap for $k$-uniform matroids of $1 - \frac{e^{-k}k^k}{k!}$, which through an LP duality argument gives an optimal CRS. A simpler and more explicit scheme was given by \citet{kashaev2023simple}. In the random order setting, \citet{arnosti2023tight} obtain the optimal prophet inequality, and through a duality argument of \citet{lee2018optimal} give an  OCRS in the random order. We note the best selectability guarantees in the offline and random order online settings are the same.

\xhdr{CRSs for general matroids.}
The original papers introducing contention resolution schemes were particularly interested in the case of matroids, and \citet{chekuri2014submodular} already obtained the optimal selectability of $1- 1/e$ in the offline case. The optimal contention resolution schemes in the online setting are obtained by \citet{lee2018optimal} through a reduction to ex-ante prophet inequalities and an LP duality argument; their analysis is a refinement of the optimal algorithm in \cite{kleinberg2012matroid} for the matroid prophet inequality problem. A recent result has established a connection between contention resolution schemes and the matroid secretary problem: \citet{dughmi2025contention}. This has inspired further interest in sample-based contention resolution schemes for matroids: \citet{fu2024sample,feldman2026nearly}.
\smallskip

\xhdr{CRSs for knapsack constraints.} The original OCRS framework includes an OCRS for knapsack constraints ~\citep{feldman2016online,feldman2021online} with selectability $3/2-\sqrt{2}$. This OCRS works even against an almighty adversary.  \citet{jiang2025tight} later introduced a tight OCRS for points $x$ in the standard knapsack relaxation, with selectability $1/(3+e^{-2})$. They allow for items to have stochastic sizes. Finally, \citet{yoshinaga2023size} consider an OCRS in a related setting where sizes are unknown to the algorithm.  

\smallskip
\xhdr{Prophet inequalities and ex-ante relaxations.}
The prophet inequality literature originates from the classical work of \citet{krengel1978semiamarts} and
\citet{samuel1984comparison}, and has since been developed for a variety of feasibility constraints,
including matroids \citep{kleinberg2012matroid}, matchings \citep{ezra2022prophet}, knapsack constraints \citep{dutting2020prophet}, and more general downward-closed systems \citep{rubinstein2016beyond, rubinstein2026big}. In addition, various variants of this problem have been studied in the literature, for example, in random order~\citep{esfandiari2017prophet,ehsani2018prophet}, with access to samples~\citep{azar2014prophet, rubinstein2020optimal,cristi2024prophet, ghuge2025single}, or with the power of recourse/cancellation~\citep{ekbatani2024prophet}. See \cite{lucier2017economic} for a comprehensive survey. 
Prophet inequalities are also tightly connected to sequential posted-price mechanisms and ex-ante relaxations in Bayesian mechanism design;
see, e.g., \citet{hartline2009simple,chawla2010multi,yan2011mechanism,correa2017posted,lucier2017economic,alaei2019optimal,alaei2022descending} and
the nearly-optimal or improved posted-price/profit bounds versus optimal online policies (the so called ``philosopher's inequality'') in \citet{anari2019nearly,braverman2025new,niazadeh2018prophet,papadimitriou2021online}.
The link between OCRSs and ex-ante prophet inequalities is made explicit in \citet{lee2018optimal},
and underlies multiple optimal OCRS constructions (including the $1/2$-OCRS for matroids).
Related to our motivation for permutation invariance, \citet{kessel2022stationary} and \revcolor{\citet{patel2024combinatorial} study
``stationary'' prophet inequalities in an infinite-horizon setting with the latter paper using offline CRSs}; we emphasize that this notion is \emph{different} from our stationary (permutation-invariant) OCRS.
Finally, timing/order incentives have been studied in closely related online-selection models,
e.g., in the secretary problem \citep{buchbinder2009secretary,kesselheim2015secretary} and online auctions \citep{hajiaghayi2004adaptive}.

\smallskip
\xhdr{Maximum entropy, Gibbs measures, and negative dependence.}
Maximum-entropy distributions go back to the classical maximum-entropy principle of \citet{jaynes1957information}. For a textbook treatment of Gibbs measures, see \citet{georgii2011gibbs}.
Algorithmically, maximum-entropy distributions and their associated exponential-family structure have been
used as rounding primitives in combinatorial optimization; see, e.g., \citet{gharan2011randomized,singh2014entropy,straszak2019maximum}.
Our matroid results use negative dependence and Rayleigh-type properties. For background on negative dependence and correlation inequalities, see the survey of \citet{pemantle2000towards}.
The strong Rayleigh framework (and its polynomial viewpoint) was developed in
\citet{borcea2009negative}, and Rayleigh matroids were studied in, e.g., \citet{choe2006rayleigh,wagner2008negatively,kahn2010negative}.
These tools motivate the KL-projection-to-a-Rayleigh-base-measure approach developed in \Cref{sec:rayleigh-matroids}. Our bipartite addability analysis uses classical correlation inequalities in the monomer--dimer model, going back to \citet{heilmann1972theory}.

\section{Beyond Maximum-Entropy: ``Weakly Rayleigh'' Matroids \& Modified Maximum-Entropy} 
\label{sec:rayleigh-matroids}
In this section, we develop a modified maximum-entropy approach for matroids. We first outline the new template in \Cref{sec:kl-template}, and then show how it can be instantiated and analyzed in \Cref{sec:kl-analysis}. Although the construction itself is conceptually simple, implementing the resulting S-OCRSs in polynomial time requires several ingredients, including solving relative-entropy (KL-divergence) optimization, sampling from certain distributions over bases, and computing conditional expectations for these distributions using counting oracles. We defer these algorithmic details to \Cref{appendix-polytime}.

\subsection{Modified template: Rayleigh measures, minimizing KL divergence and thinning}
\label{sec:kl-template}
Recall the matroid feasibility environment defined in \Cref{sec:prelim}. Let $\cM=(\ground,\Iset)$ be a rank-$r$ matroid on $\ground$ with a collection of independent sets $\Iset$ and a collection of bases $\Bset$. In this section, we focus on a broad class of matroids that admit a certain kind of negatively dependent distribution on bases, which we call \emph{weakly Rayleigh}. We begin by defining the underlying notion of a \emph{Rayleigh measure}.

\begin{definition}[Rayleigh distribution]
\label{def:rayleigh-measure}
We say that a probability measure $\mu_0$ on $\{0,1\}^{\ground}$ is \emph{Rayleigh} if for every weight vector $\wvec\in\mathbb{R}^{\ground}_{>0}$,
the ``\emph{tilted measure with respect to $\wvec$},'' defined as
\begin{equation}
\label{eq:tilted-base-measure}
\mu_{\wvec}(A)\ \propto\ \mu_0(A)\cdot \prod_{e\in A} w_e
\end{equation}
satisfies the \emph{Rayleigh inequality:} for every $e\in\ground$ and every $T\subseteq \ground\setminus\{e\}$,
\begin{equation}
\label{eq:rayleigh-ineq}
\PP_{B\sim\mu_{\wvec}}[T\subseteq B]\cdot \PP_{B\sim\mu_{\wvec}}[e\in B]
\ \ge\
\PP_{B\sim\mu_{\wvec}}[T\cup\{e\}\subseteq B].
\end{equation}
\end{definition}

\begin{remark}
\label{remark:pairwise}
It is straightforward to show  that $\mu_0$ is a Rayleigh measure\footnote{Such measures $\mu_0$ are also referred to as $\mathsf{h\text{-}NLC+}$, $\mathsf{S\text{-}MRR_2}$, or $\mathsf{NC+}$ \citep{pemantle2000towards} in the literature. \citet{choe2006rayleigh} show that this notion can also be referred to as $\mathsf{BLC[2]}$ or $\mathsf{BLC[3]}$, and \citet{kahn2010negative} later show that this is also equivalent to $\mathsf{BLC[4]}$ and $\mathsf{BLC[5]}$.} iff it satisfies the pairwise negative correlation under all tilts: for every $e,f\in\ground$ and every $\wvec$,
\[
\PP_{B\sim\mu_{\wvec}}[e \in B]\cdot \PP_{B\sim\mu_{\wvec}}[f \in B]
\ \ge\
\PP_{B\sim\mu_{\wvec}}[\{e, f\}\subseteq B].
\]
(see \Cref{apx-pairwise} for a proof of this well-known result.)
\end{remark}

\begin{definition}[Weakly Rayleigh matroid]
\label{def:weakly-rayleigh}
We say that $\cM=(\ground,\Iset)$ is \emph{weakly Rayleigh} if there exists a Rayleigh probability measure $\mu_0$ on $\Bset$
with full support (i.e., $\mu_0(B)>0$ for all $B\in\Bset$).
\end{definition}
\smallskip

We adopt the terminology of ``weakly Rayleigh" from \citep{wagner2008negatively}, who studied weakly Rayleigh set systems. Many commonly encountered classes of matroids are weakly Rayleigh. For example, all matroids representable over $\mathbb{C}$ (e.g., the class of linear matroids over $\mathbb{R}$, which includes graphic and laminar matroids) are weakly Rayleigh: if $\cM$ is represented by a matrix $A$, then the determinantal measure $\mu_0(B)\propto\det(A_B^* A_B)$ is Rayleigh. In addition, all matroids representable over $\mathbb{F}_2$ that do not have $S_8$ as a minor, all matroids of rank $\le 3$, and all matroids on at most $7$ elements are weakly Rayleigh. We highlight that not all matroids are weakly Rayleigh. We prove this in \Cref{apx:non-weakly-rayleigh} via showing that $S_8$ is a counterexample; this might be of independent interest.

We are now ready to present the main result of this section.

\begin{theorem}[Optimal S-OCRS for weakly Rayleigh matroids]
\label{thm:weakly-rayleigh}
For the weakly Rayleigh matroid environment $(E,\Fset)$ and every $\xvec\in \Ppoly$, there exists a distribution
$\mu^*\in\Delta(\Fset)$ such that Algorithm~1 (simulate-then-replace), when instantiated with $\mu^*$,
is a $\tfrac12$-selectable S-OCRS. 
Moreover, $\tfrac12$ is the best possible universal selectability constant for this class.
\end{theorem}
Note that factor $\tfrac12$ is already the best possible selectability constant for standard OCRS and rank-$1$ uniform matroids---therefore, our S-OCRS is optimal.

\xhdr{Why we need a different method.}
The method in \Cref{sec:socrs-template} constructs $\dist^*$ as the maximum-entropy distribution on feasible sets with marginals $\pvec=\alpha\xvec$ and then proves addability bounds under $\dist^*$. For matroids, this ``direct'' maximum-entropy approach can fail: the maximum-entropy distribution on independent sets may place most of its mass on configurations that make certain elements rarely addable. \revcolor{Below,} we give an explicit counterexample for a simple graphic matroid (known as the ``hat'' instance in the literature).

Let $G_n$ be the graph with terminals $u,v$ and intermediate vertices $m_1,\dots,m_n$, with edges
\[
E(G_n)=\{u m_i,  m_i v  :  i=1,\dots,n\}.
\]
Let $\mathcal{I}$ be the family of forests of $G_n$ (i.e., the independent sets of the graphic matroid of $G_n$), and let $x_e = 1/2$ for every edge $e$.
Consider the \emph{unweighted} maximum-entropy distribution on $\mathcal{I}$, namely the uniform distribution over forests.
A forest connects $u$ to $v$ iff it contains \emph{both} edges of at least one length-$2$ path $u\!-\!m_i\!-\!v$. We can count the forests as follows:
\begin{itemize}
  \item If $u$ and $v$ are disconnected, then for each $i$ we may choose $\emptyset$, $\{u m_i\}$, or $\{m_i v\}$, yielding $3^n$ forests.
  \item If $u$ and $v$ are connected, then there is a unique index $i$ whose full path is present (choose $i$ in $n$ ways), and for each $j\neq i$ we again choose among $\emptyset,\{u m_j\},\{m_j v\}$, yielding $n\cdot 3^{n-1}$ forests.
\end{itemize}
Hence, under the uniform (maximum-entropy) forest,
\[
\PP[u \text{ and } v \text{ are disconnected}]
=
\frac{3^n}{3^n+n\cdot 3^{n-1}}
=
\frac{3}{n+3}
\]
If we add a further element $e^*=(u,v)$ with $x_{e^*}$ tiny, then $e^*$ is addable precisely when $u$ and $v$ are disconnected, so $\PP[\Add(e^*)]=o(1)$ under this max-entropy distribution.
This illustrates the obstruction: the high-entropy forest distribution overwhelmingly prefers configurations that realize \emph{some} $u$--$v$ connection (there are $n$ choices), and this drives the addability probability of $e^*$ to $0$.

The same phenomenon persists for weighted Gibbs measures with any fixed edge-weight $w>0$ on every edge: the weight of forests with no $u$--$v$ path is $(1+2w)^n$, while forests with a (unique) $u$--$v$ path have total weight $n w^2(1+2w)^{n-1}$. If we fix some constant selectability that we aim for, say $\alpha$, and take $n \to \infty$, it's straightforward to show that for the maximum entropy distribution, $w \to \frac{\alpha}{2(1-\alpha)}$. Thus, the maximum entropy approach is incompatible with proving (or implementing) any constant-factor stationary OCRS.

\xhdr{A KL-divergence template on bases.}
For weakly Rayleigh matroids we instead work with distributions on \emph{bases}. Starting from a Rayleigh base measure $\mu_0$, we compute a new base distribution with certain marginals, and  then obtain an independent-set distribution by a simple \emph{thinning procedure}. In the \revcolor{preceding counterexample}, for instance, this thinning step deletes every $u$--$v$ path with probability at least $1/2$, and is essential for the new approach to work. 

Fix an activation vector $\xvec\in\Ppoly$ and any Rayleigh base measure $\mu_0$. At a high-level, our new framework for design and analysis of S-OCRS algorithms follows a \emph{modified} five-step template:

\smallskip
\begin{enumerate}[label=(\roman*), align=left, labelwidth=1em,
  labelsep=1em,
  leftmargin=!,
  itemsep=0.3em]
    \item Find $\qvec\in\Bpoly$ such that $\qvec\ge \xvec$ coordinatewise,
    \item Build a Rayleigh distribution $\mu$ on bases with marginals $\qvec$ by minimizing \emph{KL divergence} from $\mu_0$ subject to prescribed marginals $\qvec$ (a ``relative-entropy'' analogue of maximum entropy)\footnote{For several classes of matroids (graphic matroids, for instance), the uniform distribution over bases is already Rayleigh, so this step is the same as finding the maximum entropy distribution with marginals $\qvec$.},
    \item $1/2$-thin inside $\mu$ (independently across elements) and show that the resulting independent-set distribution satisfies Constraint~\ref{eq:soCRS-stationary} with caps $\qvec$,
    \item Thin further---independently coordinatewise---to match the desired activation vector $\xvec$ (i.e., to have marginals exactly equal to $\xvec/2$) and obtain a distribution $\dist^*$ on independent sets,
    \item Run the simulate-then-replace policy (\Cref{alg:sample-then-replace}) with $\dist^*$ as input.
\end{enumerate}
\smallskip

To show the correctness of our S-OCRS, we need to show that the final witness distribution $\dist^*$ is feasible for the stationary OCRS LP with $\alpha=\tfrac12$.
Note that $\dist^*$ has marginals exactly equal to $\frac{1}{2}\xvec$. Therefore, it only remains to show that $\dist^*$ satisfies \eqref{eq:soCRS-stationary}.\footnote{If $\xvec \in b\Ppoly$, we may choose $\qvec \geq \frac{\xvec}{b}$, $b/(1+b)$-thin inside $\mu$, and then thin further coordinatewise by $\frac{x_e}{bq_e}$ to match the desired activation vector $\xvec$.}

\subsection{Implementing the modified template: proof of Theorem \ref{thm:weakly-rayleigh}}
\label{sec:kl-analysis}
We now explain each step of the template and why $\dist^*$ satisfies \eqref{eq:soCRS-stationary}.

\smallskip
\xhdr{Step (i): Finding a dominating base point.}
We first note that every point in the independent-set polytope can be dominated by a point in the base polytope. This is a standard fact in matroid theory; see, e.g., \citet[Chapter 44]{schrijver2003combinatorial}. We still include a proof for completeness in \Cref{apx:ind-to-base}.

\begin{lemma}[Dominating base point]
\label{lem:dominating-base-point}
For every point $\xvec\in\Ppoly$, there exists a dominating base point $\qvec\in\Bpoly$ such that $\forall e\in \ground:~q_e\ge x_e$.
\end{lemma}
Fix such a $\qvec\in\Bpoly$ for the remainder of this subsection. 

\smallskip
\xhdr{Step (ii): KL projection onto a Rayleigh base measure.}
We now build a Rayleigh distribution $\mu$ on bases with marginals $\qvec$. When $\qvec$ lies in the relative interior of $\Bpoly$, this can be done via a KL-divergence projection. Concretely, consider this convex program over distributions $\nu$ on $\Bset$:
\begin{align}
\label{eq:KL-proj-bases}
\tag{\textsc{Min-KL}}
\min_{\nu\in\simplex{\Bset}}\quad & \KL(\nu\|\mu_0) \\
\text{s.t.}\quad & \PP_{B\sim\nu}[e\in B]=q_e \qquad \forall e\in\ground~,
\end{align}
where $\KL(\nu\|\mu_0)\defeq \sum_{B\in\Bset}\nu(B)\log\!\bigl(\nu(B)/\mu_0(B)\bigr)$.
By strict convexity and Slater's condition, the minimizer is unique; moreover, the KKT conditions (as in the proof of Lemma~\ref{lem:gibbs-addability-template}) imply that the unique minimizer $\nu^*$ has the tilted form $\nu^*=\mu_{\wvec}$ for some $\wvec\in\mathbb{R}^{\ground}_{>0}$ as in~\eqref{eq:tilted-base-measure}.\footnote{Note that a probability distribution of this tilted form is not, in general, a Gibbs distribution.} In particular, we obtain a tilted measure $\mu_{\wvec}$ satisfying \[\PP_{B\sim\mu_{\wvec}}[e\in B]=q_e \qquad \forall e\in\ground~,\]
and, since $\mu_0$ is Rayleigh, we also have the Rayleigh inequality \eqref{eq:rayleigh-ineq} under $\mu_{\wvec}$.

If $\qvec$ lies on the boundary of $\Bpoly$, we approximate it by a sequence of interior points and pass to a limit. Since $2^{\ground}$ is finite, this limiting procedure produces a probability measure on $\Bset$ with marginals $\qvec$; Rayleigh-ness is preserved under limits. Let $\mu$ denote the resulting base measure with the desired marginals and the Rayleigh property (regardless of whether $\qvec$ is on the boundary or interior of $\Bpoly$).

\xhdr{Step (iii): $\bm{1/2}$-thinning inside bases.}
Sample a base $B\sim\mu$ and then include each element of $B$ independently with probability $\tfrac12$. Let $\Sset$ denote this random set. Then $\Sset\subseteq B$ is independent, and for each $e\in\ground$,
\begin{equation}
\label{eq:q-marginals}
\PP[e\in\Sset]=\tfrac12\,\PP_{B\sim {\mu}}[e\in B]=\tfrac12\,q_e.
\end{equation}

We claim that $\Sset$ satisfies \eqref{eq:soCRS-stationary} with caps $\qvec$, meaning that for every $e\in\ground$ and every $T\subseteq\ground\setminus\{e\}$ with $\PP[\Sm{\Sset}{e}=T]>0$, we have:
\begin{equation}
\label{eq:q-stationary-cap}
\PP\!\left[e\in\Sset \given \Sm{\Sset}{e}=T\right]\le q_e.
\end{equation}
Indeed, conditioning on $\Sm{\Sset}{e}=T$ leaves only two possibilities for $\Sset$: either $\Sset=T$ or $\Sset=T\cup\{e\}$. Moreover, since the thinning is independent inside a base of fixed size $r$, for any $U\subseteq\ground$ we have $\PP[\Sset=U]=2^{-r}\cdot \PP_{B\sim\mu}[U\subseteq B]$, and therefore
\begin{align*}
\PP\!\left[e\in\Sset \given \Sm{\Sset}{e}=T\right]
&=\frac{\PP[\Sset=T\cup\{e\}]}{\PP[\Sset=T]+\PP[\Sset=T\cup\{e\}]}\\
&=\frac{\PP_{B\sim\mu}[T\cup\{e\}\subseteq B]}{\PP_{B\sim\mu}[T\subseteq B]+\PP_{B\sim\mu}[T\cup\{e\}\subseteq B]}\\
&\le \frac{\PP_{B\sim\mu}[T\cup\{e\}\subseteq B]}{\PP_{B\sim\mu}[T\subseteq B]}
\ \le\ \PP_{B\sim\mu}[e\in B]\ =\ q_e,
\end{align*}
where the penultimate inequality is the Rayleigh inequality~\eqref{eq:rayleigh-ineq} applied to $(T,e)$. This proves~\eqref{eq:q-stationary-cap}.

\xhdr{Step (iv): Coordinatewise thinning to match $\xvec$.}
The caps in \eqref{eq:q-stationary-cap} are stated in terms of $q_e$, which may exceed the desired activations $x_e$. Define thinning probabilities $s_e\in[0,1]$ by
\[
 s_e\ \defeq\
 \begin{cases}
 x_e/q_e, & q_e>0,\\
 0, & q_e=0.
 \end{cases}
\]
Let $(C_e)_{e\in\ground}$ be independent Bernoulli variables with $\PP[C_e=1]=s_e$, independent of $\Sset$, and set $R \defeq\{e\in \Sset:\ C_e=1\}$. Then $R\subseteq \Sset$ is independent and, using~\eqref{eq:q-marginals},
\[
\PP[e\in R]=\PP[e\in\Sset]\PP[C_e=1]=\tfrac12\,q_e\,s_e=\tfrac12\,x_e,
\]
which is the selectability constraint with $\alpha=\tfrac12$.

For stationary implementability, fix $e$ and $T$ with $\PP[\Sm{R}{e}=T]>0$.
Since $C_e$ is independent of the event $\{\Sm{R}{e}=T\}$, we have
\begin{equation}
\label{eq:cap-thinning-factor}
\PP\!\left[e\in R \given \Sm{R}{e}=T\right]
=s_e\cdot \PP\!\left[e\in\Sset \given \Sm{R}{e}=T\right].
\end{equation}
Let $U\defeq \Sm{\Sset}{e}$.
Conditioned on $U$, the event $\{\Sm{R}{e}=T\}$ depends only on the coins $\{C_f\}_{f\neq e}$, and is therefore conditionally independent of the event $\{e\in \Sset\}$.
Hence
\[
\PP\!\left[e\in\Sset \given U,\ \Sm{R}{e}=T\right]
=\PP\!\left[e\in\Sset \given U\right]\le q_e,
\]
where the inequality is \eqref{eq:q-stationary-cap} (applied to the realized value of $U$). Taking conditional expectations given $\{\Sm{R}{e}=T\}$ yields $\PP[e\in\Sset \given \Sm{R}{e}=T]\le q_e$. Plugging into~\eqref{eq:cap-thinning-factor} gives
\[
\PP\!\left[e\in R \given \Sm{R}{e}=T\right]\le s_e q_e = x_e,
\]
which is exactly the stationary implementability constraint.

\smallskip
\xhdr{Step (v): Conclusion.}
Let $\dist^{*}$ be the distribution of $R$. We have shown that $\dist^{*}$ satisfies the constraints of stationary OCRS LP, i.e., \eqref{eq:soCRS-selectability}--\eqref{eq:soCRS-stationary}, with $\alpha=\tfrac12$. The theorem now follows from Proposition~\ref{prop:stationary-equivalence}: for every arrival order, the simulate-then-replace procedure (\Cref{alg:sample-then-replace}) implements $\dist^{*}$ online, yielding a $\tfrac12$-selectable S-OCRS.

\subsection{\revcolor{Proof of Remark~\ref{remark:pairwise}}}
\label{apx-pairwise}
\begin{proofof}{Proof}
The implication $\eqref{eq:rayleigh-ineq}\Rightarrow$ (pairwise form) is immediate by taking $T=\{f\}$. For the converse, assume the pairwise inequality holds for every tilt: for all $e,f\in\ground$ and all $\wvec$,
\[
\PP_{B\sim\mu_{\wvec}}[e \in B]\cdot \PP_{B\sim\mu_{\wvec}}[f \in B]
\ \ge\
\PP_{B\sim\mu_{\wvec}}[\{e, f\}\subseteq B].
\]
Fix $\wvec$, $e\in\ground$, and a set $T=\{f_1,\dots,f_m\}\subseteq \ground\setminus\{e\}$.
If $\PP_{B\sim\mu_{\wvec}}[T\subseteq B]=0$, then $\PP_{B\sim\mu_{\wvec}}[T\cup\{e\}\subseteq B]=0$ as well and
\eqref{eq:rayleigh-ineq} holds trivially. Hence assume $\PP_{B\sim\mu_{\wvec}}[T\subseteq B]>0$.

For any $A\subseteq \ground$, let $\mu_{\wvec}\mid A$ denote the conditional distribution of $B\sim\mu_{\wvec}$ given $A\subseteq B$.
We first note that $\mu_{\wvec}\mid A$ is a limit of tilts of $\mu_{\wvec}$.
Indeed, for $\lambda>0$ define $\wvec^{(\lambda)}$ by
\[
w^{(\lambda)}_g \ \defeq\
\begin{cases}
\lambda\,w_g, & g\in A,\\
w_g, & g\notin A.
\end{cases}
\]
Since $\mu_{\wvec^{(\lambda)}}(B)\propto \mu_{\wvec}(B)\cdot \lambda^{|A\cap B|}$, as $\lambda\to\infty$ the mass
concentrates on sets with $A\subseteq B$, and hence $\mu_{\wvec^{(\lambda)}}$ converges to $\mu_{\wvec}\mid A$.
Because the state space is finite, this convergence implies convergence of all event probabilities.

Now fix $j\in\{1,\dots,m\}$ and write $A_{j}\defeq \{f_1,\dots,f_{j}\}$ for $j=1,\ldots,m$.
Applying the assumed pairwise inequality to the tilt $\mu_{\wvec^{(\lambda)}}$ and then taking $\lambda\to\infty$ yields that
the conditional measure $\mu_{\wvec}\mid A_{j-1}$ also satisfies pairwise negative correlation, in particular
\[
\PP[e\in B\mid A_{j-1}\subseteq B]\cdot \PP[f_j\in B\mid A_{j-1}\subseteq B]
\ \ge\
\PP[\{e,f_j\}\subseteq B\mid A_{j-1}\subseteq B].
\]
Whenever $\PP[f_j\in B\mid A_{j-1}\subseteq B]>0$, dividing both sides by this quantity gives
\[
\PP[e\in B\mid A_j\subseteq B]
\ \le\
\PP[e\in B\mid A_{j-1}\subseteq B].
\]
(If $\PP[f_j\in B\mid A_{j-1}\subseteq B]=0$, then $\PP[A_j\subseteq B]=0$ and the above inequality is again trivial.)

Iterating over $j=1,\dots,m$ yields
\[
\PP[e\in B\mid T\subseteq B]\ \le\ \PP[e\in B].
\]
Multiplying by $\PP[T\subseteq B]$ and rewriting gives exactly \eqref{eq:rayleigh-ineq}:
\[
\PP_{B\sim\mu_{\wvec}}[T\subseteq B]\cdot \PP_{B\sim\mu_{\wvec}}[e\in B]
\ \ge\
\PP_{B\sim\mu_{\wvec}}[T\cup\{e\}\subseteq B].
\]
\end{proofof}

\subsection{\revcolor{Not all matroids are weakly Rayleigh}}
\label{apx:non-weakly-rayleigh}
This follows from a simple variation on the results in \cite{branden2010half}. We start by noting that Theorem 2.1 in \cite{branden2010half} also holds when the polynomial in question is Rayleigh rather than stable:

\begin{lemma}
\label{lem:rayleigh-degenerate-quadrangle}
Let $\mathcal{M}=(E,\mathcal I)$ be a weakly Rayleigh matroid with set of bases $\mathcal{B}$ and base measure $\mu$. Let
\[
(B_1,B_2,B_3,B_4)
=
\bigl(
A\cup\{i,k\},
A\cup\{i,\ell\},
A\cup\{j,\ell\},
A\cup\{j,k\}
\bigr)
\]
be a degenerate quadrangle of bases of $\mathcal{M}$, i.e., $i, j, k, l \notin A$, and at most one of $A\cup\{i,j\}$ and $A\cup\{k,\ell\}$ is a basis. Then
\[
\mu(B_1)\mu(B_3)=\mu(B_2)\mu(B_4).
\]
\end{lemma}

\begin{proofof}{Proof}
Since $\mu$ is Rayleigh, we have that for every distinct $e,f \in E$ and every $\wvec \in \mathbb{R}_{>0}^E$,
\[
\PP_{B \sim \mu_{\wvec}}[e \in B]\cdot \PP_{B \sim \mu_{\wvec}}[f \in B]
\;\ge\;
\PP_{B \sim \mu_{\wvec}}[\{e,f\}\subseteq B].
\]
Equivalently, writing
\[
g_{\mu}(\wvec) = \sum_{B \in \mathcal{B}} \mu(B)\prod_{e \in B} w_e,
\]
we have for every distinct $e,f \in E$ and every $\wvec \in \mathbb{R}_{>0}^E$,
\[
\Delta_{ef} g_{\mu}(\wvec)
:=
\frac{\partial g_{\mu}}{\partial w_e}(\wvec)\,
\frac{\partial g_{\mu}}{\partial w_f}(\wvec)
-
g_{\mu}(\wvec)\,
\frac{\partial^2 g_{\mu}}{\partial w_e \partial w_f}(\wvec)
\;\ge\; 0,
\]

in which case we say the polynomial $g_{\mu}$ is Rayleigh. By symmetry, we may assume that $A \cup \{i,j\} \notin \mathcal{B}$.
Fix $\lambda > 0$, $\varepsilon > 0$, and define a weight vector
$\wvec^{(\lambda,\varepsilon)} \in \mathbb{R}_{>0}^E$ by
\[
w_e^{(\lambda,\varepsilon)}
:=
\begin{cases}
\lambda, & e \in A,\\
w_e, & e=i, j, k, \ell \\
\varepsilon, & e \in E \setminus (A \cup \{i,j,k,\ell\}).
\end{cases}
\]
Consider
\[
h_{\lambda,\varepsilon}(w_i,w_j,w_k,w_{\ell})
:=
\lambda^{-|A|}\, g_{\mu}\!\left(\wvec^{(\lambda,\varepsilon)}\right).
\]
Since $h_{\lambda,\varepsilon}$ is obtained from $g_{\mu}$ by rescaling and
setting values for some variables, it is Rayleigh.

Now let $\lambda \to \infty$ and $\varepsilon \to 0$. The only surviving
terms are those corresponding to bases containing $A$ and no element of
$E \setminus (A \cup \{i,j,k,\ell\})$. Because $A \cup \{i,j\} \notin \mathcal{B}$, the limit polynomial is
\[
h(w_i,w_j,w_k,w_{\ell})
=
\mu(B_1)\, w_i w_k
+
\mu(B_2)\, w_i w_{\ell}
+
\mu(B_3)\, w_j w_{\ell}
+
\mu(B_4)\, w_j w_k
+
\eta\, w_k w_{\ell},
\]
where
\[
\eta :=
\begin{cases}
\mu(A \cup \{k,\ell\}), & \text{if } A \cup \{k,\ell\} \in \mathcal{B},\\
0, & \text{otherwise.}
\end{cases}
\]
Since each $h_{\lambda,\varepsilon}$ is Rayleigh, and $\Delta_{ef}$ depends continuously on
the coefficients, the limit polynomial $h$ is also Rayleigh.

A direct calculation gives
\[
\Delta_{ik} h
=
w_{\ell}\Bigl((\mu(B_2)\mu(B_4)-\mu(B_1)\mu(B_3))\, w_j
+ \mu(B_2)\eta\, w_{\ell}\Bigr),
\]
and
\[
\Delta_{i\ell} h
=
w_k\Bigl((\mu(B_1)\mu(B_3)-\mu(B_2)\mu(B_4))\, w_j
+ \mu(B_1)\eta\, w_k\Bigr).
\]
Since $h$ is Rayleigh, both expressions are nonnegative for all
$w_i,w_j,w_k,w_{\ell} > 0$.
Fix $w_j = 1$. From $\Delta_{ik} h \ge 0$, divide by $w_{\ell}>0$ and let
$w_{\ell} \to 0$ to obtain
\[
\mu(B_2)\mu(B_4)-\mu(B_1)\mu(B_3)\ge 0.
\]
Similarly, from $\Delta_{i\ell} h \ge 0$, divide by $w_k>0$ and let
$w_k \to 0$ to obtain
\[
\mu(B_1)\mu(B_3)-\mu(B_2)\mu(B_4)\ge 0.
\]
Hence
\[
\mu(B_1)\mu(B_3)=\mu(B_2)\mu(B_4),
\]
as desired.
\end{proofof}

\xhdr{The spaces $V_{\mathcal{M}}$ and $W_{\mathcal{M}}$.} Following \cite{branden2010half}, let $V_{\mathcal{M}} \subseteq \mathbb{R}^{\mathcal{B}}$
be the linear subspace of all functions $\nu : \mathcal{B} \to \mathbb{R}$ satisfying
\[
\nu(B_1)+\nu(B_3)-\nu(B_2)-\nu(B_4)=0
\]
for every degenerate quadrangle $(B_1,B_2,B_3,B_4)$ of bases of $\mathcal{M}$.
Let $W_{\mathcal{M}} \subseteq V_{\mathcal{M}}$ be the subspace of functions of the form
\[
\nu(B)=\sum_{e \in B} v_e.
\]
for some vector $(v_e)_{e \in E} \in \mathbb{R}^E$.
We have the following analog of Theorem~2.3 in \cite{branden2010half}. The proof is exactly the same as \cite{branden2010half}.
\begin{theorem}
\label{thm:rayleigh-analogue}
Let $\mathcal{M}=(E,\mathcal{I})$ be a matroid. Assume that
\[
\dim(W_{\mathcal{M}})=\dim(V_{\mathcal{M}}).
\]
Then $\mathcal{M}$ is weakly Rayleigh if and only if the uniform measure on $\mathcal{B}$ is Rayleigh.
\end{theorem}

To finish, note that by Corollary~2.7 and Table~1 in \cite{branden2010half},
\[
\dim(W_{S_8})=\dim(V_{S_8})=8.
\]
Moreover, $S_8$ is not Rayleigh;
see \cite{choe2006rayleigh}. Therefore, $S_8$ is not weakly Rayleigh.

\subsection{\revcolor{Proof of Lemma~\ref{lem:dominating-base-point} (Dominating base point)}}
\label{apx:ind-to-base}
\begin{proofof}{Proof}
Consider the nonempty polyhedron
\[
Q  \defeq \Bigl\{\yvec\in [0,1]^{\ground}: \sum_{e\in T}y_e\le \RankOracle{T} \forall T\subseteq \ground,  \yvec\ge \xvec\Bigr\}.
\]
Let $\qvec$ maximize $\sum_{e\in\ground} y_e$ over $Q$.
If $\sum_{e}q_e=r$, then $\qvec\in\Bpoly$ and $\qvec\ge \xvec$, as desired.

Suppose by way of contradiction that $\sum_{e}q_e<r=\RankOracle{\ground}$.
Let $\mathcal{T}$ be the family of tight sets at $\qvec$:
$\mathcal{T}\defeq\left\{T\subseteq\ground:\sum_{e\in T}q_e=\RankOracle{T}\right\}$.
By submodularity of $\RankOracle{\cdot}$, $\mathcal{T}$ is closed under unions, so $T^*\defeq \bigcup_{T\in\mathcal{T}}T$ is also tight. Since $\sum_{e}q_e<\RankOracle{\ground}$, we must have $T^*\neq \ground$; pick $e\in\ground\setminus T^*$.

For sufficiently small $\eps>0$, increasing $q_e$ to $q_e+\eps$ keeps all inequalities feasible:
any violated rank inequality would have to correspond to a set $T$ with $e\in T$ that is tight at $\qvec$,
contradicting $e\notin T^*$.
This produces a feasible point in $Q$ with a larger objective value, contradicting the maximality of $\qvec$.
Therefore, $\sum_{e}q_e=r$ and $\qvec\in\Bpoly$.
\end{proofof}

\section{Beyond Maximum-Entropy: Knapsack Constraints \& Modified Maximum-Entropy} 
\label{sec:knapsack-constraints}
In this section, we develop a modified maximum-entropy approach for knapsack constraints. Recall from \Cref{sec:prelim} the knapsack constraint environment defined on a set of items $\ground$ with sizes $a_e \in (0, 1]$. For knapsack constraints, like for matroids, the direct maximum-entropy approach can fail, and \revcolor{below} we give an explicit counterexample following \citet{feldman2016online, feldman2021online}. The main issue is that maximum entropy distributions can make it very hard for very heavy items to be addable.

Consider a unit-capacity knapsack with ground set $E=[n],$ and assume the sizes are all equal to $1/n$. Let $x_i = 1$ for every item. The feasible sets are all subsets of these items.

Therefore, under any weighted Gibbs measures with fixed item-weight $w>0$, the probability we pick none of these items is $\frac{1}{(1+w)^n}$. If we fix some constant selectability that we aim for, say $\alpha$, and take $n \to \infty$, it's straightforward to show that for the maximum entropy distribution, $w \to \frac{\alpha}{1-\alpha}$. 

Now, if we add a further item $b$ with size $1$ and $x_{b}$ tiny, then $b$ is addable precisely when none of the original items are selected, so $\PP[\Add(b)]=o(1)$ under this max-entropy distribution. This illustrates the obstruction: the high-entropy  distribution overwhelmingly prefers configurations that pick \emph{some} light item (there are $n$ choices), and this drives the addability probability of the heavy item to $0$.

We therefore need a different strategy. Similar to \citet{feldman2016online, feldman2021online} and \citet{dutting2020prophet}, our idea is to split $\ground$ into \textit{heavy} items and \textit{light} items. We then use two different stationary OCRS witness distributions for the heavy items and light items, and mix the two. For the heavy items, we select at most one item. For the light items, we use the maximum-entropy approach. We defer the algorithmic details of
implementing our scheme in polynomial time to \Cref{appendix-polytime}.
\begin{theorem}[S-OCRS for knapsack constraints]
\label{thm:knapsack}
For the knapsack constraint environment $(\ground,\Fset)$ and every $\xvec\in\Ppoly$, there exists a distribution $\mu^*\in\Delta(\Fset)$ such that Algorithm~1 (simulate-then-replace), when instantiated with $\mu^*$, is a $\tfrac14$-selectable S-OCRS. 
\end{theorem}

Before presenting the proof, we make a few remarks regarding the context of our result. Following \citet{feldman2016online,feldman2021online}, we work in a setting in which items have deterministic sizes. Note, however, that our schemes, unlike theirs, do not work against almighty adversaries. \citet{jiang2025tight} work with more general stochastically sized items. It is possible to extend our result to stochastic sizes and obtain a $1/5$-selectable S-OCRS. This then yields a $1/5$-competitive prophet inequality, matching the results of \citet{dutting2020prophet}.

\begin{proofof}{Proof}

Partition the ground set into heavy and light items:
\[
    E_{H}:=\{e\in E:a_e>1/2\},
    \qquad
    E_{L}:=\{e\in E:a_e\le 1/2\},
\]
and define
\[
    X:=\sum_{e\in E_{H}}x_e,
    \qquad
    B:=\sum_{e\in E_{L}}a_e x_e.
\]

\xhdr{Step (i): The heavy-item case.} Let $c_{H}:=\frac{1}{1+X}$, and define $\mu^{H}\in\Delta(\Fset)$ by
\[
    \mu^{H}(\emptyset)=c_{H},
    \qquad
    \mu^{H}(\{e\})=c_{H}x_e \quad(e\in E_{H}),
\]
and $\mu^{H}(T)=0$ for all other feasible sets $T$. This is clearly a probability distribution over feasible sets, with each heavy item getting selectability $c_{H}$. Furthermore,
\[
    \PP_{S\sim\mu^{H}}[e\in S\mid S_{-e}=\emptyset]
    =
    \frac{c_{H}x_e}{c_{H}+c_{H}x_e}
    \le x_e,
\]
and all remaining constraints are trivial. Hence \eqref{eq:soCRS-stationary} holds. Note that this measure is also in fact a solution to a maximum entropy program.

\xhdr{Step (ii): The light-item case.} Let
\[
    \Fset_{L}:=\{T\subseteq E_{L}:a(T)\le 1\}
\]
be the light-item knapsack family. The projection $x_{L}=(x_e)_{e\in E_{L}}$ belongs to $\conv\{\mathbf{1}_T:T\in\Fset_{L}\}$. We may assume without loss of generality that $B>0$. Define $c_{L}:=\frac{1}{1+2B}$, and let $p_e:=c_{L}x_e$ for $e\in E_{L}$, and let $\mu^{L}\in\Delta(\Fset_{L})$ be the maximum-entropy distribution with marginals $p$:
\[
    \PP_{S\sim\mu^{L}}[e\in S]=p_e=c_{L}x_e.
\]
For $S\sim\mu^{L}$, define the addability event $\mathsf{Add}(e):=\{S_{-e}\cup\{e\}\in\Fset_{L}\}.$ Based on the discussion in \Cref{sec:addability}, we only need to verify the condition in~\eqref{eq:addability-lower-template}, so it suffices to show
\[
    \PP_{S\sim\mu^{L}}[\mathsf{Add}(e)]
    \ge c_{L}.\]

If $\mathsf{Add}(e)$ fails, then $ a(S_{-e})>1-a_e.$ Since $e\in E_{L}$, we have $a_e\le 1/2$, and therefore $1-a_e\ge 1/2$. By Markov's inequality,
\[
    \PP_{S\sim\mu^{L}}[\mathsf{Add}(e)^c]
    \le
    \frac{\EE[a(S_{-e})]}{1-a_e}
    \le
    \frac{\EE[a(S)]}{1-a_e}
    \le
    2\EE[a(S)]
    =
    2\sum_{f\in E_{L}}a_fp_f
    =
    2c_{L}\sum_{f\in E_{L}}a_fx_f
    =
    2c_{L}B.
\]
Therefore,
\[
    \PP_{S\sim\mu^{L}}[\mathsf{Add}(e)]
    \ge 1-2c_{L}B
    =
    \frac{1}{1+2B}
    =
    c_{L},
\]
as needed.

\xhdr{Step (iii): Mixing the two cases.} We now combine the two cases, viewing $\mu^{L}$ as a distribution in $\Delta(\Fset)$. Let
\[
    \theta:=\frac{c_{L}}{c_{H}+c_{L}},
    \qquad
    \mu:=\theta\mu^{H}+(1-\theta)\mu^{L}.
\]
An equivalent form of \eqref{eq:soCRS-stationary} is
\[
    (1-x_e)\mu(T\cup\{e\})\le x_e\mu(T),
\]
which is linear in $\mu$. Since both $\mu^{H}$ and $\mu^{L}$ satisfy \eqref{eq:soCRS-stationary}, the mixture $\mu$ also satisfies it. The marginals of heavy and light items under $\mu$ are
\[
    \PP_{S\sim\mu}[e\in S]
    =\theta c_{H}x_e \text{ for }  e\in E_{H} \quad \text{ and } \quad  \PP_{S\sim\mu}[e\in S]
    =(1-\theta)c_{L}x_e \text{ for }  e\in E_{L}.
\]
By the choice of $\theta$,
\[
    \theta c_{H}=(1-\theta)c_{L}
    =
    \frac{c_{H}c_{L}}{c_{H}+c_{L}}
    =
    \frac{1}{c_{H}^{-1}+c_{L}^{-1}}.
\]
Since $c_{H}^{-1}=1+X$ and $c_{L}^{-1}=1+2B,$ we obtain the common selectability factor $\alpha = \frac{1}{2+X+2B}.$
Finally,
\[
    X+2B
    =
    \sum_{e\in E_{H}}x_e
    +
    2\sum_{e\in E_{L}}a_ex_e
    \le
    2\sum_{e\in E}a_ex_e
    \le
    2.
\]
Thus $\alpha\ge \frac{1}{4},$ as desired.\footnote{If $\xvec \in b\Ppoly$, then $X+ 2B \leq 2b$.}\end{proofof}

To obtain an S-OCRS in the more general case where items may have stochastic sizes (that are revealed before a decision must be made on whether to accept them), we associate each item with a set of elements $E_i$ with each element corresponding to different possible size of the item. The elements within each $E_i$ are active mutually exclusively rather than independently. Let $x_e$ and $a_e$ be the probability and size of each element $e$, and assume $\sum_{e\in E_i}x_e\le1$ for every $i$ and $\sum_ea_ex_e\le1$. Feasible sets contain at most one outcome from each $E_i$, with stationarity defined by conditioning on the selections outside $E_i$. Then, we may obtain a $1/5$-selectable S-OCRS by leaving the heavy-item distribution unchanged and for the light items, taking the maximum-entropy distribution on
\[
 \{S\subseteq E_L:a(S)\le1,\ |S\cap E_i|\le1\ \forall i\}
\]
with marginals $c_Lx_e$, where $c_L=(2+2B)^{-1}$.

\xhdr{Negative resut.} We also establish an impossibility result, showing that in the tight instance in \citet{jiang2025tight}, no S-OCRS can obtain a selectability better than $3/10$. Note that this is strictly less than the selectability of the OCRS of \citet{jiang2025tight}. Recall the standard knapsack relaxation:

\[
\widetilde \Ppoly =\left\{\xvec\in[0,1]^E:\sum_{e\in E}a_ex_e\le1\right\}.
\]

\begin{lemma}[Knapsack upper-bound]
\label{lemma:knapsack-impossibility}
For every $\eps\in(0,1/4)$ there exists a knapsack instance, and a point $x \in \widetilde \Ppoly$ such that any distribution $\dist\in\simplex{\Fset}$ satisfying the stationary constraints~\eqref{eq:soCRS-selectability}--\eqref{eq:soCRS-stationary} has selectability at most $3/10+O(\eps)$. In particular, letting $\eps\to 0$ implies that every S-OCRS has selectability at most $3/10$.
\end{lemma}

\begin{proofof}{Proof}
Fix $\varepsilon \in (0, 1/4)$, let $d =1/2+\varepsilon$, and let $m$ be large. Consider an instance with an always-active small item $s$ with $(a_s,x_s)=(\varepsilon,1)$, a rarely active large item $b$ with $(a_b,x_b)=(1,\varepsilon)$, and a family $H=\{h_1,\ldots,h_m\}$ of $m$ items with size $d$ and common activation probability
\[
 p:=\frac{1-2\varepsilon}{md}.
\]
Note that $a_sx_s+a_bx_b+\sum_i a_{h_i}x_{h_i}=\varepsilon+\varepsilon+(1-2\varepsilon)=1$, so $x\in \widetilde \Ppoly.$ The only feasible sets are $\emptyset$, $\{s\}$, $\{b\}$, $\{h_i\}$, and $\{s,h_i\}$. Let $\mu$ be an $\alpha$-selectable stationary witness. For $b$, selectability and stationary implementability give $\mu(\{b\})\ge\alpha\varepsilon$ and $(1-\varepsilon)\mu(\{b\})\le\varepsilon\mu(\emptyset)$, so: \[\mu(\emptyset)+\mu(\{b\})\ge\alpha.\] Next, summing the selectability constraint over the items in $H$ gives \[\sum_i\left(\mu(\{h_i\})+\mu(\{s,h_i\})\right)\ge\alpha mp.\]

Finally, stationary implementability between $\{s\}$ and $\{s,h_i\}$ gives $(1-p)\mu(\{s,h_i\})\le p\mu(\{s\})$. Summing these inequalities and using selectability for the always-active item $s$ gives
\[
 \alpha
 \le \mu(\{s\})+\sum_i\mu(\{s,h_i\})
 \le\left(1+\frac{mp}{1-p}\right)\mu(\{s\}),
\]
and hence: 

\[\mu(\{s\})\ge\alpha/(1+mp/(1-p)).\] 
We conclude that 
\[
 1 = \mu(\emptyset)+\mu(\{b\}) +  \mu(\{s\})+  \sum_i\left(\mu(\{h_i\})+\mu(\{s,h_i\})\right) \ge\alpha\left(1+mp+\frac{1}{1+\frac{mp}{1-p}}\right).
\]
Letting $m\to\infty$ and then $\varepsilon\to0$, we have $p\to0$ and $mp=(1-2\varepsilon)/(1/2+\varepsilon)\to2$. Thus the upper bound on $\alpha$ converges to $(1+2+1/3)^{-1}=3/10$.
\end{proofof}

We note that there is also a knapsack instance and a point $x \in \Ppoly$ (i.e., in the actual convex hull rather than the relaxation) where the selectability is at most $1/3$: Consider $m = 1/\varepsilon^2$ large items with size $1$ and probability of activation $(1-\varepsilon)/m$, and $m$ small items with size $1/m$ and probability of activation $\varepsilon$, with $\varepsilon$ close to $0$.

\section{Beyond Maximum-Entropy: General Downward Closed Feasibility Environments}
\label{sec:downward-closed-constraints}
In this section, we let $(E,\Fset)$ be any finite downward-closed feasibility environment with $|E|=N$. Our goal is to establish the following result, which, to the best of our knowledge, is the first on OCRSs for general downward-closed feasibility environments.

\begin{theorem}[Optimal S-OCRS for general downward-closed feasibility environments]
\label{thm:downward-closed-socrs}
For a general downward closed feasibility environment $(E,\Fset)$ and every $\xvec\in \Ppoly$, there exists a distribution
$\mu^*\in\Delta(\Fset)$ such that Algorithm~1 (simulate-then-replace), when instantiated with $\mu^*$, is a $1/(2\sqrt{|E|}+1)$-selectable S-OCRS.
Moreover, for every positive integer $m$, there is a downward-closed feasibility environment on $N=m^2$ elements and a point in $\Ppoly$ for which every OCRS has selectability at most $2/\sqrt N$.
\end{theorem}

While previously, it was known that there exists an $O((\log N)^2)$ prophet inequality in this setting ~\citep{rubinstein2016beyond, rubinstein2026big}, these results do not extend to OCRSs, which are dual to \textit{ex-ante} prophet inequalities.

\xhdr{Positive result.} To prove the positive results, we let $q:=\frac{1}{\sqrt N},$ and partition the ground set into more likely and less likely items:
\[
    E_{H}:=\left\{e\in E:x_e>q\right\},
    \qquad
    E_{L}:=\left\{e\in E:x_e\le q\right\}.
\]
Since $x\in\Ppoly$, there exists a distribution $\nu\in\Delta(\Fset)$ such that
\[
  \PP_{I\sim\nu}[e\in I]=x_e \qquad \forall e\in E.
\]

We will use two different stationary OCRS witness distributions for the more likely items and less likely items, and mix the two. For the more likely items, we will sample a feasible set from $\nu$, and then keep each element with probability $q$. For the light items, we will select at most one item.

\begin{proofof}{Proof of \Cref{thm:downward-closed-socrs}}We start by working with the more likely item case.

\xhdr{Step (i): The more likely item case.} Define $\mu^{H}$ as follows. Sample $I\sim\nu$, discard all items in $E_L$,
and independently keep each item of $I\cap E_H$ with probability $q$. Let
$R^{H}$ be the resulting set, and let $\mu^{H}$ be the distribution of $(R^{H})$.
Since $R^{H}\subseteq I$ and $\Fset$ is downward closed, $\mu^{H}$ is supported on
$\Fset$. Moreover,
\[
    \PP_{S\sim\mu^{H}}[e\in S]
    =
    \begin{cases}
        qx_e, & e\in E_H,\\
        0, & e\in E_L.
    \end{cases}
\]

We claim that $\mu^{H}$ satisfies \eqref{eq:soCRS-stationary}. If $e\in E_L$, this is trivial, since
$e$ is never selected. If $e\in E_H$, let $D_e$ be the independent retention
coin of $e$. Conditioning on an event of the form $(R^{H})_{-e}=T$ does not
reveal $D_e$. Hence, whenever the conditioning event has positive probability,
\[
    \PP[e\in R^{H}\mid (R^{H})_{-e}=T]
    =
    q\PP[e\in I\mid (R^{H})_{-e}=T]
    \le q
    \le x_e,
\]
where the last inequality uses $e\in E_H$. Thus $\mu^{H}$ is
stationary-implementable with respect to $x$.

\xhdr{Step (ii): The less likely item case.} Define $\mu^{L}$ by
\[
    \mu^{L}(\{e\})= \frac{qx_e}{1+q} \quad(e\in E_L),
    \qquad
    \mu^{L}(\emptyset)=1-\frac{q\sum_{e\in E_L}x_e}{1+q},
\]
and $\mu^{L}(T)=0$ for all other feasible sets $T$. This is a probability
distribution, because
\[
    \frac{q\sum_{e\in E_L}x_e}{1+q}
    \le
    \frac{q^2}{1+q}\cdot |E_L|
    \le \frac{1}{1+q}.
\]
The marginals of $\mu^{L}$ are
\[
    \PP_{S\sim\mu^{L}}[e\in S]
    =
    \begin{cases}
        \dfrac{q}{1+q}x_e, & e\in E_L,\\
        0, & e\in E_H.
    \end{cases}
\]

We now check \eqref{eq:soCRS-stationary}. The only important case is when
$e\in E_L$ and $T=\emptyset$. Since $\mu^{L}(\emptyset)\ge \frac{q}{1+q},$ we have
\[
    \PP_{S\sim\mu^{L}}[e\in S\mid S_{-e}=\emptyset]
    =
    \frac{qx_e}{(1+q)\mu^{L}(\emptyset)+qx_e}
    \le
    \frac{qx_e}{(1+q)\mu^{L}(\emptyset)}
    \le x_e.
\]
Thus $\mu^{L}$ is stationary-implementable with respect to $x$.

\xhdr{Step (iii): Mixing the two cases.} Let
\[
    \mu:=\frac{1}{2+q}\mu^{H}+\frac{1+q}{2+q}\mu^{L}.
\]
Since \eqref{eq:soCRS-stationary} is linear when written in the form:
\[
    (1-x_e)\mu(T\cup\{e\})\le x_e\mu(T),
\]
and since both $\mu^{H}$ and $\mu^{L}$ satisfy \eqref{eq:soCRS-stationary}, the mixture $\mu$ also
satisfies \eqref{eq:soCRS-stationary}. For $e\in E_H$,
\[
    \PP_{S\sim\mu}[e\in S]
    =
    \frac{1}{2+q} qx_e
    =
    \frac{x_e}{2\sqrt N+1}.
\]
For $e\in E_L$,
\[
    \PP_{S\sim\mu}[e\in S]
    =
    \frac{1+q}{2+q}\cdot \frac{q}{1+q}x_e
    =
    \frac{x_e}{2\sqrt N+1}.
\]Thus the construction gives a $1/(2\sqrt N+1)$-selectable stationary OCRS. \end{proofof}

\xhdr{Negative resut.} The next result shows that the previous lower bound has the right order of magnitude even for usual OCRSs. This is a stronger impossibility statement than an upper bound for just stationary OCRSs. The proof is closely related to the ~\citet{kleinberg2012matroid} proof of impossibility of beating $O\left(\frac{1}{\sqrt q}\right)$ for prophet inequalities for the intersection of $q$ matroids.

\begin{lemma}[General downward closed feasibility environments upper-bound]
\label{lemma:downward-impossibility}
Let $m$ be a positive integer. There is a downward-closed feasibility
environment $(E,\Fset)$ with $|E|=N=m^2$ and a point $x\in\Ppoly$ such that every OCRS has selectability at most $2/\sqrt N$ on this instance.
\end{lemma}
\begin{proofof}{Proof}
Partition the ground set into $m$ blocks
\[
  E=E_1\sqcup E_2\sqcup\cdots\sqcup E_m,
   \quad |E_j|=m.
\]
Let
\[
  \Fset:=\{S\subseteq E:S\subseteq E_j\text{ for some }1\le j\le m\}.
\]
Thus a feasible set may contain any subset of one block, but it may not mix
elements from different blocks. This family is downward closed. Set $x_e=\frac1m \ \forall e\in E.$
Choosing a block uniformly and taking it whole has marginals $x$, so
$x\in\Ppoly$. Now fix an arbitrary OCRS and consider revealing the elements of $E$ in any fixed order. Let $A$ denote the final
selected set of the OCRS under this order. For each block $j$, define
\[
  Y_j:=|A\cap E_j|,
   \quad
  D_j:=\{Y_j>0\}.
\]
By $\alpha$-selectability under this particular order,
\begin{equation}\label{eq:ocrslower-block}
  \EE[Y_j]
  =\sum_{e\in E_j}\PP[e\in A]
  \ge \sum_{e\in E_j}\alpha x_e
  = \alpha.
\end{equation}

We next upper-bound the expected number of selected elements from a block,
conditional on selecting at least one element from that block. On the event
$D_j$, let $e_j^\star$ be the first selected element of $E_j$ in the within-block
arrival order. After the arrival of $e_j^\star$, any additional selected element
from $E_j$ must be an active element among the remaining at most $m-1$ elements
of $E_j$. Conditional on the history up to and including $e_j^\star$, these
future activations are still independent Bernoulli random variables with success
probability $1/m$. Hence the expected number of additional selected elements
from $E_j$ is at most $1$. Therefore
\begin{equation}\label{eq:cond-block-size}
  \EE[Y_j\mid D_j]\le 2.
\end{equation}
Combining \eqref{eq:ocrslower-block} and \eqref{eq:cond-block-size}, we get that $\PP[D_j]
  \ge
  \frac{\EE[Y_j]}{2}
  \ge
  \frac{\alpha}{2}.$
Finally, the events $D_1,\ldots,D_m$ are disjoint, because a feasible selected
set cannot contain elements from two different blocks. Thus
\[
  1
  \ge \sum_{j=1}^m \PP[D_j]
  \ge
  m\alpha/2,
\]
so $\alpha\le 2/m=2/\sqrt N$, as needed. \end{proofof}

\section{Deferred Proofs and Technical Details}
\label{apx:missing-proofs}
\subsection{Proof of Proposition~\ref{prop:socrs-to-chain} (S-OCRS \texorpdfstring{$\Longrightarrow$}{=>} recurring OCRS)}
\label{apx:socrs-recurring-equivalence}

\begin{proofof}{Proof}
Maintain a feasible set $\widehat{\Sset}\in\Fset$ as the state of the system, representing the set of elements that are currently selected, together with simulations for all elements that have never appeared in the sequence. We aim to keep this state distributed as $\dist$, a solution to the S-OCRS LP, at all times.
Initialize by sampling $\widehat{\Sset}\sim\dist$.
Whenever element $e$ renews (its current epoch ends and the next epoch starts), first ``forget'' its previous membership by setting $T\defeq \Sm{\widehat{\Sset}}{e}$.
Next, if the new epoch of $e$ is inactive, keep $\widehat{\Sset}\leftarrow T$.
If the new epoch is active, accept it with probability
\(\PP_{\Sset\sim\dist}[e\in\Sset\given \Sm{\Sset}{e}=T]/x_e\), and set $\widehat{\Sset}\leftarrow T\cup\{e\}$ upon acceptance.
This is exactly the simulate-then-replace update that preserves the distribution of $\widehat{\Sset}$.
Since the imaginary state distribution stays equal to $\dist$ after every renewal (regardless of the renewal order), each active epoch of $e$ is accepted with probability
\(\PP_{\Sset\sim\dist}[e\in\Sset]/x_e\ge \alpha\). Moreover, the algorithm is feasible in every realization.
\end{proofof}

\subsection{Proof of Proposition~\ref{prop:stationary-equivalence} (S-OCRS \texorpdfstring{$\Longleftrightarrow$}{<=>} S-OCRS LP)}
\label{apx:socrs-LP-equivalence}

\begin{proofof}{Proof} The proof consists of two steps: 

\emph{(Stationary OCRS $\Rightarrow$ LP).}
Let a stationary OCRS at $\xvec$ be given, and let $\dist$ denote the common output distribution (Definition~\ref{def:stationary-ocrs}).
Selectability of the OCRS is exactly~\eqref{eq:soCRS-selectability}.
To see why \eqref{eq:soCRS-stationary} holds, fix $e\in\ground$ and $T\in\Fset$ with $e\notin T$ and consider running the same OCRS
under an arrival order in which $e$ arrives \emph{last}.
By stationarity, the final selected set still has law $\dist$.
Moreover, in this run the event $\{\Sm{\Sset}{e}=T\}$ is fully determined before $e$ is revealed, hence it is independent
of the activation of $e$.
Since the OCRS can select $e$ only when $e$ is active (an event of probability $x_e$), we obtain
$\PP[e\in\Sset \given \Sm{\Sset}{e}=T]\le x_e$, establishing~\eqref{eq:soCRS-stationary}.

\medskip
\emph{(LP $\Rightarrow$ Stationary OCRS).}
Assume $\dist\in\simplex{\Fset}$ satisfies~\eqref{eq:soCRS-stationary}.
We describe an online procedure that (for \emph{any} arrival order $\pi$) produces an output set $\Sset$ with
distribution $\dist$, while never selecting inactive elements. Process the elements in the arrival order $\pi$, and let $\mathcal{H}$ denote the (random) history of past selection
decisions.
When element $e$ arrives, define the target conditional inclusion probability
\[
q_e\ \defeq\ \PP_{\Sset\sim\dist}[e\in\Sset \given \mathcal{H}],
\]
where the conditioning event $\mathcal{H}$ is interpreted as the event that $\Sset$ agrees with the already-realized
past selections on the elements that have arrived so far.
$q_e$ is always bounded by $x_e$; indeed, $\PP[e\in\Sset\given \mathcal{H}]$ is an average over configurations of $\Sm{\Sset}{e}$, each of which has
conditional probability at most $x_e$ by~\eqref{eq:soCRS-stationary}.

Now, upon observing whether $e$ is active, act as follows:
if $e\notin\Aset$, reject it; if $e\in\Aset$, accept it with probability $q_e/x_e$ (using fresh internal randomness).
Since $q_e\le x_e$, this is well-defined, and it ensures that
$\PP[e\in\Sset \given \mathcal{H}]=q_e$, i.e., the online process matches the conditionals of $\dist$. Therefore the final output set has distribution $\dist$ for every $\pi$, hence the procedure is stationary. Finally,~\eqref{eq:soCRS-selectability} guarantees $\alpha$-selectability of the resulting stationary OCRS.
\end{proofof}

\subsection{Maximum entropy S-OCRS for rank-$L$ hypergraph matchings}
\label{apx:hypergraph}
Recall from \Cref{sec:prelim} the rank-$L$ hypergraph matching environment defined on a hypergraph $H=(V,\ground)$ of rank at most $L$. Our next application of the maximum-entropy template gives a simple S-OCRS for this setting. For completeness, recall that a set $S\subseteq \ground$ is feasible if it is a (hypergraph) matching, i.e., its hyperedges are pairwise vertex-disjoint. Accordingly,
\[
\Fset \ \defeq\ \Bigl\{S\subseteq \ground:\ \forall v\in V,\ \bigl|\{e\in S:\ v\in e\}\bigr|\le 1\Bigr\}.
\]
Note that for every $\xvec$ in the feasibility polytope $\Ppoly=\conv\{\mathbf{1}_S:S\in\Fset\}$, we must have:
\begin{equation}
\sum_{e\in \delta(v)} x_e\ \le\ 1 \qquad \forall v\in V,
\end{equation}
where $\delta(v)$ denotes the set of hyperedges incident to $v$.

\begin{theorem}[S-OCRS for rank-$L$ hypergraph matchings]
\label{thm:hg-socrs}
For the rank-$L$ hypergraph matching environment $(\ground,\Fset)$ and every $\xvec\in\Ppoly$, there exists a Gibbs distribution $\dist^*$ such that Algorithm~\ref{alg:str-gibbs}, when instantiated with $\dist^*$, is a $\tfrac{1}{L+1}$-selectable S-OCRS.
\end{theorem}

\begin{proofof}{Proof}
Fix any $\xvec\in\Ppoly$ and set $\alpha\defeq \tfrac{1}{L+1}$ and $\pvec\defeq \alpha\,\xvec$.
Let $\dist^{*}$ be the maximum-entropy distribution---that is, the solution to~\eqref{eq:maxent-primal-template}---with marginals $\pvec$.
Based on the discussion in \Cref{sec:addability}, we only need to verify the condition in~\eqref{eq:addability-lower-template}.
Fix a hyperedge $e\in\ground$.
The addability event
\vspace{-2mm}
\[
\Add(e)\ =\ \{\Sm{\Sset}{e}\cup\{e\}\in\Fset\}
\]
is exactly the event that no selected hyperedge in $\Sm{\Sset}{e}$ intersects $e$.

For each vertex $v\in V$, define
\vspace{-2mm}
\[
\Occ(v)\ \defeq\ \Bigl\{\exists f\in \Sset\ \text{with}\ v\in f\Bigr\}.
\]
Since $\Sset$ is always a matching, for any fixed vertex $v$ the events $\{f\in\Sset\}$ over distinct $f\in\delta(v)$ are disjoint. Therefore,
\vspace{-3mm}
\begin{equation}
\label{eq:hg-occ-bound}
\PP_{\Sset\sim\dist^{*}}\!\big[\Occ(v)\big]
\ =\ \sum_{f\in\delta(v)}\PP_{\Sset\sim\dist^{*}}[f\in\Sset]
\ =\ \sum_{f\in\delta(v)} p_f.
\end{equation}
Using $p_f=\alpha x_f$ and~\eqref{eq:hg-vertex-packing}, we get
\vspace{-2mm}
\begin{equation}
\label{eq:hg-occ-alpha}
\PP_{\Sset\sim\dist^{*}}\!\big[\Occ(v)\big]
\ =\ \alpha\sum_{f\in\delta(v)} x_f
\ \le\ \alpha.
\end{equation}

Finally, $\Add(e)$ fails only if at least one vertex of $e$ is occupied by some hyperedge in $\Sm{\Sset}{e}$.
Thus, by a union bound and~\eqref{eq:hg-occ-alpha},
\vspace{-1mm}
\[
\PP_{\Sset\sim\dist^{*}}[\Add(e)]
\ \ge\ 1-\sum_{v\in e}\PP\big[\Occ(v)\big]
\ \ge\ 1-|e|\alpha
\ \ge\ 1-L\alpha
\ =\ \alpha.
\]
This is exactly~\eqref{eq:addability-lower-template}. The conclusion follows.
\end{proofof}

\subsection{Proof of Lemma~\ref{correlation_lemma} (Gibbs positive correlation for bipartite matchings)}
\label{apx:monomer-dimer}
Fix the weight vector $\wvec=(w_e)_{e\in\ground}$ induced by a Gibbs distribution $\dist$.
For any subgraph $H$ of $G$, define the partition function
\vspace{-2mm}
\[
\Zpart(H)\ \defeq\ \sum_{M\in\Match(H)}\ \prod_{e\in M} w_e,
\]
where $\Match(H)$ denotes the set of matchings in $H$.
For a vertex $z\in U\cup V$, write $G-z$ for the graph obtained by deleting $z$ and all incident edges, and similarly define $G-u-v$.

Under the Gibbs distribution $\dist$, the probability that a vertex is unmatched is a ratio of partition functions:
\vspace{-2mm}
\begin{equation}
\label{eq:unmatched-ratio}
\PP_{\Sset\sim\dist}[\overline{\Occ(z)}]
\ =\ \frac{\Zpart(G-z)}{\Zpart(G)}.
\end{equation}
Indeed, $z$ is unmatched if and only if $\Sset$ uses no edge incident to $z$, i.e., if and only if $\Sset$ is a matching of $G-z$.
Similarly,
\vspace{-2mm}
\begin{equation}
\label{eq:two-unmatched-ratio}
\PP_{\Sset\sim\dist}[\overline{\Occ(u)} \cap \overline{\Occ(v)}]
\ =\ \frac{\Zpart(G-u-v)}{\Zpart(G)}.
\end{equation}

Thus, the lemma reduces to showing that for any bipartite $G=(U\cup V,\ground)$, any $u\in U$, $v\in V$, and any weights $w_e>0$, we have
\vspace{-2mm}
\begin{equation}
\label{eq:log-supermodular}
\Zpart(G)\,\Zpart(G-u-v)\ \ge\ \Zpart(G-u)\,\Zpart(G-v).
\end{equation}
This inequality is well known for monomer--dimer systems (see for instance Lemma 22.15 in \cite{molloy2002hard}); for completeness, we give a simple combinatorial proof.

\begin{proofof}{Proof of inequality~\eqref{eq:log-supermodular}}
Let $w(M)\defeq \prod_{e\in M} w_e$ denote the weight of a matching.
We construct an injective map
\vspace{-2mm}
\[
\Phi:\ \Match(G-u)\times \Match(G-v)\ \longrightarrow\ \Match(G)\times \Match(G-u-v),
\]
with the property that for every $(M_u,M_v)\in\Match(G-u)\times\Match(G-v)$, we have:
$$
w(M_u)w(M_v)=w(\Phi_1(M_u,M_v))\,w(\Phi_2(M_u,M_v)).
$$
Summing over all pairs then gives
\vspace{-2mm}
\begin{align*}
\Zpart(G-u)\,\Zpart(G-v)
&= \sum_{(M_u,M_v)} w(M_u)w(M_v)
= \sum_{(M_u,M_v)} w(\Phi_1(M_u,M_v))\,w(\Phi_2(M_u,M_v)) \\
&\le \Zpart(G)\,\Zpart(G-u-v),
\end{align*}
which is exactly~\eqref{eq:log-supermodular}.

To define $\Phi$, consider the symmetric difference
\vspace{-2mm}
\[
H\ \defeq\ M_u\triangle M_v\ =\ (M_u\setminus M_v)\cup (M_v\setminus M_u).
\]
Every vertex has degree at most $2$ in $H$, so each connected component of $H$ is either a cycle or a path.
Let $C_v$ be the (possibly empty) connected component of $H$ containing the vertex $v$.
Since $M_v$ is a matching in $G-v$, the vertex $v$ has degree at most $1$ in $C_v$, and hence is an endpoint of a (possibly empty) path.
This path, starting at $v$, alternates between edges from $M_u$ and $M_v$.
\revcolor{This path cannot contain $u$: since $M_u$ has no edge incident to $u$, if the path contained $u$ it would have to end there with an $M_v$-edge, while starting at $v$ with an $M_u$-edge, which gives an even length path between $u$ and $v$.}

Define $\Phi(M_u,M_v)=(M,N)$ by swapping the matchings on $C_v$:
\vspace{-2mm}
\begin{align*}
M\ &\defeq\ (M_v\setminus C_v)\cup (M_u\cap C_v),\\
N\ &\defeq\ (M_u\setminus C_v)\cup (M_v\cap C_v).
\end{align*}
Since we swap the two matchings on the entirety of $C_v$, both $M$ and $N$ are matchings.
By construction, $N$ contains no edge incident to $v$.
In addition, because $C_v$ avoids $u$, the matching $N$ contains no edge incident to $u$ either.
Therefore $M\in\Match(G)$ and $N\in\Match(G-u-v)$. We now need to show the two following properties of $\Phi$ to finish the proof:

\smallskip
\noindent\textbf{(i) Weight preservation:}
On components other than $C_v$ we do not change the matchings; on $C_v$ we swap which matching receives each edge.
Thus $w(M_u)w(M_v)=w(M)w(N)$.

\smallskip
\noindent\textbf{(ii) Injectivity:}
Given an image pair $(M,N)$, let $C_v$ be the connected component containing $v$ in the symmetric difference $M\triangle N$.
Swapping back on $C_v$ recovers a unique preimage $(M_u,M_v)$, so $\Phi$ is injective.
\end{proofof}

\subsection{Proof of Lemma~\ref{lemma:bipartite-impossibility} (Bipartite matching upper bound)}
\label{apx:bipartite-impossiblity}
\begin{proofof}{Proof}
Fix $\eps\in(0,1/2)$ and let $n\defeq 1/\eps$ (assume $1/\eps$ is an integer for simplicity).
Construct the bipartite graph $G=(U\cup V,\ground)$ with
\[
U=\{a\}\cup\{u_1,\dots,u_n\}, \quad
V=\{v_1,\dots,v_n\},
\]
and $2n$ edges $e_i\defeq \{a,v_i\}$, and $g_i\defeq\{u_i,v_i\}$.
Define the activation vector $\xvec$ by
\[
x_{e_i}=\eps, \quad x_{g_i}=1-\eps.
\]
Then $\xvec$ is a fractional matching (and hence in the bipartite matching polytope $\Ppoly$): the degree constraint at $a$ is tight since $\sum_i x_{e_i}=n\eps=1$, and each $v_i$ has $x_{e_i}+x_{g_i}=1$.

Let $\dist$ be any distribution on matchings satisfying stationary implementability, i.e., constraint \eqref{eq:soCRS-stationary}, for $\xvec$. Moreover, suppose it attains selectability $\alpha$, i.e.,
\[
\PP_{\Sset\sim\dist}[e\in\Sset]\ \ge\ \alpha x_e  \quad\forall e\in\ground.
\]
Let $X_i\defeq \Ind{e_i\in\Sset}$ and $A_i\defeq 1-\Ind{g_i\in\Sset}$.
By selectability, we know that
$\EE[X_i]\ge \alpha\eps$, and
$\EE[A_i]\le 1-\alpha(1-\eps).$
Since at most one of the edges $\{e_i\}$ can be chosen,
\begin{equation}
\label{eq:a-matched-lb}
\PP[\text{$a$ is matched in $\Sset$}]
=\sum_{i=1}^n \EE[X_i]
\ \ge\ n\alpha\eps\ =\ \alpha.
\end{equation}
Let $N\defeq \sum_{i=1}^n A_i$. Then
\begin{equation}
\label{eq:ES-ub}
\EE[N]
=\sum_{i=1}^n \EE[A_i]
\ \le\ n\bigl(1-\alpha(1-\eps)\bigr).
\end{equation}

Our strategy is to show, using the stationarity constraint, that
\[
\PP[\text{$a$ is matched in $\Sset$}]\ \le\ \frac{\rho \EE[N]}{1+\rho \EE[N]},
\text{ where } \rho\defeq \frac{\eps}{1-\eps}.
\]
Combined with~\eqref{eq:a-matched-lb} and~\eqref{eq:ES-ub}, this immediately implies
\[
\alpha\ \le\ \frac{1-\alpha(1-\eps)}{2-\eps-\alpha(1-\eps)},
\]
and letting $\eps\to 0$ yields $\alpha^2-3\alpha+1\ge 0$, hence $\alpha\le (3-\sqrt{5})/2=\alphaStar$.

To prove the inequality above, fix $i\in[n]$ and condition on the selection status of all edges other than $e_i$.
The stationary constraint~\eqref{eq:soCRS-stationary} gives
\[\PP[X_i=1\given \Sm{\Sset}{e_i}]\le \eps.\]
Since $X_i=1$ implies $A_i=1$, we may write
\begin{align}
\label{eq:Xi-given-others}
\PP[X_i=1\given \Sm{\Sset}{e_i}]
&\le \rho A_i \PP[\text{$a$ unmatched}\given \Sm{\Sset}{e_i}].
\end{align}
Here, the inequality above follows from noting that if any $e_j$ with $j\neq i$ is selected, then both sides are $0$, and
if no $e_j$ with $j\neq i$ is selected, then $\text{$a$ unmatched}$ is exactly the event $X_i=0$.

Now condition only on $A=(A_1,\dots,A_n)$ and take expectations in~\eqref{eq:Xi-given-others} to obtain
\begin{equation}
\label{eq:Xi-given-A}
\EE[X_i\given A]
\ \le\ \rho A_i \PP[\text{$a$ unmatched}\given A].
\end{equation}
Summing~\eqref{eq:Xi-given-A} over $i$ and using disjointness of the events $\{X_i=1\}$ gives, for $p(A)\defeq \PP[\text{$a$ matched}\given A]$,
\[
p(A)=\sum_{i=1}^n \EE[X_i\given A]
\ \le\ \rho N (1-p(A)).
\]
Solving yields $p(A)\le f(N)\defeq \rho N/(1+\rho N)$.
Since $f$ is concave in $N\ge 0$, Jensen's inequality gives
\[\PP[\text{$a$ matched}]=\EE[p(A)]\le \EE[f(N)]\le f(\EE[N]),\]
which completes the proof.
\end{proofof}

\subsection{Maximum-entropy barrier on $K_4$}
\label{apx:K4-maxent-barrier}
\begin{proposition}[Maximum-entropy barrier on $K_4$]
\label{prop:K4-maxent-barrier}
Let $G$ be the complete graph on vertex set $\{1,2,3,4\}$.
Write the edges as a 4-cycle (``outer'' edges) $E_{\mathrm{out}} \defeq \{12,23,34,41\}$ together with the two diagonals $E_{\mathrm{diag}} \defeq \{13,24\}$. Fix a parameter $\eps\in(0,1/2)$ and define an activation vector $\xvec\in[0,1]^{\ground}$ by
\begin{equation}
\label{eq:K4-x}
x_e  = \frac{1-\eps}{2}\quad \forall e\in E_{\mathrm{out}} \quad, \quad x_e  = \eps\quad \forall e\in E_{\mathrm{diag}}~,
\end{equation}
where $\ground  \defeq E_{\mathrm{out}}\cup E_{\mathrm{diag}}$. Then $\xvec$ lies in the (general) matching polytope of $G$.
For $\alpha\in(0,1)$, let $\pvec\defeq \alpha\xvec$ and let $\dist^{*}$ be the unique
maximum entropy distribution \eqref{eq:maxent-primal-template} over matchings with marginals $\pvec$.
If $\dist^{*}$ satisfies stationary implementability \eqref{eq:soCRS-stationary} with respect to $\xvec$,
then
\[
\limsup_{\eps\to 0} \alpha  \le \frac{\sqrt{3}-1}{2}\approx 0.366.
\]
\end{proposition}
\begin{proofof}{Proof}
Fix $\eps\in(0,1/2)$ and $\alpha\in(0,1)$, set $\pvec=\alpha\xvec$, and let $\dist^{*}$ be the maximum-entropy distribution with marginals $\pvec$.
Since $\dist^{*}(S)\propto \prod_{e\in S} w_e$ over matchings $S$, symmetry implies that all outer edges share a common weight $t>0$ and both diagonals share a common weight $s>0$:
\[
w_e=t \  \forall e\in E_{\mathrm{out}}, \quad w_e=s \ \forall e\in E_{\mathrm{diag}}.
\]
Every matching in $K_4$ has size $0$, $1$, or $2$.
There is one empty matching, six single-edge matchings (four outer, two diagonal), and three perfect matchings: two consisting of opposite outer edges and one consisting of both diagonals.
Therefore the partition function is
\begin{equation}
\label{eq:K4-Z}
\Zpart(t,s)
 = 1+4t+2s+2t^2+s^2.
\end{equation}
For any fixed outer edge $e\in E_{\mathrm{out}}$, the matchings containing $e$ are $\{e\}$ and $\{e,e'\}$ where $e'$ is the opposite outer edge, hence
\[\PP_{\Sset\sim\dist^{*}}[e\in\Sset]
 =\
\frac{t+t^2}{\Zpart(t,s)}\]
Similarly, for any diagonal $d\in E_{\mathrm{diag}}$,
\[\PP_{\Sset\sim\dist^{*}}[d\in\Sset]
 =\
\frac{s+s^2}{\Zpart(t,s)}\]
By construction of $\dist^{*}$,
\[
\PP_{\Sset\sim\dist^{*}}[e\in\Sset] = p_e = \alpha(1-\eps)/2
\quad\forall e\in E_{\mathrm{out}},
 \quad
\PP_{\Sset\sim\dist^{*}}[d\in\Sset] = p_d = \alpha\eps
\quad\forall d\in E_{\mathrm{diag}}.
\]
Simple algebra therefore gives the identities:
\begin{equation}
\label{eq:ratio}
\frac{s(1+s)}{t(1+t)}
 =\
\frac{\alpha\eps}{\alpha(1-\eps)/2}
 =\
\frac{2\eps}{1-\eps},  \quad 
\alpha
 =\
\frac{2}{1-\eps}\cdot \frac{t(1+t)}{\Zpart(t,s)}.
\end{equation}

Fix a diagonal $d\in E_{\mathrm{diag}}$. By Lemma~\ref{lem:gibbs-addability-template} and the stationary implementability constraint in the LP, i.e., \eqref{eq:soCRS-stationary}, whenever $d$ is addable to $\Sm{\Sset}{d}$,
\[
\PP_{\Sset\sim\dist^{*}}\big[d\in\Sset \given \Sm{\Sset}{d}=T\big]  = \rho_d = \frac{s}{1+s} \leq \eps.
\]
Equivalently, $s\le \eps/(1-\eps)$, and hence
\[
s(1+s) \le \frac{\eps}{1-\eps}\cdot \frac{1}{1-\eps}
 = \frac{\eps}{(1-\eps)^2}.
\]
Combining this bound with \eqref{eq:ratio} gives
\begin{equation}
\label{eq:t-cap}
t(1+t) \le \frac{1}{2(1-\eps)}.
\end{equation}
Letting $\eps\to 0$ in~\eqref{eq:t-cap} yields $t(1+t)\le 1/2$.
Moreover, when $\eps\to 0$ we have $s\to 0$, so $\Zpart(t,s)\to 1+4t+2t^2$ and also 
\[\alpha\to \frac{2t(1+t)}{1+4t+2t^2} \defeq f(t).\]

A direct calculation shows that $f(t)$ is increasing for $t\ge 0$, so the largest possible $\alpha$ occurs at the largest feasible $t$. Let $t_0$ be the positive root of $t^2+t=1/2$, namely $t_0=(\sqrt{3}-1)/2$. Therefore, $\alpha \leq f(t_0)$. Evaluating $f$ at $t_0$ gives:

\[
f(t_0)
=
\frac{2t_0(1+t_0)}{1+4t_0+2t_0^2}
=
\frac{1}{1+\sqrt{3}}
=
\frac{\sqrt{3}-1}{2}.
\]
Therefore $\limsup_{\eps\to 0}\alpha\le (\sqrt{3}-1)/2$, as claimed.
\end{proofof}

\subsection{Proof of Lemma~\ref{lem:poisson-comparison} (Poisson comparison)}
\label{apx:poisson-comparison}
\begin{proofof}{Proof}
Write $a_t\defeq \PP[\revcolor{W}=t]$ and $b_t\defeq \PP[Q=t]$, and define the corresponding truncated distributions on
$\{0,1,\dots,k\}$:
\[
\widehat a_t\ \defeq\ \PP[\revcolor{W}=t \given \revcolor{W}\le k]\ =\ \frac{a_t}{\PP[\revcolor{W}\le k]},
\qquad
\widehat b_t\ \defeq\ \PP[Q=t \given Q\le k]\ =\ \frac{b_t}{\PP[Q\le k]}.
\]

It is well known in the combinatorics and probability literature that the coefficient sequence $(a_t)$ is \emph{ultra log-concave}~\citep{liggett1997ultra,pemantle2000towards}, meaning that for all $t\ge 1$,
\begin{equation}
\label{eq:ulc-poibin}
a_t^2\ \ge\ a_{t-1}a_{t+1}\cdot \frac{t+1}{t}.
\end{equation}
Clearly, the same inequality holds with $a_t$
replaced by $\widehat a_t$ for $t=1,\dots,k-1$.
For a Poisson distribution, a direct calculation gives equality:
\begin{equation}
\label{eq:poisson-equality}
\widehat b_t^2\ =\ \widehat b_{t-1}\widehat b_{t+1}\cdot \frac{t+1}{t}
\qquad \text{for all } t=1,\dots,k-1.
\end{equation}
Dividing \eqref{eq:ulc-poibin} (for $\widehat a_t$) by \eqref{eq:poisson-equality} yields that the ratio
\[
r_t\ \defeq\ \frac{\widehat a_t}{\widehat b_t}\qquad(t=0,1,\dots,k)
\]
is log-concave:
$r_t^2\ge r_{t-1}r_{t+1}$ for $t=1,\dots,k-1$.
In particular, $(r_t)_{t=0}^k$ is unimodal. Assume now for the sake of  contradiction that $\widehat a_k>\widehat b_k$.
Then $r_k>1$.
Since $(r_t)$ is unimodal, the set $\{t:r_t > 1\}$ is an interval containing $k$,
so there exists an index $r\in\{0,1,\dots,k\}$ such that $r_t\le 1$ for $t<r$ and $r_t > 1$ for $t\ge r$.
Define $\delta_t\defeq \widehat a_t-\widehat b_t$.
Then $\delta_t\le 0$ for $t<r$, $\delta_t > 0$ for $t\ge r$, and $\sum_{t=0}^k \delta_t=0$.
Therefore,
\[
\sum_{t=0}^k t\,\delta_t
\ \ge\
r\sum_{t=r}^k \delta_t + (r-1)\sum_{t=0}^{r-1}\delta_t
\ =\
\sum_{t=r}^k \delta_t
\ >\ 0,
\]
This shows
\[
\EE[\revcolor{W} \given \revcolor{W}\le k]\ =\ \sum_{t=0}^k t\,\widehat a_t
\ >\
\sum_{t=0}^k t\,\widehat b_t\ =\ \EE[Q \given Q\le k],
\]
contradicting the assumption in the statement of Lemma~\ref{lem:poisson-comparison}. Thus we must have $\widehat a_k\le \widehat b_k$.
\end{proofof}

\subsection{Proof of Theorem~\ref{thm:constant-scaling-k-uniform} (S-OCRS for rank-$L$ hypergraph matchings with capacity $k$)}
\label{apx:constant-scaling-k-uniform}
\begin{proofof}{Proof}

Recall that we fix $\varepsilon\in [0,1]$, and let:
\[
Q\sim\operatorname{Poisson}((1-\varepsilon)k),\quad
\beta:=\PP[Q=k\given Q\le k],\quad
\gamma:=(1-\varepsilon)(1-\beta).
\]
Let $\dist$ be the Gibbs distribution on $\Fset$ with parameters $\rho_e=\gamma x_e$. 
Note that $\PP_{\Sset\sim\dist}[e\in\Sset \given \Sm{\Sset}{e}=T]$ is equal to either $\rho_e$ or $0$. Since we have explicitly chosen $\rho_e= \gamma x_e$ with $\gamma = (1-\varepsilon)(1 - \beta) \leq 1$, \ref{eq:soCRS-stationary} is satisfied automatically, so we concentrate on checking \ref{eq:soCRS-selectability} with $\alpha = \gamma (1 - L\beta)$, i.e.,
\[\PP_{\Sset\sim\dist}[e\in\Sset] \ge \gamma (1 - L\beta)x_e.\]

Now, we know that \[\PP[e\in\Sset] = \PP[\Add(e)]\cdot \rho_e =  \PP[\Add(e)]\cdot \gamma x_e.\] Following the notation of Theorem~\ref{thm:uniform-matroid}, let $(Y_e)_{e\in\ground}$ be independent Bernoulli random variables with $\PP[Y_e=1]=\rho_e= \gamma x_e$. Letting $\revcolor{W_v}\defeq\sum_{e\in \delta(v)} Y_e$, we know $e$ is addable if none of the vertices that are a part of $e$ are at capacity. So by the union bound it follows that:
\[\PP[\Add(e)] \geq 1 - \sum_{v \in e}\PP[\revcolor{W_v} = k | (Y_e)_{e\in\ground} \in \Fset].\]

So it suffices to show that $\PP[\revcolor{W_v} = k | (Y_e)_{e\in\ground} \in \Fset] \leq \beta$, since then \[
 \PP[\Add(e)] \ge 1-|e|\beta \ge 1-L\beta,
\]and the scheme is $\gamma(1-L\beta)$-selectable, as desired. We will show in \Cref{prop:efron} that: $\PP[\revcolor{W_v} = k | (Y_e)_{e\in\ground} \in \Fset] \leq \PP[\revcolor{W_v} = k | \revcolor{W_v} \leq k]$. Assuming this result, note that $\revcolor{W_v}$ is Poisson--binomial, and
\[
\EE[\revcolor{W_v}\mid \revcolor{W_v}\le k] \leq \EE[\revcolor{W_v}] = \sum_{f\in \delta(v)} \rho_f = \gamma\sum_{f\in\delta(v)} x_f
 \le \gamma k
 = (1-\varepsilon)(1-\beta)k.
\]
On the other hand, for $Q\sim\mathrm{Poisson}((1-\varepsilon)k)$ we have the identity
\[
\EE[Q\mid Q\le k]
=(1-\varepsilon)k  \PP[Q<k\mid Q\le k]
=(1-\varepsilon)(1-\beta)k.
\]
Applying Lemma~\ref{lem:poisson-comparison} therefore gives
\[
 \PP[\revcolor{W_v}=k\mid \revcolor{W_v}\le k] \le  \PP[Q=k\mid Q\le k] = \beta,
\]as needed. For the concrete bound, note that  standard Chernoff and Stirling bounds give
\[
\PP[Q \geq k]\le e^{-\frac{k\varepsilon^2}{2}},
\quad
\PP[Q=k]=e^{-(1-\varepsilon )k}\frac{((1-\varepsilon )k)^k}{k!}\le \frac{e^{k(\varepsilon+ \log(1-\varepsilon ))}}{\sqrt{2\pi k}}\leq \frac{e^{-\frac{k\varepsilon^2}{2}}}{\sqrt{2\pi k}}. 
\]
Hence for $L\ge2$ and $k>2\log L$, taking $\varepsilon =\sqrt{2\log L/k}$, for which $e^{-k\varepsilon ^2/2}=1/L$ therefore gives:
\[
\beta\le \frac{1}{\sqrt{2\pi k}}\cdot \frac{1/L}{1-1/L}
=
\frac{1}{\sqrt{2\pi k}}\cdot \frac{1}{L-1},
\qquad
L\beta\le \frac{L}{L-1}\cdot \frac{1}{\sqrt{2\pi k}}
\le \sqrt{\frac{2}{\pi k}}.
\]
With this choice, the selectability lower bound becomes
\[
\gamma(1-L\beta) = (1- \varepsilon)(1-\beta)(1-L\beta)\geq
1-\sqrt{\frac{2\log L}{k}} -O\left(\sqrt{\frac{1}{k}}\right),
\]
as claimed. (The result also works for $L=1$, since with $\varepsilon = 0$, we have $\beta = \sqrt{\frac{2}{\pi k}} + O\left(\frac{1}{k}\right)$.)
\end{proofof}
It remains to show that $\PP[\revcolor{W_v} = k | (Y_e)_{e\in\ground} \in \Fset] \leq \PP[\revcolor{W_v} = k | \revcolor{W_v} \leq k]$. Intuitively speaking, this ought to be true because conditioning on the stricter condition $(Y_e)_{e\in\ground} \in \Fset$ ought to decrease the values of the $Y$ variables, since vertices $v'$ other than $v$ will also need $\revcolor{W_{v'}}\leq k$. This in turn should reduce the chance that $\revcolor{W_v} = k$. Formally, we establish the result in the following proposition using Efron monotonicity \citep{efron1965increasing}:

\begin{proposition}
\label{prop:efron}
Let $(Y_e)_{e\in\ground}$ be independent Bernoulli random variables with $\PP[Y_e=1]=\rho_e= \gamma x_e$. Fix a vertex $v\in V$, and let $\revcolor{W_v} = \sum_{e\in \delta(v)} Y_e$. Then, 
\[\PP[\revcolor{W_v} = k | (Y_e)_{e\in\ground} \in \Fset] \leq \PP[\revcolor{W_v} = k | \revcolor{W_v} \leq k].\]
\end{proposition}
\begin{proofof}{Proof}
Consider for the vertex $v$ the edges which are neighbors, and the rest:
\[N\defeq\{f\in\delta(v):Y_f=1\}
\quad\text{and}\quad
R\defeq\{f\in\ground\setminus\delta(v):Y_f=1\},
\]
so that $|N|=\revcolor{W_v}$. Condition on a realization $R=r$. Then
\begin{align*}
\PP\left[\revcolor{W_v}=k\given (Y_f)_{f\in\ground}\in\Fset,\ R=r\right]
&= \frac{\PP[\revcolor{W_v}=k,\ N\cup r\in\Fset\given R=r]}
{\PP[N\cup r\in\Fset\given R=r]}\\
&=
\frac{\PP[N\cup r\in\Fset\given \revcolor{W_v}=k,R=r]\PP[\revcolor{W_v}=k]}
{\displaystyle\sum_{t=0}^k
\PP[N\cup r\in\Fset\given \revcolor{W_v}=t,R=r]\PP[\revcolor{W_v}=t]}.
\end{align*}
Here we have used the independence of $R$ and $N$. The discrete version of Efron's monotonicity theorem states that, for independent Bernoulli random variables, the conditional expectation of any coordinatewise increasing function, conditional on their sum being $t$, is increasing in $t$ \citep[Proposition~2.3]{goldschmidt2008fragmenting}; see also \citet{efron1965increasing} for the original theorem. Given $R=r$, the indicator of the event $N\cup r\in\Fset$ is coordinatewise decreasing in $(Y_f)_{f\in\delta(v)}$, since $\Fset$ is downward-closed. Applying the theorem to the complementary indicator therefore gives, for every $t\leq k$,
\[
\PP[N\cup r\in\Fset\given \revcolor{W_v}=t,R=r]
\geq
\PP[N\cup r\in\Fset\given \revcolor{W_v}=k,R=r].
\]
Substituting this inequality into the denominator above gives
\begin{align*}
\PP\left[\revcolor{W_v}=k\given (Y_f)_{f\in\ground}\in\Fset,\ R=r\right]
&\leq
\frac{\PP[N\cup r\in\Fset\given \revcolor{W_v}=k,R=r]\PP[\revcolor{W_v}=k]}
{\displaystyle\sum_{t=0}^k
\PP[N\cup r\in\Fset\given \revcolor{W_v}=k,R=r]\PP[\revcolor{W_v}=t]}\\
&=
\frac{\PP[\revcolor{W_v}=k]}{\sum_{t=0}^k\PP[\revcolor{W_v}=t]}
=\PP[\revcolor{W_v}=k\given \revcolor{W_v}\leq k].
\end{align*}
Averaging over $r$ under the conditional distribution of $R$ given $(Y_f)_{f\in\ground}\in\Fset$ gives the desired result.\end{proofof}

\subsection{Proof of Lemma~\ref{prop:uniform-tightness} (Uniform matroid upper bound)}
\label{apx:uniform-tightness}
\begin{proofof}{Proof}
Let $\dist\in\simplex{\Fset}$ be any feasible distribution for the symmetric instance satisfying stationary implementability~\eqref{eq:soCRS-stationary}, and suppose it attains some selectability $\alpha$. Averaging $\dist$ over all permutations of $\ground$ preserves these properties. Thus, we may assume $\dist$ is exchangeable: $\dist(S)$ depends only on $|S|$.
Define
\[
\pi_\ell\ \defeq\ \PP_{\Sset\sim\dist}[\,|\Sset|=\ell\,]\qquad(\ell=0,1,\dots,k).
\]
Then every subset of size $\ell$ has probability $\pi_\ell/\binom{n}{\ell}$.

Fix $e\in\ground$ and fix any $T\subseteq \ground\setminus\{e\}$ with $|T|=\ell\le k-1$.
Conditioning on $\Sm{\Sset}{e}=T$ leaves only the possibilities $\Sset=T$ and $\Sset=T\cup\{e\}$, so
\[
\PP[e\in\Sset \given \Sm{\Sset}{e}=T]
=
\frac{\pi_{\ell+1}/\binom{n}{\ell+1}}{\pi_\ell/\binom{n}{\ell}+\pi_{\ell+1}/\binom{n}{\ell+1}}
=
\frac{(\ell+1)\pi_{\ell+1}}{(n-\ell)\pi_\ell+(\ell+1)\pi_{\ell+1}}.
\]
The constraint from the stationary OCRS LP---that is,  \eqref{eq:soCRS-stationary} (with $x_e=q$)---therefore implies, for all $\ell=0,\dots,k-1$,
\begin{equation}
\label{eq:recurrence-ineq-uniform}
(1-q)(\ell+1)\pi_{\ell+1}\ \le\ q(n-\ell)\pi_\ell.
\end{equation}

Since the instance is symmetric, $\PP[e\in\Sset]=\EE[|\Sset|]/n$ for every $e$ and hence the selectability $\alpha$
satisfies
\[
\alpha
\ \le\
\frac{\PP[e\in\Sset]}{q}
\ =\
\frac{\EE[|\Sset|]}{nq}
\ =\
\frac{\EE[|\Sset|]}{k}.
\]
Thus, upper-bounding $\alpha$ reduces to upper-bounding $\EE[|\Sset|]=\sum_{\ell=0}^k \ell \pi_\ell$ subject to
\eqref{eq:recurrence-ineq-uniform} and $\sum_{\ell=0}^k \pi_\ell=1$.
At an optimum, all constraints \eqref{eq:recurrence-ineq-uniform} must be tight:
if for some $\ell$ the inequality is strict, then shifting an infinitesimal mass from $\pi_\ell$ to $\pi_{\ell+1}$
increases $\EE[|\Sset|]$ and preserves feasibility for sufficiently small shifts.
Hence, for $\ell=0,\dots,k-1$,
\begin{equation}
\label{eq:recurrence-eq-uniform}
\frac{\pi_{\ell+1}}{\pi_\ell}
\ =\
\frac{q}{1-q}\cdot \frac{n-\ell}{\ell+1}.
\end{equation}
The right-hand side equals the ratio of successive binomial probabilities for $\mathrm{Bin}(n,q)$, so $|\Sset|$ has the distribution of $B\sim \mathrm{Bin}(n,q)$ conditioned on $\{B\le k\}$.
Therefore
\[
\EE[|\Sset|]
\ =\ \EE[B \given B\le k]
\ =\
\frac{\EE[B\Ind{B\le k}]}{\PP[B\le k]}.
\]
Using the standard binomial identity $\EE[B\Ind{B\le k}]=nq\cdot \PP[\mathrm{Bin}(n-1,q) < k]$ and $nq=k$,
we obtain
\[
\alpha
\ \le\
\frac{\EE[|\Sset|]}{k}
\ =\
\frac{\PP[\mathrm{Bin}(n-1,q)\le k-1]}{\PP[\mathrm{Bin}(n,q)\le k]},
\]
which is \eqref{eq:binomial-ratio}. Finally, as $n\to\infty$ with $q=k/n$, the Poisson limit theorem gives
$\mathrm{Bin}(n,q)\Rightarrow \mathrm{Poisson}(k)$ and $\mathrm{Bin}(n-1,q)\Rightarrow \mathrm{Poisson}(k)$, so the ratio
\eqref{eq:binomial-ratio} converges to 
$$\frac{\PP[\mathrm{Poisson}(k) < k]}{\PP[\mathrm{Poisson}(k)\le k]}=\alpha_k,$$
as desired.
\end{proofof}

\subsection{Proof of convergence of set of admitted proposals (S-OCRS to admission control)}
\label{apx-recurring-implies-admission}
In this section, we prove the convergence claim used in
\Cref{lem:recurring-implies-admission}. Recall that we fix a distribution over clones $\mu_n$, and type-$i$ proposals assigned to a clone $e^j$ arrive at rate $y_{ie}/n$. For every time $t$, we let ${S}(t) \in \Fset_n$ be the set of currently admitted proposals just before time $t$ with $S(0) = \emptyset$. If $e_j \in S(t)$ or $S(t) \cup\{e^j\}\notin\Fset_n$,  the proposal is rejected. Otherwise, the proposal is admitted with probability
\[ f(S(t), e^j) = \frac{n\mu_n(S(t)\cup\{e^j\})}{x_e\mu_n(S(t))} \leq 1,\]
where $x_e = \sum_{i\in\mathcal I}\ell_{ie}y_{ie}$, and $e^j$ is removed from $S(t)$ after time $\ell_{ie}$ passes. The goal is to show that $S(t)$ converges in distribution to $\mu_n$.

The set-valued process $S(t)$ is not Markov on its own, because the durations for which clones are occupied depends on the type of the
proposal occupying it. Therefore, for the purpose of analysis, we label the type of the proposal occupying each clone and write
\[
Z(t)
\defeq
\left(S(t),(I_c(t))_{c\in S(t)}\right),
\]
where $I_c(t) \in \mathcal{I}$ is the type of request using the clone $c$.  For every product $e$ and type $i$, define
\[
\omega_{ie}
\defeq
\frac{y_{ie}\ell_{ie}}{x_e}.\]
For a state $(A,\tau_A)$, where $A\in\Fset_n$ and
$\tau_c\in\mathcal I$ denotes the type occupying $c\in A$, define
\[
\Pi(A,\tau_A)
\defeq
\mu_n(A)
\prod_{c=e^j\in A}\omega_{\tau_c e}.\]
The occupied-set marginal of $\Pi$ is $\mu_n$. Now $Z(t)$ is a continuous-time Markov chain on $\operatorname{supp}(\Pi)$, with finite rates of transition between states. For any state in $\operatorname{supp}(\Pi)$, there is a positive rate at which it transitions to the empty state (if all occupied units return before there are any new arrivals of requests). Furthermore, for any state in $\operatorname{supp}(\Pi)$, there is a positive rate at which we transition to that state from the empty state (if we have arrivals of the appropriate types of requests and we choose the appropriate clones; note that $\mu_n(A)>0$ and $c\in A$, then $\mu_n(A\setminus\{c\})>0$). We conclude that $Z(t)$ is a non-explosive positive recurrent CTMC on $\operatorname{supp}(\Pi)$, and so converges to a unique stationary distribution (see Chapter 3 of \citep{norris1998markov}). We will show that $\Pi$ is an invariant distribution for $Z(t)$, and therefore that $S(t)$ converges in distribution to $\mu_n$, as needed.

Now fix two adjacent states in $\operatorname{supp}(\Pi)$ of the form $z=(B,\tau_B)$ and $z'=(B\cup\{c\},(\tau_B,i)),$
where $c=e^j\notin B$. The transition from $z$ to $z'$ occurs when a type-$i$ proposal is
assigned to clone $c$ and is admitted. Its transition rate is therefore
\[q(z,z')
=\frac{y_{ie}}{n}f(B,c).
\]
The reverse transition occurs when the type-$i$ occupation of $c$ ends. Since its duration is exponential with mean $\ell_{ie}$, the reverse
transition rate is $q(z',z) =\frac{1}{\ell_{ie}}.$ 
Using the definitions of $f$, $\omega_{ie}$, and $\Pi$, we obtain
\begin{align*}
\Pi(z)q(z,z')
&=
\mu_n(B)
\left(
\prod_{b=g^k\in B}\omega_{\tau_b g}
\right)
\frac{y_{ie}}{n}f(B,c)\\
&=
\mu_n(B)
\left(
\prod_{b=g^k\in B}\omega_{\tau_b g}
\right)
\frac{y_{ie}}{n}
\frac{n\mu_n(B\cup\{c\})}
{x_e\mu_n(B)}\\
&=
\mu_n(B\cup\{c\})
\left(
\prod_{b=g^k\in B}\omega_{\tau_b g}
\right)
\frac{y_{ie}}{x_e}\\
&=
\mu_n(B\cup\{c\})
\left(
\prod_{b=g^k\in B}\omega_{\tau_b g}
\right)
\omega_{ie}\frac{1}{\ell_{ie}}\\
&=
\Pi(z')q(z',z).
\end{align*}
Thus every pair of adjacent states satisfies detailed balance. It follows that $\Pi$ is invariant for $Z(t)$ as required.

\subsection{Implementation using posted pricing}
\label{apx-posted-pricing}
We now extend our reusable-product allocation model to the posted pricing setting. In this setting, it will be convenient to assume that each duration distribution is deterministic and equals $\ell_{ie}$ (we can extend to more general distributions by creating new types for each possible duration of use). When a request arrives, a \textit{randomized posted pricing policy} must post nonnegative prices $p_{ed}(M,U)$ for each product $e$, where $M$ denotes the current multiset of allocated products, $d$ denotes the duration of use, and $U$ denotes a random variable independent of the type of the arriving request. A type-$i$ request chooses a product maximizing its \textit{utility}, $\reward_{ie}-p_{e, \ell_{ie}}(M,U)$, in case this maximum is at least zero. This product must then be immediately allocated. We will show that our guarantees extend to this restricted class of policies:

\begin{theorem}
\label{prop:reusable-pricing}
Suppose that for every $n$-split of $(\ground,\mathcal N)$, there is an $\alpha$-selectable S-OCRS. Then, for every $\eps>0$, there is a randomized posted pricing policy $\pi$ such that
\[
\operatorname{R}(\pi)\geq(\alpha-\eps)\operatorname{OPT}.
\]
\end{theorem}

\begin{proofof}{Proof} Let $\arrivalRate\defeq\sum_{i\in\mathcal I}\arrivalRate_i$ be the total arrival rate. Among the optimal solutions to $\operatorname{FLP}$, choose $(y^\circ,\xvec^\circ)$ minimizing the total allocation rate $\sum_{i,e}y^\circ_{ie}$, and set $z^\circ_{ie}\defeq y^\circ_{ie}/\arrivalRate$. Such a solution can be computed by solving an LP (in polynomial time under the separation-oracle assumption of \Cref{sec:fluid}). If we follow the scheme from \Cref{sec:fluid}, the probability that an arriving request proposes product $e$ for duration $d$ is
\[
c^\circ_{ed}\defeq\sum_{i:\ell_{ie}=d}z^\circ_{ie}.
\]
Then, our admission rule in the proof of \cref{lem:recurring-implies-admission} reduces these probabilities according to the multiset of currently allocated products. Even though the admission probabilities formally depend on clones, we can determine an effective admission probability $f_e(M)$ for a request for product $e$ that depends only on the allocated products $M$, by symmetrizing $\mu_n$ and averaging over clones. The probability that an arriving request is admitted to product $e$ for duration $d$ when the multiset of allocated products is $M$ is then $\hat{c}_{ed} = f_e(M)c^\circ_{ed} \leq c^\circ_{ed}$.

Our goal is to identify randomized prices that induce these allocation probabilities for each product and duration (not necessarily for each product and request type), while simultaneously ensuring that the resulting reward remains large. For $0\leq c\leq c^\circ$, consider the following strengthening of the fluid LP, written in terms of allocation probabilities $z_{ie}=y_{ie}/\arrivalRate$ and expected reward per arrival:
\begin{align}
\operatorname{LP}(c)\defeq\
\max_{z_{ie}\geq0}\quad
&\sum_{i\in\mathcal I}\sum_{e\in\ground}\reward_{ie}z_{ie}\notag\\
\textnormal{s.t.}\quad
&\sum_{e\in\ground}z_{ie}\leq\frac{\arrivalRate_i}{\arrivalRate}
&&\forall i\in\mathcal I,\notag\\
&\sum_{i:\ell_{ie}=d}z_{ie}\leq c_{ed}
&&\forall(e,d)\in\mathcal J.
\label{eq:pricing-duration-rates}
\end{align}
where $\mathcal J\defeq\{(e,\ell_{ie}):i\in\mathcal I,\ e\in\ground\}$ indexes the product--duration pairs. Note that the expected reward per arrival of our policy in \cref{lem:recurring-implies-admission} when the set of allocated products is $M$ is bounded by $\operatorname{LP}(\hat{c})$, since its allocation probabilities are feasible for $\operatorname{LP}(\hat{c})$. Thus, if for every $0\leq c\leq c^\circ$, we can show there exists a nonnegative randomized price that is independent of the arriving type, induces allocation probabilities exactly $c$, and obtains expected reward $\operatorname{LP}(c)$ per arrival, our proof would be complete (by coupling the performance of our policy with the policy in \cref{lem:recurring-implies-admission}).

It is natural to try to identify such prices by considering the dual of $\operatorname{LP}(c)$:
\begin{align*}
\operatorname{DLP}(c)\defeq\min_{u_i,p_{ed}\geq0}\quad
&\sum_{i\in\mathcal I}\frac{\arrivalRate_i}{\arrivalRate}u_i+\sum_{(e,d)\in\mathcal J}c_{ed}p_{ed}\\
\textnormal{s.t.}\quad
&u_i+p_{e,\ell_{ie}}\geq\reward_{ie}
&&\forall i\in\mathcal I,\ e\in\ground.
\end{align*}
Suppose $z$ is an optimal solution to $\operatorname{LP}(c)$ and
$(u,p)$ an optimal solution to $\operatorname{DLP}(c)$. Here $p_{ed}$, the dual variable for~\eqref{eq:pricing-duration-rates}, should be interpreted as a possible price for product $e$ used for duration $d$. Note that $u_i$ must equal $\max_{e \in E} (r_{ie} - p_{e,\ell_{ie}})_+$, the maximum possible utility $i$ may obtain. 

Suppose we use the vector $p$ as prices. By complementary slackness, any $e$ with $z_{ie} > 0$ is in the set of products of maximal utility for type $i$ requests. Furthermore, if $\sum_{e \in E}z_{ie} < \arrivalRate_i/\arrivalRate$, then $u_i$ equals $0$, so not allocating anything to a type $i$ request also maximizes its utility. Thus, every choice prescribed by $z$ maximizes utility at prices $p$.

However, there are two problems with using $p$ as prices. Firstly, it is not clear that there is a \textit{type independent} way to break ties among utility maximizing choices while ensuring that the allocation probabilities match those prescribed by $z$. Secondly, even if we matched the probabilities prescribed by $z$, the constraints in~\eqref{eq:pricing-duration-rates} need not be tight for $z$, but recall that we require the allocation probability for each product--duration pair to equal $c_{ed}$ for our coupling argument.

We would therefore like to show that there exists an optimal solution to $\operatorname{DLP}(c)$ in which all the prices $p_{ed}$ are positive. This would solve the second issue, since by complementary slackness, we would know that for any optimal solution $z$ to $\operatorname{LP}(c)$, all the constraints in~\eqref{eq:pricing-duration-rates} are tight. It would also allow us to make progress on the first issue, since we may apply a small, common, random perturbation to all the prices while keeping them positive.

\textbf{Positive optimal dual solution.} Let $(u^\circ,p^\circ)$ be an optimal solution to $\operatorname{DLP}(c^\circ)$ in a strictly complementary optimal pair for $\operatorname{LP}(c^\circ)$  (see the Goldman--Tucker theorem~\citep[Theorem~II.3]{roos2005interior}).
Since $z^\circ$ is optimal for $\operatorname{LP}(c^\circ)$, every optimum of $\operatorname{LP}(c^\circ)$ makes all the constraints in~\eqref{eq:pricing-duration-rates} tight, since otherwise, we would obtain a fluid optimum with a smaller total allocation rate, contradicting our choice of $y^\circ$. Strict complementarity thus gives $p^\circ_{ed}>0$.

Note that these positive optimal prices can be computed in polynomial time: solve $\operatorname{LP}(c^\circ)$, then maximize $t$ over solutions to $\operatorname{DLP}(c^\circ)$ whose objective equals $\operatorname{LP}(c^\circ)$, subject to
\[
p_{ed}\geq t
\quad\forall(e,d)\in\mathcal J,
\qquad
0\leq t\leq1.
\]
The resulting LP has polynomial size and a positive optimal value.

Now consider $0\leq c\leq c^\circ$ and let $(u,p)$ be an optimal solution to $\operatorname{DLP}(c)$ (which can also be computed in polynomial time). Write $\wedge$ and $\vee$ for the coordinatewise minimum and maximum. We will show that $p\vee p^\circ>0$ is also an optimal solution to $\operatorname{DLP}(c)$. 

Note that the pairs
$(u\wedge u^\circ,p\vee p^\circ)$ and $(u\vee u^\circ,p\wedge p^\circ)$ are both feasible for $\operatorname{DLP}(c)$ since, in each constraint, one variable from an original feasible pair is retained and the other is increased. Denote the objective of $\operatorname{DLP}(c)$ by
\[
\operatorname{D}_c(u,p)\defeq
\sum_{i\in\mathcal I}\frac{\arrivalRate_i}{\arrivalRate}u_i
+\sum_{(e,d)\in\mathcal J}c_{ed}p_{ed}.
\]
Since $c\leq c^\circ$,
\[\operatorname{D}_c(u\wedge u^\circ,p\vee p^\circ)
+\operatorname{D}_{c^\circ}(u\vee u^\circ,p\wedge p^\circ)
\leq\operatorname{D}_c(u,p)+\operatorname{D}_{c^\circ}(u^\circ,p^\circ).\]
This is because the utility terms have the same sum on both sides, while larger prices are paired with smaller coefficients on the left hand side. On the other hand, the optimality of $(u, p)$ and $(u^\circ,p^\circ)$  gives the reverse inequality. This means that $p^c\defeq p\vee p^\circ>0$ is an optimal solution to $\operatorname{DLP}(c)$.

Next, we would like to prescribe a type-independent random perturbation of $p^c$ that can ensure that the probabilities of allocations match $c$.

\xhdr{Matching the allocation probabilities with randomized prices.}  Let $Q$ be the convex hull of the product--duration probability vectors obtained by deterministically assigning each type one of its utility-maximizing choices at $p^c$, weighting type $i$ by its arrival probability $\arrivalRate_i/\arrivalRate$. Since $p^c>0$, complementary slackness implies that every optimum of $\operatorname{LP}(c)$ attains all the bounds~\eqref{eq:pricing-duration-rates} and uses only these choices. Hence $c\in Q$.

Note that we can maximize any linear function over $Q$ in polynomial time. To see this, assign a weight $w_{ed}$ to each product--duration pair $(e,d)\in\mathcal J$, and weight zero to making no purchase. For $q\in Q$, the sum $\sum_{(e,d)\in\mathcal J}w_{ed}q_{ed}$ is maximized by letting each type select an alternative with largest weight among its utility-maximizing choices at $p^c$, since the choices for different types can be made independently.

Applying \citet[Theorem~6.5.11]{gls88} to $Q$ and $c$ therefore computes, in polynomial time, at most $|\mathcal J|+1$ vertices $q^k$ of $Q$ and probabilities $\theta_k$ such that
\[
\sum_k\theta_k q^k=c,
\quad
\sum_k\theta_k=1,
\quad
\theta_k\geq0.
\]

For each vertex $q^k$, the construction in the proof of \citet[Theorem~6.5.8]{gls88} computes, in polynomial time, weights $w^k$ such that $q^k$ uniquely maximizes $\sum_{(e,d)\in\mathcal J}w^k_{ed}q_{ed}$ over $q\in Q$. For these weights, each type has a unique largest-weight alternative among its utility-maximizing choices at $p^c$. Choose $\delta>0$ sufficiently small that, for every $k$, the prices
\[
p^c_{ed}-\delta w^k_{ed}
\]
remain positive and preserve every strict utility comparison at $p^c$, including comparisons with making no purchase. Such a $\delta$ can be computed in polynomial time. The price change adds $\delta w^k_{ed}$ to the utility of product $e$ used for duration $d$, while leaving the utility of making no purchase at zero. Thus, among choices tied at $p^c$, each arriving request uniquely prefers the alternative with largest weight under $w^k$. These prices therefore induce the allocation probability vector $q^k$.

Post the price vector $p^c-\delta w^k$ with probability $\theta_k$, independently of the arriving type. The resulting product--duration probabilities are $\sum_k\theta_k q^k=c$. Every selected choice still maximizes utility at $p^c$. Writing $u_i^c$ for that maximum utility, the expected reward per arrival is therefore
\[
\sum_{i\in\mathcal I}\frac{\arrivalRate_i}{\arrivalRate}u_i^c
+\sum_{(e,d)\in\mathcal J}p^c_{ed}c_{ed}
=\operatorname{LP}(c),
\]
as needed.
\end{proofof}

\section{Discussion of Time Complexity of Algorithms}
\label{apx:time-complexity}
\label{appendix-polytime}
This appendix discusses how to implement (in polynomial time, up to a prescribed accuracy) the stationary OCRS policies
constructed in the paper.

\smallskip
\xhdr{Computational requirements for polynomial-time running of a stationary OCRS.}
Fix $\xvec\in\Ppoly$ and let $\dist$ be any distribution on $\Fset$ that satisfies the stationary constraints
\eqref{eq:soCRS-selectability}--\eqref{eq:soCRS-stationary}. The discussion after Proposition~\ref{prop:stationary-equivalence} gives an explicit implementation via a simulate-then-replace policy. This discussion showed that to run the stationary OCRS corresponding to $\dist$, it suffices to have a polynomial-time procedure
that samples (or approximately samples) from $\dist$, and is able to (exactly or approximately) compute: \[\PP_{\Sset\sim\dist} \big[e\in\Sset \given \Sm{\Sset}{e}=T\big]~,\]
for any $e$ and any set $T$ that does not contain $e$.
In our constructions, $\dist$ has additional structure (maximum-entropy for the first three environments, a Gibbs distribution for hypergraph matchings with vertex capacities, a mixture of maximum-entropy distributions for the knapsack constraint, and a
KL-divergence based construction for weakly Rayleigh matroids), which reduces these requirements to
simpler primitives. \revcolor{We spell this out in the following sections.}

\subsection{The maximum-entropy constructions}
\label{app:polytime-maxent-template}

In the three feasibility environments whose proofs follow the maximum-entropy method in
Section~\ref{sec:socrs-template} (matchings (Theorem~\ref{thm:graph-socrs}), bipartite matchings
(Theorem~\ref{thm:bipartite-tight}), and the rank-$k$ uniform matroid (Theorem~\ref{thm:uniform-matroid})), $\dist$ is the maximum-entropy distribution $\dist^*$ with prescribed marginals
$\pvec=\alpha\xvec$ (Section~\ref{sec:maxent}).
By Lemma~\ref{lem:gibbs-addability-template}, $\dist^*$ is a Gibbs distribution with parameters
$\wvec=(w_e)_{e\in\ground}$ and constants $\rho_e=w_e/(1+w_e)$. In this case, simulate-then-replace takes the particularly simple form given by Algorithm \ref{alg:str-gibbs}.

The only online computations are (a) determining whether $T\cup\{e\}\in\Fset$ and
(b) sampling a Bernoulli draw with parameter $\rho_e/x_e$. Thus, once the constants $\{\rho_e\}$ and an initial sample $\widehat{\Sset}\sim\dist^*$ are available, each arrival can be processed in polynomial time. We thus need to explain:
\begin{enumerate}[leftmargin=2em,label=(\roman*)]
\item how we compute (approximately) the Gibbs parameters $\wvec$ and hence $\rho_e=w_e/(1+w_e)$ in polynomial time, so that
$\dist^*$ has marginals $\pvec=\alpha\xvec$; and
\item how we can sample (approximately) an initial set $\widehat{\Sset}\sim\dist^*$ in polynomial time.
\end{enumerate}

Formally, by constructing polynomial time implementations for the above two steps, we will prove the following theorem:

\begin{theorem}[Polynomial-time S-OCRS for matchings, $\mathbf{k}$-uniform matroids]
\label{thm:polytime-matchings-uniform-matroids}
For every $\varepsilon>0$: 
\begin{enumerate}
    \item There is a $(\tfrac13-\varepsilon)$-selectable S-OCRS for the matching feasibility environment $(\ground,\Fset)$ that runs in time $\mathrm{poly}(|\ground|,1/\varepsilon)$.
    \item There is a $(\frac{3-\sqrt{5}}{2}-\varepsilon)$-selectable S-OCRS for the bipartite matching feasibility environment $(\ground,\Fset)$ that runs in time $\mathrm{poly}(|\ground|,1/\varepsilon)$.
    \item There is a $(\alpha_k-\varepsilon)$-selectable S-OCRS for the $k$-uniform matroid feasibility environment $(\ground,\Fset)$ that runs in time $\mathrm{poly}(|\ground|,\log \left(1/\varepsilon\right))$.
\end{enumerate}
\end{theorem}

 To compute Gibbs parameters as in step (i) above, we need to solve a maximum-entropy program. Before proceeding to the proof of \Cref{thm:polytime-matchings-uniform-matroids}, we summarize the approach of \cite{singh2014entropy, straszak2019maximum} to achieve this goal in a convenient form in the next subsection.  
 
\subsection{Solving max-entropy/ minimum KL divergence programs}
\label{app:polytime-gibbs-dual}
An initial observation is that if we fix $\pvec$ in the relative interior of $\Ppoly$, the maximum-entropy problem
\eqref{eq:maxent-primal-template} has a convex dual in variables $\thetavec\in\mathbb{R}^{\ground}$:
\begin{equation}
\label{eq:maxent-dual}
\min_{\thetavec\in\mathbb{R}^{\ground}}
h(\thetavec)
\ \defeq\
\log \Zpart(\thetavec)\ -\ \ip{\thetavec}{\pvec},
\qquad
\Zpart(\thetavec)=\sum_{S\in\Fset}\exp\!\Big(\sum_{e\in S}\theta_e\Big).
\end{equation}
The gradient of $h$ is the marginal mismatch:
\[
(\nabla h(\thetavec))_e
=
\PP_{\Sset\sim \dist_{\thetavec}}[e\in\Sset]\ -\ p_e,
\]
where $\dist_{\thetavec}$ is the Gibbs distribution proportional to
$\exp(\sum_{e\in S}\theta_e)\Ind{S\in\Fset}$.
Thus, in any environment where we can evaluate or approximate the Gibbs marginals
$\PP_{\Sset\sim \dist_{\thetavec}}[e\in\Sset]$ we should be able to use standard convex optimization algorithms (ellipsoid methods or first-order methods) and compute an
$\varepsilon$-approximate minimizer $\thetavec$ in time polynomial in $|\ground|$ and $\log \left(1/\varepsilon\right)$, given a good bound on the search region for $\thetavec$.\footnote{In our applications we take
$\pvec=\alpha\xvec$ with a fixed $\alpha<1$, which places $\pvec$ in the relative interior of $\Ppoly$ and ensures that
an optimal dual solution exists with finite coordinates.} From $\thetavec$ we recover the optimal weights $\wvec$ and $\rho_e$ via
$w_e=\exp(\theta_e)$ and $\rho_e=w_e/(1+w_e)$.

However, establishing a good bound on the search region for $\thetavec$ is not always easy. This is the principal contribution of \cite{singh2014entropy}. Unfortunately, their running time is still polynomial in the inverse of the distance of $\pvec$ from the boundary of $\Ppoly$ (which can happen, for instance, if $\xvec$ has a particularly small coordinate). Furthermore, their results are not stated for the case of minimizing KL divergence, which is useful for us when working with weakly Rayleigh matroids. \cite{straszak2019maximum} remove this dependence on the distance from the boundary, and generalize to minimizing KL divergence.

Formally, fix a reference distribution $\dist_0\in\simplex{\Fset}$ with full support, i.e.\ $\dist_0(S)>0$ for all $S\in\Fset$,
and fix a target marginal vector $\pvec\in\Ppoly$.
Consider the convex program
\begin{equation}
\min\Bigl\{ \KL(\nu\|\mu_0):\ \nu\in\simplex{\Fset},\ \PP_{S\sim\nu}[e\in S]=p_e\ \ \forall e\in\ground\Bigr\},
\end{equation}
where $\KL(\nu\|\mu_0)\defeq \sum_{S\in\Fset}\nu(S)\log\!\bigl(\nu(S)/\mu_0(S)\bigr)$. 
Let $\dist^*$ denote an optimal solution.
When $\dist_0$ is uniform on $\Fset$, the objective differs from $\Entropy(\dist)$ by an additive constant,
so this is equivalent to the usual maximum-entropy program \eqref{eq:maxent-primal-template}.

For $\thetavec\in\mathbb{R}^{\ground}$ define the tilt of $\dist_0$ by
\vspace{-2mm}
\[
\dist_{\thetavec}(S)
\ \defeq\
\frac{\dist_0(S)\exp\!\bigl(\sum_{e\in S}\theta_e\bigr)}{\sum_{T\in\Fset}\dist_0(T)\exp\!\bigl(\sum_{e\in T}\theta_e\bigr).}
\qquad (S\in\Fset),
\]
and let $g_{\dist_0}(\wvec) = \sum_{S\in\Fset}\dist_0(S)\prod_{f\in S}w_f$, the weighted base generating polynomial. The work of \cite{straszak2019maximum} shows that access to a counting oracle for $g_{\dist_0}$ suffices to solve the convex program.

\begin{definition}[Approximate generalized counting oracle with cost $\mathsf{T}(\eta)$]
\label{def:approx-counting-oracle}
For $e\in\ground$ define the marginal sum
\vspace{-2mm}
\[
g_{\dist_0,e}(\wvec)\ \defeq\ \sum_{\substack{S\in\Fset:\\ S\ni e}}\dist_0(S)\prod_{f\in S}w_f.
\]
An oracle $\mathcal{O}$ takes as input $(\wvec,\eta)$ with $\wvec\in\mathbb{R}^{\ground}_{>0}$ and $\eta\in(0,1/2)$,
and returns numbers $\widehat g$ and $\{\widehat g_e\}_{e\in\ground}$ such that
\vspace{-2mm}
\[
(1-\eta)\,g_{\dist_0}(\wvec)\ \le\ \widehat g\ \le\ (1+\eta)\,g_{\dist_0}(\wvec),
\qquad
(1-\eta)\,g_{\dist_0,e}(\wvec)\ \le\ \widehat g_e\ \le\ (1+\eta)\,g_{\dist_0,e}(\wvec)\quad\forall e\in\ground,
\]
and runs in time $\mathsf{T}(\eta)$ (suppressing polynomial factors in $|\ground|$ and the bit-complexity of $\wvec$).
\end{definition}

We need two further technical definitions to state the formal result.

\begin{definition}[Unary facet complexity]
Let $\Ppoly\subseteq\mathbb{R}^{\ground}$ be a polytope with integer vertices.
Its \emph{unary facet complexity}, denoted $\mathrm{fc}(\Ppoly)$, is the smallest $M\in\mathbb{N}$
for which there exist finitely many vectors $a^{(j)}\in\mathbb{Z}^{\ground}$ with
$\|a^{(j)}\|_\infty\le M$, scalars $b_j\in\mathbb{R}$, and a linear subspace $H\subseteq\mathbb{R}^{\ground}$
such that
\[
\Ppoly=\Bigl(\{x\in\mathbb{R}^{\ground}:\ \ip{a^{(j)}}{x}\le b_j\ \text{for all }j\}\Bigr)\cap H.
\]
\end{definition}

\begin{definition}[Log-mass parameter]
Define
\[
L_{\dist_0}\ \defeq\ \max_{S\in\Fset}\ |\log \dist_0(S)|.
\]
\end{definition}

\begin{theorem}
\label{thm:compute-maxent-from-counting}
Let $\Fset\subseteq 2^{\ground}$ and let $\Ppoly=\conv\{\mathbf{1}_S:S\in\Fset\}$.
Let $M\defeq \mathrm{fc}(\Ppoly)$.
Fix any $\dist_0\in\simplex{\Fset}$ with full support and define $L_{\dist_0}$ as above. There exists an algorithm which, given:
(i) $\pvec\in\Ppoly$,
(ii) $\eps\in(0,1)$, and
(iii) oracle access to $\mathcal{O}$ for $\dist_0$,
outputs a vector $\thetavec\in\mathbb{R}^{\ground}$ such that
\[
\|\dist_{\thetavec}-\dist^*\|_1\ <\ \eps,
\]
where $\dist_{\thetavec}$ is the tilted Gibbs distribution above. Moreover, the algorithm guarantees a norm bound of the form
\[
\|\thetavec\|_2\ \le\ \mathrm{poly}\!\Bigl(|\ground|,\ M,\ L_{\dist_0},\ \log(1/\eps)\Bigr).
\]
The algorithm makes $\mathrm{poly}(|\ground|,M,L_{\dist_0},\log(1/\eps))$ oracle calls, all with a common accuracy parameter
\[
\eta\ =\ \frac{1}{\mathrm{poly}(|\ground|,1/\eps)},
\]
and hence runs in total time
\[
\mathrm{poly}\!\Bigl(|\ground|,\ M,\ L_{\dist_0},\ \log(1/\eps)\Bigr)\cdot \mathsf{T}(\eta).
\]
\end{theorem}

Note that while the results in \cite{straszak2019maximum} are stated for an evaluation oracle that can compute $g_{\mu_0}(\wvec)$ exactly, examination of the argument reveals that it also works for an approximate oracle, as long as we can also access values of $g_{\mu_0, e}(\wvec)$. This type of approximate counting oracle is what is used in \cite{singh2014entropy}.

Theorem~\ref{thm:compute-maxent-from-counting} makes
$\mathrm{poly}(|\ground|,M,L_{\dist_0},\log(1/\varepsilon))$ oracle calls, all at a common accuracy
$\eta = 1/\mathrm{poly}(|\ground|,1/\varepsilon)$.
Consequently:
\begin{itemize}[leftmargin=2em]
\item If the generalized counting oracle can be answered to relative error $\eta$ in
time $\mathrm{poly}(|\ground|,\log(1/\eta))$ (a high-precision, or evaluation, oracle model), then the
Gibbs parameters can be computed in time $\mathrm{poly}(|\ground|,\log(1/\varepsilon))$. We note that in this case, it actually suffices to have an oracle for $g_{\mu_0}$ alone, since $g_{\mu_0, e}$ can be calculated by interpolation, see \cite{straszak2019maximum} for details.
\item If the best available oracle is only an FPRAS-type primitive with running time
$\mathrm{poly}(|\ground|,1/\eta)$, then the same reduction yields
running time polynomial in $1/\varepsilon$.
\end{itemize}

\subsection{Proof of Theorem~\ref{thm:polytime-matchings-uniform-matroids} for matchings and uniform matroids, and discussion of hypergraph matchings with vertex capacities}

We now discuss for matchings and uniform matroids the (non-)existence of the needed approximate counting oracles, and our ability to sample from the max-entropy distribution. We will also indicate which oracle regime applies in each feasibility environment.

\xhdr{Matchings, bipartite matchings, and hypergraph matchings.} For general matchings (and hence bipartite matchings) classical MCMC methods give an FPRAS for the matching partition function, and hence for the functions $g_{\dist_0}$ and $g_{\dist_0,e}$ (see chapter 12, \cite{jerrum_sinclair_mcmc}). The running time is polynomial in the problem size and
 $1/\eta$ for relative error~$\eta$. Plugging this into
Theorem~\ref{thm:compute-maxent-from-counting} yields a randomized polynomial-time procedure for computing
the Gibbs parameters needed by Algorithm~\ref{alg:str-gibbs}, with overall running time polynomial in
$1/\varepsilon$ and $\log \left(1/\delta\right)$ where $\delta$ is the allowed probability of error (since the unary facet complexity of the matching polytope is just $1$, and the log-mass parameter is polynomially bounded). Note that the counting oracle being randomized is not an issue since we only make polynomially many calls to the oracle  (the success probability can always be boosted appropriately).

Once we have access to the Gibbs parameters, there is a rapidly mixing Markov chain whose stationary distribution is the Gibbs distribution; see Chapter 12 in \cite{jerrum_sinclair_mcmc}.
Thus we can sample $\widehat{\Sset}\sim\dist^*$ up to total-variation error $\varepsilon$ in time polynomial in the
instance size and $\log(1/\varepsilon)$.

Putting these pieces together yields an OCRS for matchings that obtains a selectability of $1/3 - \varepsilon$ (for general matchings) or $(3-\sqrt{5})/2 - \varepsilon$ (for bipartite matchings), that runs in time polynomial in the problem size and $1/\varepsilon$. Note that most of the time is spent `preprocessing' and calculating the Gibbs parameters, and both sampling the Gibbs distribution, and actually running the algorithm are fast.

In the general case of hypergraph matchings, there is no efficient counting oracle, nor is the unary facet complexity necessarily polynomially bounded, so we do not claim a polynomial time algorithm. In special subclasses of hypergraphs, there are sometimes evaluation oracles and bounds on the facet complexity, in which case solving the maximum entropy program in polynomial time is possible. For such classes, we can also utilize a standard counting to sampling reduction of \citet{jerrum_valiant_vazirani_1986} to sample from the Gibbs distribution and then run our algorithm.

\xhdr{$\boldsymbol{k}$-uniform matroids.} In this case, a simple DP based method provides an evaluation oracle, and we can then apply Theorem~\ref{thm:compute-maxent-from-counting}. But \cite{chen1994weighted} also provide a direct method to compute the  Gibbs parameters needed by Algorithm~\ref{alg:str-gibbs} in polynomial time to any needed precision, i.e., to an error of $\varepsilon$ in time polynomial in $\log(1/\varepsilon)$. They also show how to sample from the resulting Gibbs distribution in polynomial time. Thus, we have an OCRS for $k$-uniform matroids with a selectability of $\alpha_k - \varepsilon$ that runs in time polynomial in the problem parameters and $\log (1/\varepsilon)$. 

\xhdr{Hypergraph matchings with vertex capacities.} For our S-OCRS for hypergraph matchings with vertex capacities discussed in \Cref{sec:NRM}, $\dist$ is a Gibbs distribution with constants $\rho_e=\gamma x_e$, where $\gamma$ was an explicit constant. Like in the previous feasibility environments, simulate-then-replace takes the particularly simple form given by Algorithm \ref{alg:str-gibbs}. There are no online computations, since $\rho_e/x_e$ is constant. Thus, the only requirement to implement the algorithm efficiently is that we sample (approximately) an initial set $\widehat{\Sset}\sim\dist$ in polynomial time. Unfortunately, even knowing $\rho_e$ there is no general efficient way to sample from this Gibbs distribution, so we do not claim a polynomial time algorithm (though we give a different poly-time OCRS in \Cref{sec:NRM}). Once again, if we have access to an appropriate counting oracle, and bounds on the facet complexity, we can utilize the standard counting to sampling reduction \citep{jerrum_valiant_vazirani_1986} to sample from the Gibbs distribution and then run our algorithm.

\subsection{\revcolor{Knapsack constraints}}
\label{app:polytime-knapsack}
For the knapsack constraint, we have the following theorem:

\begin{theorem}[Polynomial-time S-OCRS for knapsack constraint]
\label{thm:polytime-knapsack}
For every $\varepsilon>0$, there is a $(\tfrac14-\varepsilon)$-selectable S-OCRS for the knapsack feasibility environment $(\ground,\Fset)$ that runs in time $\mathrm{poly}(|\ground|,1/\varepsilon)$.
\end{theorem}

The S-OCRS for knapsack constraints discussed in \cref{sec:knapsack-constraints} is a mixture of two different maximum-entropy distributions, $\mu^H$ and $\mu^L$. Accordingly, simulate-then-replace involves tossing a coin, and choosing to run Algorithm \ref{alg:str-gibbs} with either the large item S-OCRS, or the small item S-OCRS. For each of these S-OCRS, we need to be able to sample from an initial distribution, and compute appropriate conditional probabilities.

Unfortunately for the knapsack polytope, we cannot invoke \Cref{thm:compute-maxent-from-counting} directly, since there is no polynomial-sized bound on the unary facet complexity. Instead, we apply Theorem 2.8 from \citet{singh2014entropy}. This result is analogous to \Cref{thm:compute-maxent-from-counting}, but does not require a bound on the facet complexity; instead there is an assumption on the marginals lying in the relative interior of the polytope, and the runtime of the algorithm is dependent on the distance of the marginals from the boundary.

To get around this dependence, we start by discarding each active item with probability $\varepsilon$; this effectively scales the $x_e$ down by a factor of $1 -\varepsilon$ and impacts the selectability by a factor of at most $1 -\varepsilon$, but lets us assume that $x_e \leq 1- \varepsilon$. We will also add the items with $x_e \leq \frac{\varepsilon}{|E|}$ to the heavy items; this does not change any of the calculations in \cref{sec:knapsack-constraints}, except for the bound on $X+2B$, which is now bounded by $2 + \varepsilon$ rather than $2$.

For the heavy item S-OCRS, we select at most one item, and we calculated in \Cref{sec:knapsack-constraints} that:
$\PP_{S\sim\mu^{H}}[e\in S\mid S_{-e}=\emptyset] = \frac{x_e}{1+x_e}.$ It is thus straightforward to sample an initial state from this distribution, and to process each arrival in polynomial time. For the light-item S-OCRS, we  have:
\[ E_{L}:=\{e\in E:a_e\le 1/2, x_e > \varepsilon/|E|\}, \quad \Fset_{L}:=\{T\subseteq E_{L}:a(T)\le 1\}.
\] The construction in \Cref{sec:knapsack-constraints} chose $c_{L}:=\frac{1}{1+2\sum_{e\in E_{L}}a_e x_e}$, let $p_e=c_{L}x_e$ for $e\in E_{L}$, and $\mu^{L}\in\Delta(\Fset_{L})$ was the maximum-entropy distribution with marginals $p$. Once the constants $\{\rho_e\}$ associated with this maximum-entropy distribution and an initial sample $\widehat{\Sset}\sim\dist^L$ are available, each arrival can be processed in polynomial time, just as discussed in \Cref{app:polytime-maxent-template}. Therefore it remains to show how to compute these parameters and sample the initial state in polynomial time. 

Now we know that $p_e = c_Lx_e$, and $c_L \leq 1$. Moreover, $B\le1$ implies $c_L\ge1/3$. So we conclude that:
\[
\frac{\varepsilon}{3|E|}\leq p_e \le x_e \le 1 - \varepsilon.
\]
Thus, $\pvec$ lies in the relative interior of the knapsack polytope, and the distance from the boundary is polynomial in $|E|$ and $\frac{1}{\varepsilon}$. Next, we must find approximate counting oracles for the knapsack polytope. For weights $\wvec=(w_e)_{e\in E_L}$, define
\[g(\wvec)
:=
\sum_{S\in\Fset_L}\prod_{e\in S}w_e,
\quad
g_e(\wvec)
:=
\sum_{\substack{S\in\Fset_L\\e\in S}}\prod_{f\in S}w_f.
\]
Let $(Y_e)_{e\in E_L}$ be independent random variables with $\PP[Y_e=1]=\frac{w_e}{1+w_e}$, and write $R=\{e:Y_e=1\}$. Then
\[
g(\wvec)
=
\left(\prod_{e\in E_L}(1+w_e)\right)
\PP[a(R)\le1], \quad g_e(\wvec)
=
\left(\prod_{f\in E_L}(1+w_f)\right)
\PP[e\in R,\ a(R)\le1].
\]
Theorem~3.6 of \citet{gopalan2011fptas} computes an approximation to $\PP[a(R)\le1]$ in time polynomial in $|E|$, the input encoding length, and the inverse relative error. The same theorem computes an approximation for $g_e$ after fixing $Y_e=1$ and reducing the capacity by
$a_e$.

The approximate counting oracle, together with bound on the distance of $\pvec$ from the boundary of the polytope mean we can apply Theorem 2.8 in \citet{singh2014entropy} to compute the needed Gibbs distribution to accuracy $\varepsilon$ in time polynomial in $|E|$ and $1/\varepsilon$. Once the approximate Gibbs parameters $\wvec$ are available, we can approximately sample from the
corresponding Gibbs distribution using the standard counting-to-sampling self-reduction
\citep{jerrum_valiant_vazirani_1986}.

\subsection{\revcolor{Weakly Rayleigh matroids}}
\label{app:polytime-weakly-rayleigh}
The weakly Rayleigh matroid construction differs from the first three environments: the distribution $\dist^{*}$ is \emph{not} the maximum-entropy distribution on independent sets. Instead, the construction proceeds by (i) finding a distribution on bases with prescribed marginals (a KL projection of a Rayleigh base measure), and (ii) applying independent thinning inside a sampled base. We therefore discuss its algorithmic implementation separately. 

Implementing our algorithm requires certain natural assumptions---satisfied by a broad subclass of weakly Rayleigh matroids. The following theorem consolidates the  assumptions we need to obtain a polynomial-time computational guarantee for the modified template that we use for weakly Rayleigh matroids.
 
\begin{theorem}[Polynomial-time S-OCRS for weakly Rayleigh matroids]
\label{thm:polytime-weakly-rayleigh}
Let $\cM=(\ground,\Iset)$ be a weakly Rayleigh matroid (Definition~\ref{def:weakly-rayleigh}) of rank~$r$ on $n:=\lvert E\rvert$~elements, with Rayleigh base measure $\dist_0$. Suppose the following conditions hold:
\begin{enumerate}[leftmargin=2em,label=(\roman*)]
\item \textbf{Rank oracle:} There is an oracle that, given $T\subseteq \ground$, returns $\RankOracle{T}$ in polynomial time.
\item \textbf{High precision/ evaluation oracle:} There is an oracle that, given $\wvec\in\mathbb{R}^{\ground}_{>0}$ and accuracy parameter $\eta\in(0,1/2)$, returns a multiplicative $(1\pm\eta)$-approximation to the weighted base generating polynomial, that is, 
\[
g_{\dist_0}(\wvec)\ \defeq\ \sum_{B\in\mathcal B}\dist_0(B)\prod_{f\in B}w_f~,
\]
in time polynomial in $n, \log(1/\eta)$, and the bit complexity of $\wvec$.
\item \textbf{Bounded log-mass:} The log-mass parameter $L_{\dist_0}\defeq \max_{B\in\Bset}|\log \dist_0(B)|$ is polynomially bounded in $n$.
\end{enumerate}
Then, for every $\xvec\in\Ppoly$ and every $\varepsilon>0$, there is a $(\tfrac12-\varepsilon)$-selectable S-OCRS for the matroid feasibility environment $(\ground,\Fset)$ that runs in time $\mathrm{poly}(n,\log(1/\varepsilon))$.\footnote{Note the assumption (i) is not strictly speaking necessary, since a rank oracle can be implemented using an evaluation oracle under the assumption of bounded log-mass.}
\end{theorem}
 
\begin{remark}[$\mathbb{C}$-representable matroids satisfy the oracle assumptions]
\label{rem:specific-matroids}
Any subclass of weakly Rayleigh matroids for which an efficient evaluation oracle exists and for which the log-mass parameter is polynomially bounded falls within the scope of Theorem~\ref{thm:polytime-weakly-rayleigh}. The conditions of Theorem~\ref{thm:polytime-weakly-rayleigh} are satisfied in particular by
$\mathbb{C}$-representable matroids. 

If $\cM$ is represented by a matrix $A\in\mathbb{C}^{r\times n}$, there is a natural base measure that is \textit{determinantal} (see \cite{lyons_determinantal} for a discussion): $\dist_0(B)\propto\det(A_B^*A_B)$, and this measure is Rayleigh. The Cauchy--Binet formula gives that for every $\wvec\in\R^n_{>0}$, we have $g_{\dist_0}(\wvec)=\det(A\,\mathrm{diag}(\wvec)\,A^*)/\det(AA^*)$, computable in $\mathrm{poly}(n,\log(1/\eta))$ time via standard determinant computation. The log-mass bound follows from Hadamard's inequality. This class includes graphic matroids and laminar matroids.

For graphic matroids in particular, we can let $\mu_0$ be uniform on spanning trees. The evaluation oracle reduces to computing the weighted spanning tree polynomial, which equals a determinant of a weighted Laplacian by Kirchhoff's matrix-tree theorem.
\end{remark}

\smallskip
 
The remainder of this subsection is devoted to the proof of Theorem~\ref{thm:polytime-weakly-rayleigh}.

\begin{proofof}{Proof of Theorem~\ref{thm:polytime-weakly-rayleigh}} 

We start by recalling the construction of $\dist^*$.
 Fix $\xvec\in\Ppoly$. First, compute a dominating base point $\qvec\in\Bpoly$ with $\qvec\ge \xvec$ (Lemma~\ref{lem:dominating-base-point}).
Next, construct a probability measure $\mu$ on bases $\Bset$ such that
\[
\PP_{B\sim\mu}[e\in B]=q_e \quad \forall e\in\ground,
\qquad\text{and}\qquad
\mu \text{ is Rayleigh (Definition~\ref{def:weakly-rayleigh}).}
\]
When $\qvec$ lies in the relative interior of $\Bpoly$, such a measure $\mu$ can be obtained as the unique minimizer of the
KL projection program~\eqref{eq:KL-proj-bases}, and it has the exponential-family form
$\mu=\mu_{\wvec}$ for some $\wvec\in\mathbb{R}^{\ground}_{>0}$.
When $\qvec$ lies on the boundary of $\Bpoly$, an exact exponential tilt need not exist; in this case we take any sequence
$\qvec^{(t)}$ in the relative interior of $\Bpoly$ with $\qvec^{(t)}\to \qvec$, form the corresponding tilted measures
$\mu^{(t)}=\mu_{\wvec^{(t)}}$, and pass to a limit along a convergent subsequence.
Since $\Bset$ is finite, this produces a well-defined limiting base measure $\mu$ with marginals $\qvec$; moreover the
Rayleigh inequalities are preserved under this limit, so $\mu$ is Rayleigh. (Note that when we try to compute $\wvec$, we will not take this limit, but just solve the program approximately even when $\qvec$ lies on the boundary.)
 
Finally, define the retention probabilities
\[
\tau_e\ \defeq\
\begin{cases}
\dfrac{x_e}{2q_e}, & q_e>0,\\[6pt]
0, & q_e=0,
\end{cases}
\]
and generate $\Sset\sim\dist^*$ by drawing $B\sim\mu$ and then keeping each element $e\in B$ independently with
probability $\tau_e$.
 
\medskip
 
\textbf{Simulate-then-replace for $\bm{\dist^*}$.}
Pre-sample $\widehat{\Sset}\sim\dist^*$.
When $e$ arrives, let $T=\Sm{\widehat{\Sset}}{e}$ and define the target conditional inclusion probability
\[
\pi_e(T)\ \defeq\ \PP_{\Sset\sim\dist^*}\!\big[e\in\Sset \given \Sm{\Sset}{e}=T\big]
\ =\ \frac{\dist^*(T\cup\{e\})}{\dist^*(T)+\dist^*(T\cup\{e\})}.
\]
Since $\dist^*$ is supported on $\Iset$, we have $\pi_e(T)=0$ whenever $T\cup\{e\}\notin\Iset$.
Now:
\begin{itemize}[leftmargin=2em]
\item If $e\notin\Aset$, set $\widehat{\Sset}\leftarrow T$.
\item If $e\in\Aset$ and $T\cup\{e\}\notin\Iset$, set $\widehat{\Sset}\leftarrow T$.
\item If $e\in\Aset$ and $T\cup\{e\}\in\Iset$, accept $e$ with probability $\pi_e(T)/x_e$ and update
\[
\widehat{\Sset}\leftarrow
\begin{cases}
T\cup\{e\}, & \text{if accepted},\\
T, & \text{if rejected}.
\end{cases}
\]
\end{itemize}
 
Clearly, the online computations are (a) determining whether $T\cup\{e\}\in\Iset$ (an independence test in~$\cM$) and
(b) sampling a Bernoulli draw with parameter $\pi_e(T)/x_e$.
Thus, once an initial sample $\widehat{\Sset}\sim\dist^*$ is available and we can evaluate (or approximate) the
conditionals $\pi_e(T)$ for the encountered sets~$T$, each arrival can be processed in polynomial time.
 
The remainder of the proof is devoted to explaining:
\begin{enumerate}[leftmargin=2em,label=(\roman*)]
\item how we compute a dominating base point $\qvec\in\Bpoly$ and how we approximate a Rayleigh base measure $\mu$ on $\Bset$ with
marginals $\qvec$; and
\item how we can approximately sample from $\mu$ (and hence from $\dist^*$) and evaluate the ratios
$\dist^*(T\cup\{e\})/\dist^*(T)$ needed to compute $\pi_e(T)$.
\end{enumerate}
 
\xhdr{Step (i): Finding a dominating base point.} Let us start by noting that the first step of calculating an $\varepsilon$-approximate base point $\qvec\in\Bpoly$ which dominates $\xvec$ can be done in time polynomial in $\log(1/\varepsilon)$. Indeed, it suffices to obtain a maximizer $\sum_e y_e$ over the polytope:
\[
Q \ \defeq\ \Bigl\{\yvec\in [0,1]^\ground:\ \sum_{e\in T}y_e\le \RankOracle{T}\ \forall T\subseteq \ground,\ \ \yvec\ge \xvec\Bigr\},
\]
and this is possible using the rank oracle, via the ellipsoid method (see for instance \cite{gls88}). Note that we have already shown in \Cref{apx:ind-to-base} that the optimizer necessarily satisfies $\sum_e y_e = r$.
 
\xhdr{Step (ii): KL projection onto a Rayleigh base measure.} Here, we use access to our evaluation oracle, and Theorem~\ref{thm:compute-maxent-from-counting} in a black box manner, to find an $\varepsilon$-approximate Rayleigh measure $\mu$ with the (approximately) correct marginals in running time polynomial in $\log(1/\varepsilon).$ (Note that the matroid base polytope has unary facet complexity 1, Theorem~\ref{thm:compute-maxent-from-counting} accepts points on the boundary of the polytope as input, and the discussion after Theorem~\ref{thm:compute-maxent-from-counting} clarifies that a high-precision or evaluation oracle for $g_{\mu_0}$ suffices to evaluate $g_{\mu_0, e}$ using interpolation.) At the end of this step, we have access to approximate Gibbs parameters $\wvec$. 
 
\xhdr{Step (iii): Sampling $\bm{B\sim \mu}$ and sampling from $\bm{\mu^{*}}$.} 
To sample from $\mu$, we follow a standard counting-to-sampling self-reduction template (see \cite{jerrum_valiant_vazirani_1986}). For disjoint sets $I,J\subseteq E$, define the constrained function
\[
g_{\mu_0, J, I}(\wvec) :=\ \sum_{B\in\mathcal B:\ I\subseteq B,\ B\cap J=\emptyset}\ \mu_0(B)\prod_{f\in B} w_f.
\]
Consider the bivariate polynomial
\[
F_{I,J}(t,u)\ :=\ g_{\mu_0}(\wvec(t,u)),
\quad\text{where}\quad
w_i(t,u)=
\begin{cases}
t w_i,& i\in I,\\
u w_i,& i\in J,\\
w_i,& \text{otherwise}.
\end{cases}
\]
Since $g$ is multi-affine, $F_{I,J}(t,u)$ has $\deg_t\le |I|$ and $\deg_u\le |J|$, we can check
directly that
\[
g_{\mu_0, J, I}\ =\ [t^{|I|}u^0]\, F_{I,J}(t,u).
\]
Let $d:=|I|$ and $m:=|J|$.  Using univariate interpolation at the positive points
$\{1,2,\dots,m+1\}$ in the $u$-variable and a $d$-th forward-difference identity in the $t$-variable,
one may write $g_{\mu_0, J, I}$ as the following explicit linear combination of evaluations of
$g_{\mu_0}(\cdot)$ at positive points:
\begin{equation}
    \label{eq:explicit-interp-formula}
    g_{\mu_0,J,I}(\wvec)
\;=\;
\frac{1}{d!}\sum_{a=0}^{d}(-1)^{d-a}\binom{d}{a}
\sum_{b=1}^{m+1}(-1)^{b-1}\binom{m+1}{b}\;
g_{\mu_0}\!\bigl(\wvec(1+a,b)\bigr),
\end{equation}
Thus, with access to an oracle that can evaluate $g_{\mu_0}(\cdot)$ to relative error $\eta$ in time polynomial in the problem parameters and 
$\log(1/\eta)$, each constrained value $g_{\mu_0,J,I}(\wvec)$ can also be computed to relative error $\eta$ in the same polynomial time. 

To achieve this, first of all, note that we can detect whether the polynomial is $0$ using the rank oracle for the matroid. If it is non-zero, the bound on the log-mass guarantees that the value cannot be less than exponentially small. We can therefore choose the evaluation accuracy used for the terms $g_{\mu_0}(\wvec(1+a,b))$ sufficiently small so that the resulting error in the linear combination
translates to the desired relative accuracy for $g_{\mu_0,J,I}(\wvec)$.
Requesting this higher precision increases the running time only polynomially, due to the fact that the sum of the coefficients of $g_{\mu_0,J,I}(\wvec)$ is bounded exponentially in the size of the matroid, and we use a high-precision oracle. 

Note that a similar argument is used in \cite{straszak2019maximum} to calculate marginals.
 
Now fix any ordering $1,2,\dots,n$ and run the following procedure:
initialize $I\gets\emptyset$ and $J\gets\emptyset$.
For $k=1,2,\dots,n$, compute
\[
Z_1 \ :=\ g_{\mu_0,J,I\cup\{k\}}(\wvec),\qquad
Z_0 \ :=\ g_{\mu_0,J\cup\{k\},I}(\wvec),
\]
(using only evaluations of $g_{\mu_0}(\cdot)$ via \eqref{eq:explicit-interp-formula}).
Define
\[
\pi_k\ :=\ \PP_{\;B\sim\mu}\!\bigl[k\in B\mid I\subseteq B,\ B\cap J=\emptyset\bigr]
\ =\ \frac{Z_1}{Z_1+Z_0}
\ =\ \frac{Z_1}{g_{\mu_0,J,I}(\wvec)}.
\]
Include $k$ in $I$ with probability $\pi_k$ and otherwise add it to $J$.
Lastly, output $B:=I$.  If, at each step $k$, we use an approximate conditionals $\tilde\pi_k$ satisfying $|\tilde\pi_k-\pi_k|\ \le\ \varepsilon/n$,
the distribution of the output base $\tilde B$ is within $\varepsilon$ in total variation distance of $\mu$ by a standard step-by-step coupling argument.
Thus we obtain an approximate sample of $\mu$ in time polynomial in $\log(1/\varepsilon)$.\footnote{We note that in some special cases, we can also sample using a random walk on the bases, see \cite{anari2021log}, for instance.}
 
Then, sample from $\mu^{*}$ by keeping each element of the base with probability $\tau_e = \frac{x_e}{2q_e}$.
 
\xhdr{Step (iv): Compute $\bm{\mu^{*}}$(T) for any set:} Fix $T\subseteq[n]$.  Clearly we have:
\[
\mu^*(T)
=\sum_{B\supseteq T} \mu(B)\Big(\prod_{i\in T} \tau_i\Big)\Big(\prod_{i\in B\setminus T}(1-\tau_i)\Big).
\]
(Recall that $\tau_i = \frac{x_i}{2q_i}$.) To calculate $\mu^*(T)$, we follow a similar interpolation strategy as above. Define a weight vector $\wvec^{(T)}(t)\in\R^n_{>0}$ by
\vspace{-2mm}
\[
w^{(T)}_i(t)\ :=\
\begin{cases}
t\,w_i, & i\in T,\\
w_i(1-\tau_i), & i\notin T,
\end{cases}
\qquad (t>0).
\]
The polynomial $F_T(t):=g_{\mu_0}(\wvec^{(T)}(t))$ has degree at most $|T|$ and 
\[
[t^{|T|}]\, g_{\mu_0}(\wvec^{(T)}(t))
\ =\ \sum_{B\supseteq T}\mu_0(B)\Big(\prod_{i\in T}w_i\Big)\Big(\prod_{i\in B\setminus T} w_i(1-\tau_i)\Big),
\]
which means
\vspace{-2mm}
\[
\mu^{*}(T)
\ =\ \frac{\prod_{i\in T}\tau_i}{g_{\mu_0}(\wvec)}  \cdot [t^{|T|}]\, g_{\mu_0}(\wvec^{(T)}(t)).
\]

Let $d:=|T|$. Since $F_T(t)=g_{\mu_0}(\wvec^{(T)}(t))$ has degree at most $d$, we can calculate its  coefficients from positive evaluations using the same interpolation trick as before:
\[
[t^{d}]\,F_T(t)
\;=\;
\frac{1}{d!}\sum_{a=0}^{d}(-1)^{d-a}\binom{d}{a}\,F_T(1+a),
\]
so computing $\mu^{*}(T)$ reduces to evaluations of $g_{\mu_0}$. We can therefore compute $\mu^{*}(T)$ (and the ratio $\frac{\mu^{*}(T\cup\{e\})}{\mu^{*}(T)+\mu^{*}(T\cup\{e\})}$ to accuracy $\varepsilon$ in time polynomial in $\log(1/\varepsilon))$. This is exactly the acceptance probability used by the online simulate-then-replace policy. Combining Steps~(i)--(iv) completes the proof of Theorem~\ref{thm:polytime-weakly-rayleigh}: under the stated oracle assumptions, we obtain a $(\tfrac{1}{2}-\varepsilon)$-selectable S-OCRS for weakly Rayleigh matroids, running in time $\mathrm{poly}(n,\log(1/\varepsilon))$.
\end{proofof}

\end{document}